\documentclass[10pt,a4paper]{article}
\usepackage{graphpap,epsfig,graphicx,textcomp,xcolor}
\begin{document}

\begin{center}
{\Large {\bf Nonequilibrium tricritical behaviour in anisotropic XY ferromagnet driven by elliptically polarised propagating magnetic field wave}}\end{center}

\vskip 0.5cm

\begin{center}{\it Olivia Mallick$^1$ and Muktish Acharyya$^{2,*}$}\\
{\it Department of Physics, Presidency University,}\\
{\it 86/1 College Street, Kolkata-700073, INDIA}\\
{$^1$E-mail:mallickolivia0@gmail.com}\\
{$^2$E-mail:muktish.physics@presiuniv.ac.in}\end{center}

\vskip 0.5cm

\noindent {\bf Abstract:} 
The three-dimensional anisotropic XY ferromagnet, driven by an elliptically polarized propagating magnetic field wave, has been extensively investigated by Monte Carlo simulation with the Metropolis single spin flip algorithm. Both the effects of the bilinear exchange type and the single-site anisotropies are thoroughly investigated. The time-averaged magnetization (over the complete cycle of the elliptically polarized propagating magnetic field wave) components play the role of the dynamic order parameter. For a fixed set of values of the strength of anisotropy and the field amplitudes, the system has been found to get dynamically ordered at a pseudocritical temperature. The pseudocritical temperature of such a dynamic nonequilibrium phase transition has been found to depend both on the strength of anisotropy and the amplitudes of the elliptically polarized propagating magnetic field wave. A comprehensive phase diagram is represented here in the form of an image plot of the pseudocritical temperature in the plane formed by the strength of anisotropy and field amplitudes. Interestingly, this nonequilibrium phase transition has been found to be discontinuous (first order) for higher values of the field amplitude. On the other hand, the continuous (second order) transition has been noticed for lower values of the field amplitude. Such an interesting nonequilibrium {\it tricritical} behavior has been observed in driven XY ferromagnet. The order of such a nonequilibrium phase transition has been confirmed by the thermal variation (near the transition) of the statistical distribution of the order parameter and by the thermal variation of the fourth-order Binder cumulant. In the plane formed by field amplitude and anisotropy, a {\it tricritical} line has been shown as the accompanying (and complementary) phase diagram. The dependence of the pseudocritical temperature, on the frequency and wavelength of the elliptically polarized propagating magnetic field wave, has also been reported. 
\vskip 0.5cm


\vskip 0.5cm

\vskip 1cm

\noindent $^*$ Corresponding author
\newpage

\noindent {\bf {I. Introduction:}}

\noindent The specially `ordered' phase with lack of long range magnetic ordering in SO(2) symmetric XY ferromagnet in two dimensions, is a breakthrough in the research of thermodynamic conventional phase transition\cite{stanley}. This specially `ordered' Kosterlitz-Thouless(KT) phase\cite{kosterlitz1,kosterlitz2} has drawn the intense attention of modern researchers in this field. The KT transitions are observed in a wide variety of physical systems, namely, superconducting films\cite{superconductors} , liquid crystals\cite{liquid1,liquid2} etc.

To understand the equilibrium phase transition in a three-dimensional XY ferromagnet and to know its universality class, the critical behavior of the XY ferromagnet has been studied \cite{hasenbusch1,hasenbusch2} in three dimensions by 
Monte Carlo (MC) simulation.  The XY phases show interesting behaviors for various kinds of interactions. The additional antinematic  interaction\cite{zukuvic1} along with ferromagnetic interaction, an interesting peculiar low temperature canted ferromagnetic phase (resulting from the competition between the collinear ferromagnetic and noncollinear antinematic ordering tendencies), has been observed. A complex canted antiferromagnetic phase has been identified in the Monte Carlo study\cite{zukovic2} of the geometrically frustrated XY model in triangular lattice. The antiferromagnetic XY model (with higher order interactions) on a triangular lattice has also been investigated\cite{zukovic3} by Monte Carlo simulation to find the low-temperature phases having chiral long-range ordering. The layered XY antiferromagnet (with the intra planar ferromagnetic and interplanar antiferromagnetic interaction) has shown rich phase diagrams in a recent Monte Carlo study\cite{erol1}.

 So far, we have discussed the interesting phases observed in XY model (pure) with various kinds of interactions. What will be the role of the quenched disorder in such XY phases? The disorder can be implemented by random anisotropy, random impurity, random field etc. Very recently, the role of such disorder, has been investigated\cite{olivia1} in three dimensional anisotropic XY ferromagnet through extensive Monte Carlo simulation. The random bilinear exchange kind of disorder compels the system to get ordered at lower temperatures. The ordering at lower temperatures, in the random field anisotropic (constant bilinear exchange type) XY model has also been observed\cite{olivia2}. A compensating field (to preserve the critical temperature of the three dimensional isotropic XY ferromagnet) has been found as a linear function of the strength of the constant bilinear exchange anisotropy. The randomly quenched nonmagnetic impurity also plays a role of disorder in the system. A recent Monte Carlo study\cite{olivia3} on three-dimensional anisotropic XY ferromagnet shows that the critical temperature increases linearly with the strength of anisotropy (both bilinear exchange type and single site type). It may be noted here that the nonlinear increase of the critical temperature has been observed earlier for single site anisotropy in finite temperature quantum field theoretic calculation \cite{dtc}.

 The kinetic aspects \cite{redner} of cooperatively interacting many body systems show many interesting behaviors that cannot be observed in equilibrium responses discussed so far. Let us briefly discuss here the kinetic behavior of XY models. The dynamical behaviors of XY models have been studied in the last few decades. Historically, the critical dynamical behavior of the two-dimensional XY model has been investigated\cite{ludovic} by spin-wave analysis and the Monte Carlo method. The initial condition-dependent aging dynamics and violation of the fluctuation-dissipation theorem were observed. The nonequilibrium steady state of the XY chain has been studied\cite{aschbacher} for two different endpoint temperatures. Recently, the dynamic critical behavior of XY chain (with long-range interaction) has been investigated\cite{yang} through Monte Carlo simulation and found the 
mean field universality class, as expected. However, they suggested in their MC study that the small value of the trial angle range has to be used to detect the dynamical critical behavior properly. Recently, the entanglement witnesses have been used\cite{igloi} to detect the entanglement in XY chain in thermal equilibrium. They have found that the energy-based entanglement witnesses are efficient in detecting the nearest-neighbor entanglement in spin chains in various circumstances. The nonequilibrium state of positive entropy production has been found. The off-equilibrium dynamics of two-dimensional\cite{abriet1} and three-dimensional \cite{abriet2} systems have been studied by Monte Carlo simulation to measure the extent of violation of fluctuation-dissipation from the two-time autocorrelation function and linear response functions. The XY model on a small world network has been investigated \cite{kateryna} by kinetic Monte Carlo simulation to find that the dynamic universality class is of mean field type. The generic hydrodynamic equations of the nonequilibrium XY model have been derived\cite{sartori} for different parameters ($N$). They have found the long-range order of the flocking phase of dry polar active fluids for $N=1$ and the quasi-long-range order of the equilibrium XY model in low temperature for higher values of $N$. The domain growth and aging are studied\cite{puri} in the two and three dimensional random field XY model by Monte Carlo simulation. The asymptotic logarithmic growth law has been proposed with positive exponents.

 The off-equilibrium dynamical aspects of many body thermodynamic systems have a typical scale of time called `relaxation time'. This is a characteristic time to determine how fast (or slow) the system restores its equilibrium state after the removal of external perturbation (responsible for setting the system out of equilibrium). What will happen if such an external perturbation is time-dependent? Such time-dependent external perturbation always keeps the system far from equilibrium. The system gets no chance to achieve its original equilibrium state. A prototypical example may be the Ising ferromagnet in the presence of an oscillating magnetic field. An interesting non-equilibrium phase transition has been found to take place. The time average magnetization (being the order parameter) has been found to become non-zero at a finite temperature, causing the dynamic phase transition. A considerable amount of literature has been developed\cite{rmp,rv} through systematic investigations on such non-equilibrium phase transitions in various magnetic models, and the work in this field is still going on. Although the nonequilibrium phase transition in discrete symmetric spin models (Ising, Blume-Capel, etc.) has been studied rigorously, the continuous symmetric spin models (XY, Heisenberg, etc.) have drawn the attention of the researchers. Let us briefly mention here a few studies of the non-equilibrium phase transitions in the continuous symmetric models. The classical anisotropic Heisenberg ferromagnet, driven by a sinusoidally oscillating (in time but uniform over space) magnetic field, showed both off-axial\cite{ofx} and axial\cite{avx} non-equilibrium phase transitions depending on the strength of anisotropy. The anisotropic XY ferromagnet driven by a sinusoidally oscillating magnetic field has been analyzed\cite{yasui} by the Ginzburg-Landau mean field model and found to have a dynamic phase transition depending on the frequency and amplitude of the driving magnetic field. The thin ferromagnetic film of the Heisenberg spin model with bilinear exchange anisotropy driven by a periodically pulsed magnetic field has been investigated\cite{hall1} by Monte Carlo simulation. It is observed that the thin films with competing surface fields showed\cite{hall1} separate and distinct dynamic phase transitions for the bulk and surface layers of the film. Moreover, the results showed\cite{hall2} that the films exhibit a single discontinuous dynamic phase transition as a function of the anisotropy of the bilinear exchange interaction in the Hamiltonian. Multiple non-equilibrium phase transitions have been observed\cite{mdt} in the three-dimensional anisotropic Heisenberg ferromagnet driven by {\it elliptically} polarized time-dependent (with no space dependence at any instant of time) magnetic field. The comprehensive phase diagram shows\cite{multi} both continuous and discontinuous nonequilibrium phase transitions. Various interesting nonequilibrium phases have been found recently\cite{arnab} in two dimensional XY ferromagnet with uniaxial anisotropy driven by time periodic (no spatial variation at any particular instant of time) analyzed by the Glauber Monte Carlo algorithm and phenomenological mean field equation. The recent studies\cite{tharnier} on the nonequilibrium phase transitions in transverse field anisotropic (bilinear exchange type) XY ferromagnet (where the two ends of the XY chain are kept in contact with different magnetic potentials to keep the system in an off-equilibrium state) are worth mentioning here as a prototypical example of a non-equilibrium phase transition in an open quantum system. The existence of the nonequilibrium phase transitions in magnetic samples is also verified experimentally\cite{wang,berger1,berger2,berger3,berger4}.

In the above paragraph, we have discussed the non-equilibrium phase transitions in continuous symmetric models driven by a time-dependent magnetic field (except for reference\cite{tharnier}). In those cases, the system was kept far from equilibrium by an external, time dependent magnetic field. In the extended system, the spatio-temporal variation of the external drive sets different types of nonequilibrium states just by breaking the synchronized oscillations of the spins throughout the lattice. It may be a naive question, {\it how does the system respond to such spatio-temporal variation of external drive?} Keeping this in mind, the nonequilibrium responses of the system described by continuous symmetric spin models have been studied recently by Monte Carlo simulations. The driven spin wave mode showed\cite{maxy} non-equilibrium phase transitions in three dimensional anisotropic XY ferromagnet irradiated by propagating and standing magnetic field waves. The travelling magnetic field wave through the anisotropic Heisenberg nano-rod showed\cite{mahei} non-equilibrium phase transitions. In both of the above-mentioned cases, the polarization of the propagating magnetic field was linear. What will happen if the anisotropic XY ferromagnet is driven by {\it elliptically} polarized propagating magnetic field wave? This question has been addressed in the present study through a detailed Monte Carlo investigation. The manuscript has been organized as follows: The Hamiltonian of such an anisotropic (both bilinear exchange type as well as single site type) XY ferromagnet irradiated by an elliptically polarized propagating magnetic field wave, has been introduced in the next section (section-II). The simulational results are reported in Section III. The paper ends with a summary in Section IV. 

\vskip 1cm

\noindent {\bf {II.MODEL AND SIMULATION METHOD:}}
\vspace{1.0 mm}

\noindent The time-dependent Hamiltonian of anisotropic  XY ferromagnet driven by an elliptically polarised magnetic field is given by,

\begin{equation}
{H(t)}= -J \sum_{<i,j>} [(1+\Omega)S_i^xS_j^x + (1-\Omega)S_i^yS_j^y] -\sum_{i} \vec{h_i}(t) \cdot \vec{S_i}.
\end{equation}

\noindent for bilinear exchange anisotropy ($\Omega$), and

\begin{equation}
{H(t)}= -J \sum_{<i,j>} [S_i^xS_j^x + S_i^yS_j^y] 
-D\sum [(S_i^x)^2-(S_i^y)^2]
-\sum_{i} \vec{h_i}(t)\cdot \vec{S_i}.
\end{equation}

\noindent for single-site anisotropy ($D$).

The first term in the Hamiltonian (in Eqn-1 and Eqn-2) represents the nearest neighbor spin-spin interaction with ferromagnetic interaction strength($J>0$). Equation-1 includes the bilinear exchange anisotropy controlled by parameter $\Omega$.
Equation-2 includes single-site anisotropy controlled by parameter $D$. Physically, the bilinear exchange anisotropy 
($\Omega$) brings the nonuniformity in the effective strengths of relative ferromagnetic interactions for x and y components of the spin vectors.  The single site anisotropy is characterized by $D$ (in the second term in Eqn-2). The single site anisotropy represents the 
effective and intrinsic crystal field.  We have considered only the positive values of $\Omega$ (between 0 and 1) and $D$ (between 0 and 2.5)throughout the present simulational study. The term ${\vec h_i}(t) \cdot {\vec S_i}$ corresponds to the Zeeman term or spin-field interaction term for both (Eqn-1 and Eqn-2) cases. $\vec h_{i}(t)$ is an externally applied, elliptically polarized, propagating magnetic field wave. The spatio-temporal variations of the elliptically polarized magnetic field wave have the following mathematical form: $\vec h= {\hat x} h_x + {\hat y} h_y= h_{0x}\cos(2\pi f t-\frac{2\pi}{\lambda}z)\hat{x}+h_{0y}\sin(2\pi f t-\frac{2\pi}{\lambda}z)\hat{y}$ where $f$ and $\lambda$ represents the frequency and wavelength of the elliptically polarized propagating magnetic field wave. The field is propagating along $z$ direction. Without loss of generality, one can easily get that $\frac{{h_{x}}^2}{{h_{0x}}^2}+\frac{{h_{y}}^2}{{h_{0y}}^2}=1$, for an elliptically polarized field in the X-Y plane (in general, $h_{0x}\ne h_{0y}$). This propagating magnetic field keeps the system away from the equilibrium. The model is defined for a simple cubic lattice (size $L^3$) with periodic boundary conditions applied in all three lattice directions. The propagating magnetic field and the anisotropies are measured in the unit of $J$.

We study the temporal evolution and non-equilibrium behaviors of an anisotropic XY model (for $L=20$). Initially, the system is prepared in a high-temperature paramagnetic phase by assigning the random initial orientation of each spin. This configuration sets the magnetization components to zero. This represents a high-temperature disordered phase. We evolve the system via the Monte Carlo Simulation (by Metropolis single spin flip algorithm) to have the non-equilibrium steady state (NESS) at any given temperature ($T$). At any finite temperature $T$ (measured in unit of $J/k_{B}$, where $k_{B}$ is Boltzmann constant) and for the fixed values of $T$, $f$, $\lambda$, $h_{0x}$,$h_{0y}$, $\omega$, $\Omega$ and $D$ a spin with initial configuration $\theta_{i}(x,y,z,t)$ is chosen.
\textcolor{blue}{In the random updating scheme, we have chosen a  lattice site $(x,y,z)$, randomly. The angular orientaion of the classical spin vector at that randomly chosen site $(x,y,z)$ is $\theta_i(x,y,z)$. Then we have proposed a new angular orientation $\theta_f(x,y,z)$ at this site $(x,y,z)$. The probability of acceptance of this proposed angular orientation is determined by the Metropolis transition rate\cite{binder}, $P_f$=Min $[{\rm exp}(\frac{-\delta H(t)}{k_{B}T}) ,1]$}.

Here $\delta H$ refers to the change in energy resulting due to the change in the orientation of the spin, $\theta_{i}$ $\rightarrow$ $\theta_{f}$. A uniformly distributed random number ($r=[0,1]$) is chosen. The chosen site is assigned to the new spin configuration $\theta_{f}(x,y,z,t')$ (for the next instant $t'$) if $r<P_{f}$. As usual, one 
Monte Carlo Step per Site (MCSS) refers to $L^3$ number of such updates. The simulation spans a duration of $1.4 \times 10^5$ MCSS, with an initial transient segment of $0.8 \times 10^5$ MCSS excluded to achieve the assumed ergodic limit. If the frequency $f$ of the propagating magnetic field wave is taken as $f=0.01$ (as we have considered throughout the simulation), the time period $(\tau)$ for a single oscillating cycle of the magnetic field is 100 MCSS. All the physical quantities are averaged over $0.6 \times 10^5$ MCSS (i.e., averaged over 600 number of such cycles of propagating magnetic field wave).

We have calculated the following quantities: The instantaneous components of the magnetization are,
(i) $m_x(t)={{1} \over {L^3}}\sum S_i^x$ and $m_y(t)={{1} \over {L^3}}\sum S_i^y$.
The  dynamical order parameter ${\vec Q} = Q_x{\hat x} + Q_y{\hat y}$, where $Q_x = {{\omega} \over {2\pi}} \oint m_x(t) dt$ and $Q_y = {{\omega} \over {2\pi}} \oint m_y(t) dt$. Those are the time averages of magnetization components over a full cycle of the propagating magnetic field wave. The components of the dynamical order parameter will vary in different cycles of propagating magnetic field wave. We measured all the thermodynamic observables by averaging (denoted by $\langle...\rangle$) over different cycles of propagating magnetic field. The variance of the component of the dynamic order parameter has also been calculated as Var$(Q_x)=L^3(\langle Q_x^2 \rangle-\langle Q_x \rangle^2)$. The fourth order Binder cumulant is $U_L = 1- {{\langle Q_x^4 \rangle} \over {3 \langle Q_x^2 \rangle^2}}$.
The dynamic energy density has been calculated as $E = {{\omega} \over {2\pi L^3}} \oint H(t) dt$. The dynamic specific heat $C$ has been defined as ${{d \langle E \rangle} \over {dT}}$, where $\langle E \rangle$ is the average (over different cycles of propagating magnetic field wave) dynamic energy density. The specific heat $C$ has been calculated by numerical differentiation \cite{scarborough}(using central difference method) of the average energy density $\langle E \rangle$ with respect to temperature ($T$).

\vskip 0.5 cm

\noindent {\bf {III.  Results:}}\\

\noindent In this section, we are providing the simulational results for the
bilinear exchange anisotropy (in subsection-A) and single site
anisotropy (subsection-B).

\vskip 0.5cm
\noindent {\bf {A. Bilinear Exchange Anisotropy ($\Omega$):}}\\

\noindent The system (with fixed $\Omega$, $h_{0x}$, $h_{0y}$, $f$ and $\lambda$) is gradually cooled down from a high temperature random phase
(thermally characterised by a finite but high value of temperature $T$). In the high temperature random phase the order parameter is
zero (${\vec Q} = Q_x {\hat x}+ Q_y {\hat y} = 0$). As the system is getting gradually cooled, a remarkable change  is observed 
(at any lower temperature) in the dynamical order parameter ($\langle Q_x \rangle$). The order parameter component $\langle Q_x \rangle$ acquires nonzero value (with $\langle Q_y \rangle =0$) at any finite temperature. This temperature is called pseudo-critical temperature ( dynamic transition temperature
or dynamic pseudo-critical temperature). Below a certain pseudocritical temperature (or dynamic transition temperature), the system was found to get dynamically ordered 
($\langle Q_x \rangle \neq 0$) for the $x$-component of the dynamic order parameter ($\langle Q_x \rangle$) only (whereas the $y$-component 
($\langle Q_y \rangle$) remains zero). This kind of dynamic ordering may be 
called {\it partial}. In this context, it would be worth mentioning that the temporal variation of the instantaneous magnetization components $m_x(t)$ (in a chosen cycle of propagating magnetic field wave) is oscillatory around nonzero value of $m_x(t)$. This provides the symmetry broken oscillation (SBO). Above the transition temperature, $m_x(t)$ has been found to oscillate symmetrically around zero (resulting in the dynamically disordered phase $Q_x=0$).  This provides
the symmetry restoring oscillation (SRO). On the other hand, $m_y(t)$ always oscillates symmetrically about zero. This prompted us to state that this dynamic phase transition is accompanied by a partial breaking of the dynamical symmetry of the order parameter ${\vec Q}$.

\begin{figure}[h!]

\includegraphics[angle=0, height=6cm,width=6.5cm]{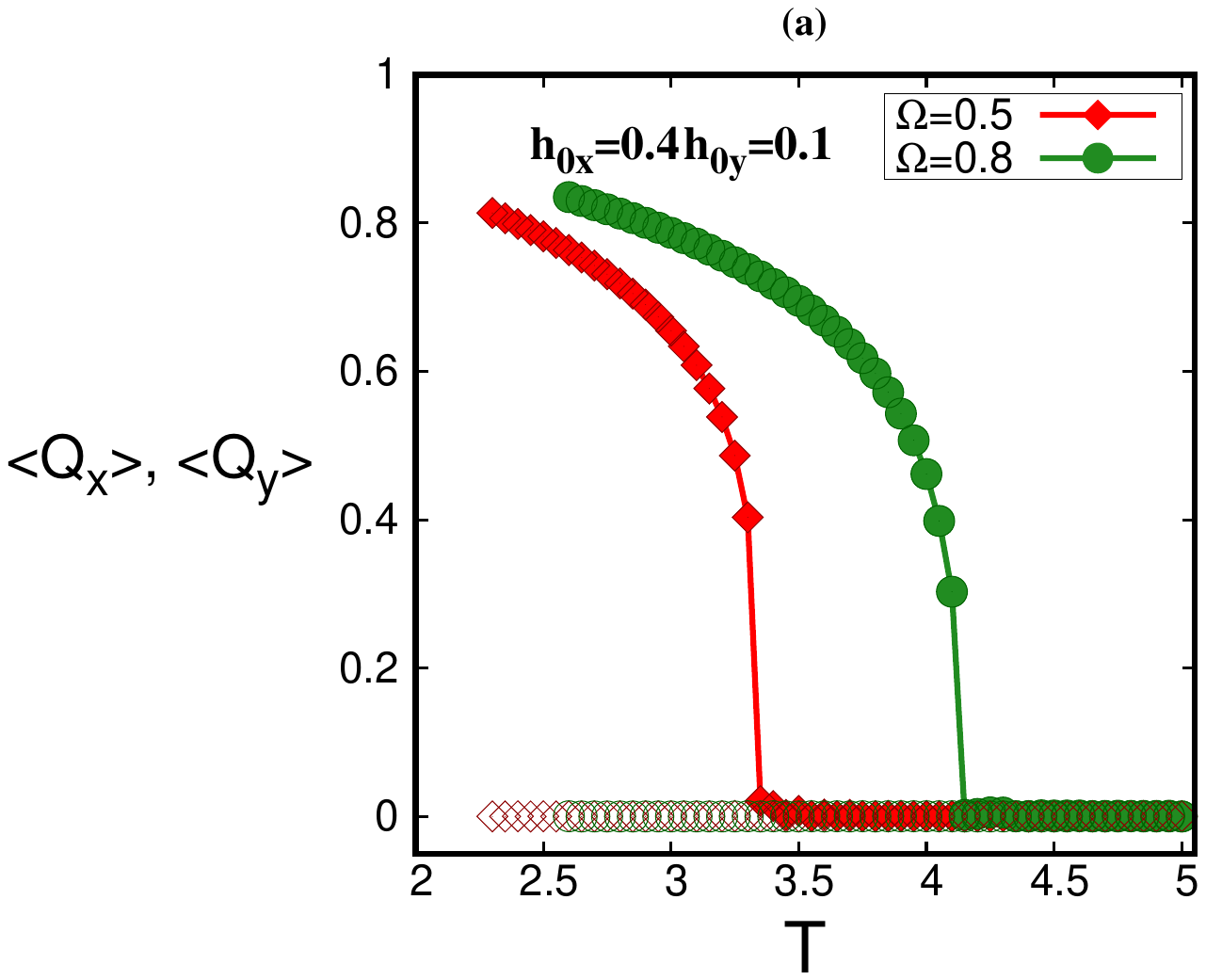}
\includegraphics[angle=0,height=6cm,width=6.5cm]{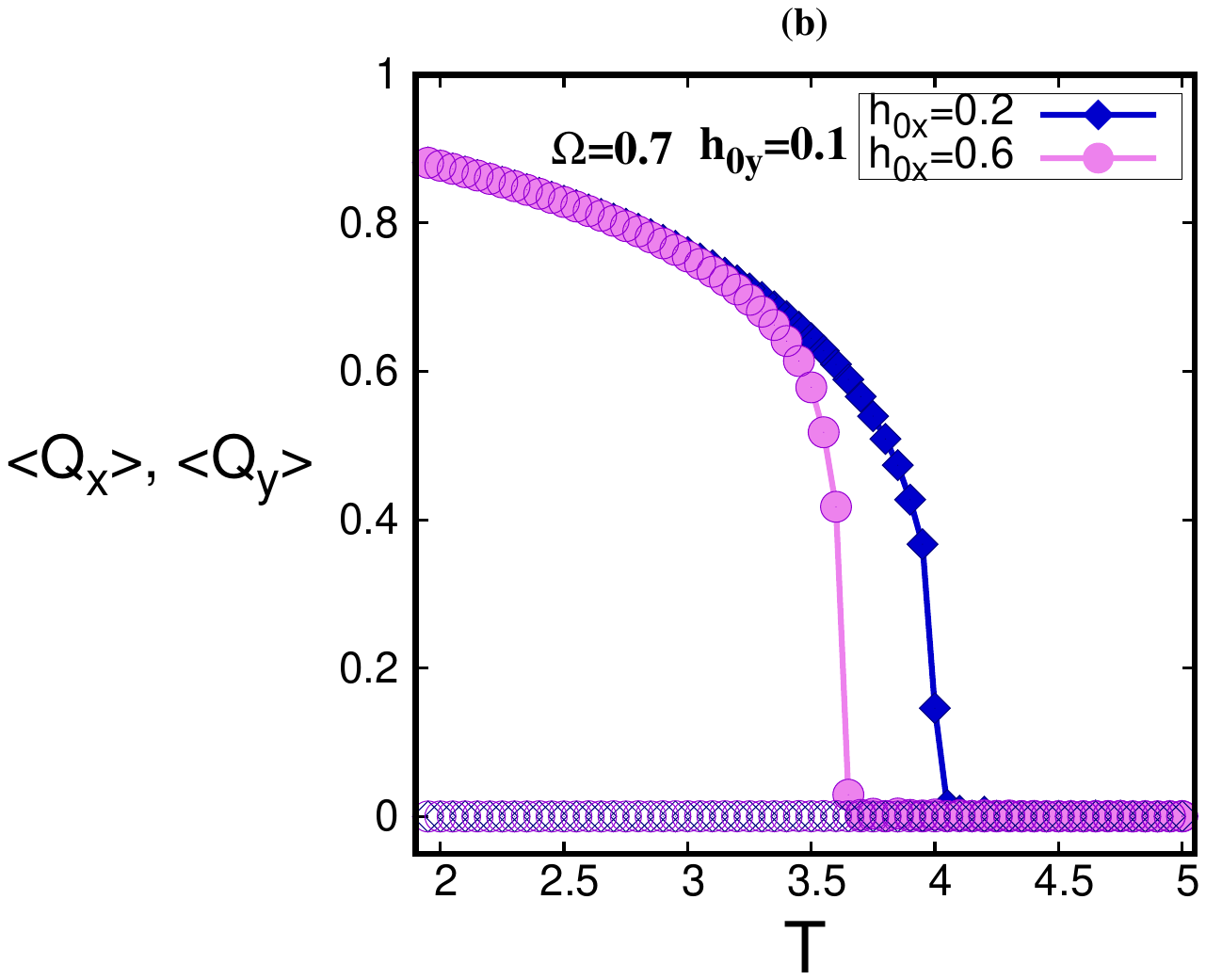}\\
\includegraphics[angle=0,height=6cm,width=6.5cm]{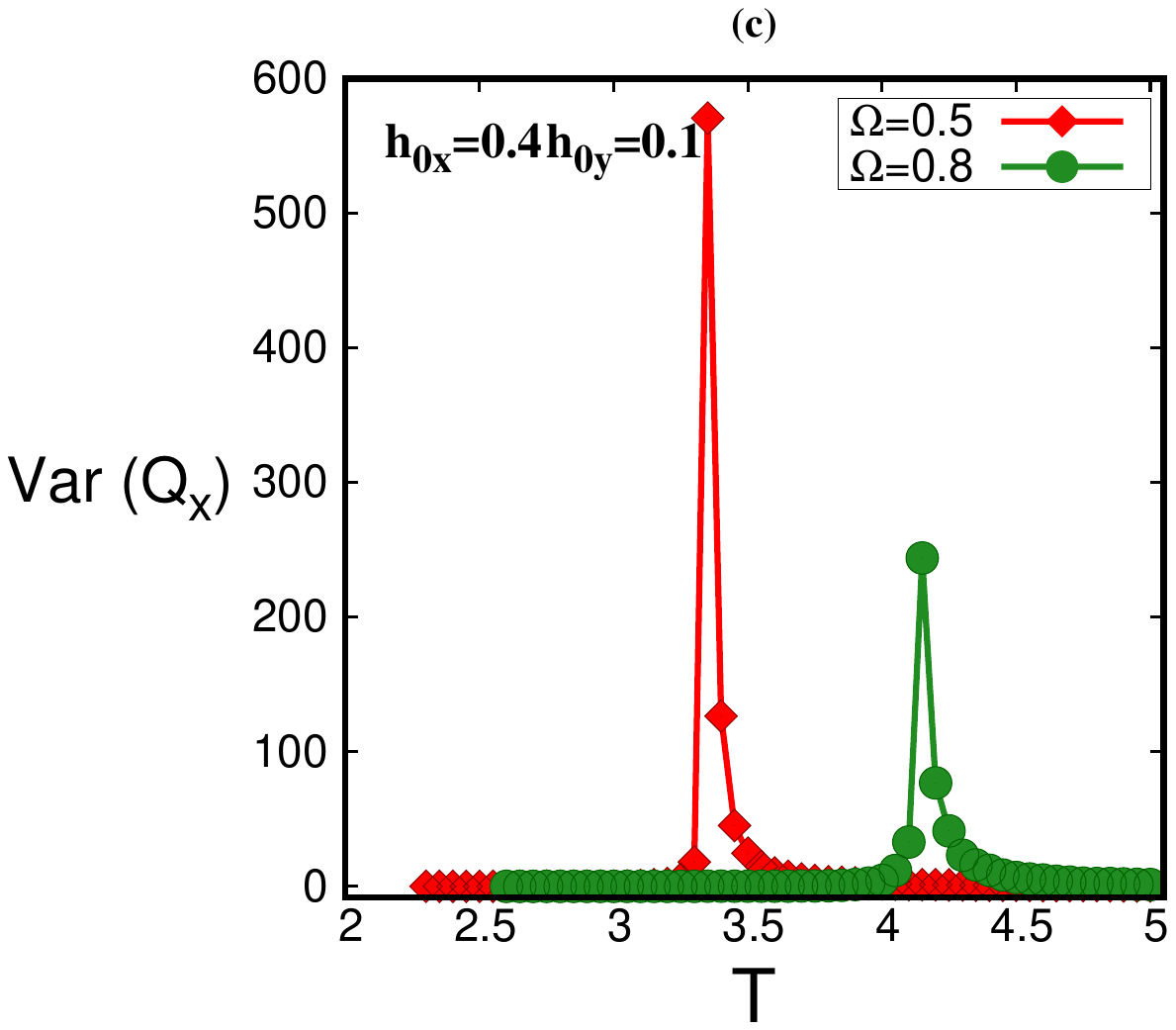}
\includegraphics[angle=0,height=6cm,width=6.5cm]{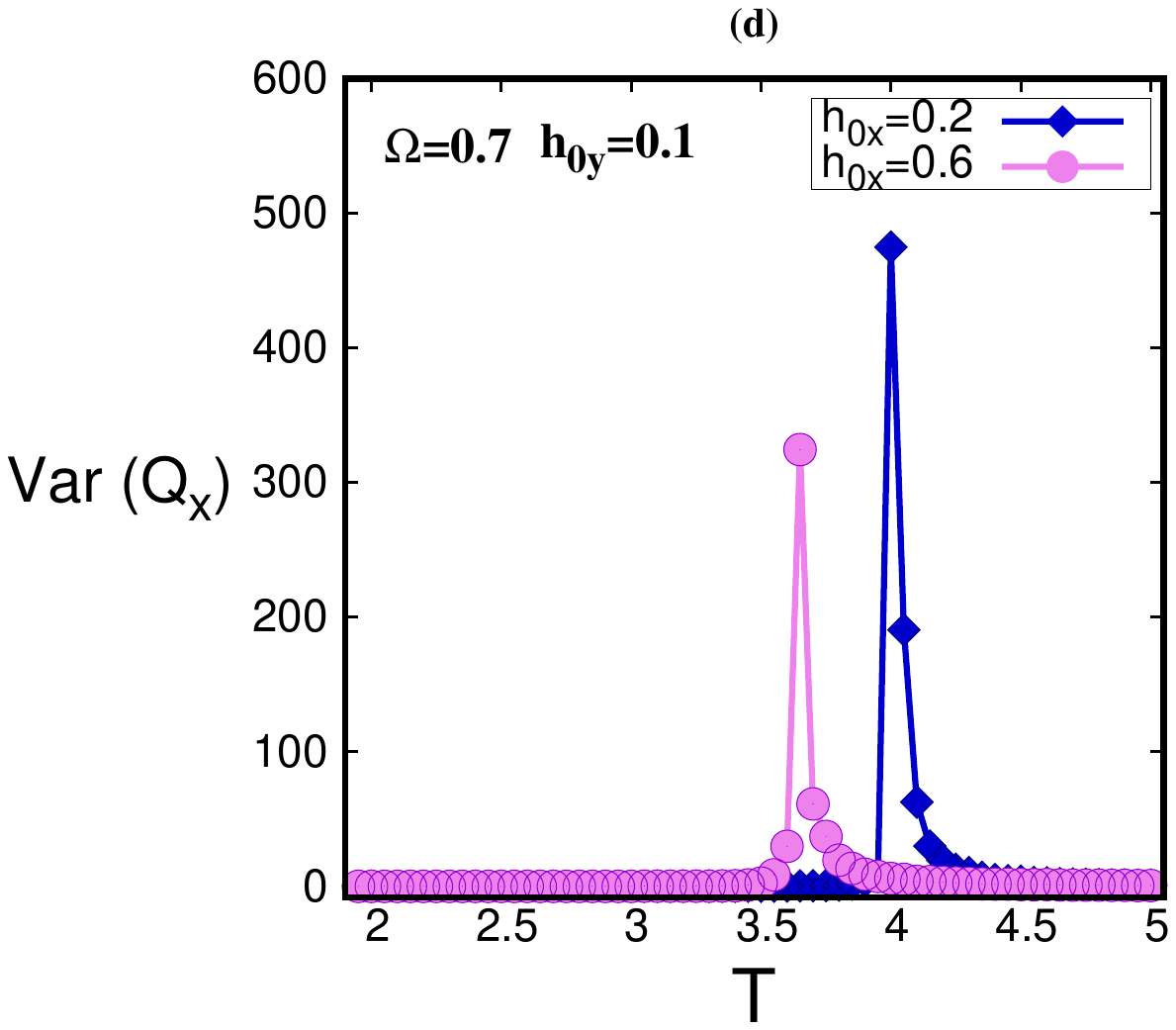}\\
\includegraphics[angle=0,height=6cm,width=6.5cm]{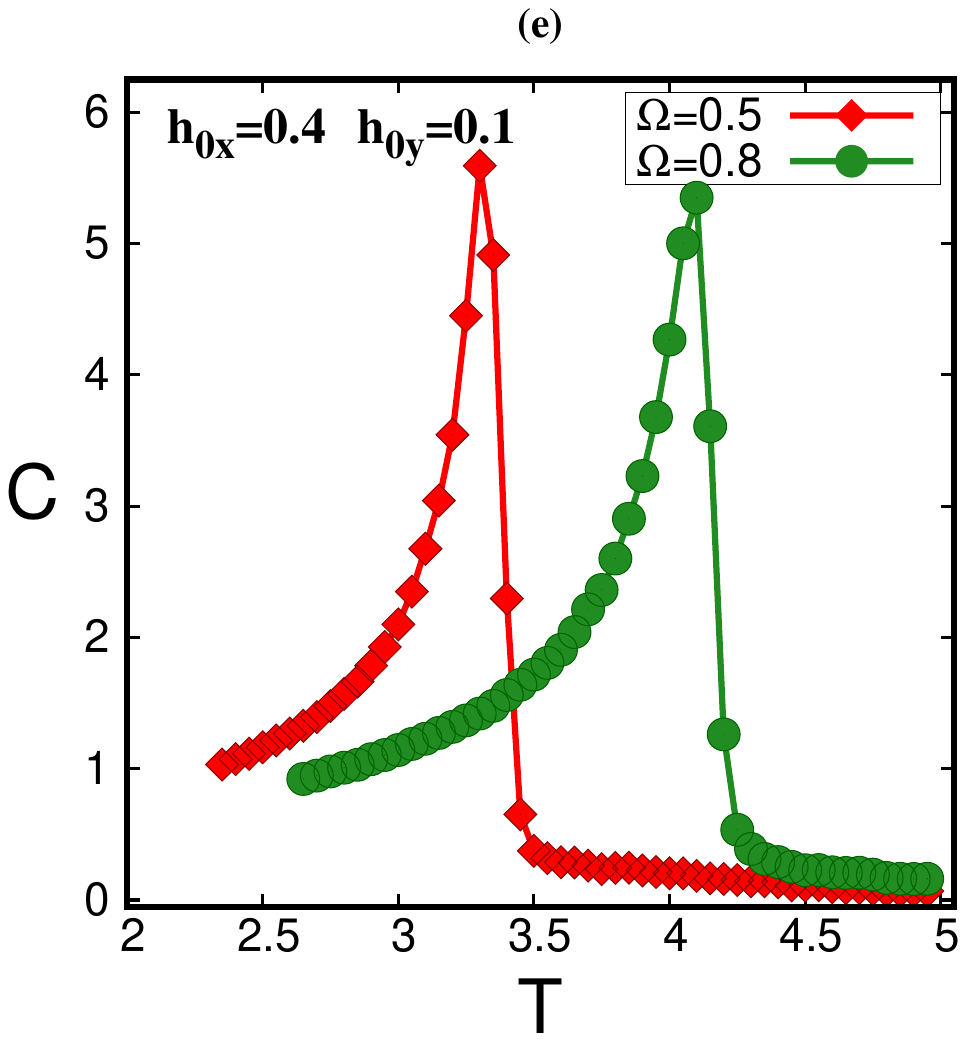}
\includegraphics[angle=0,height=6cm,width=6.5cm]{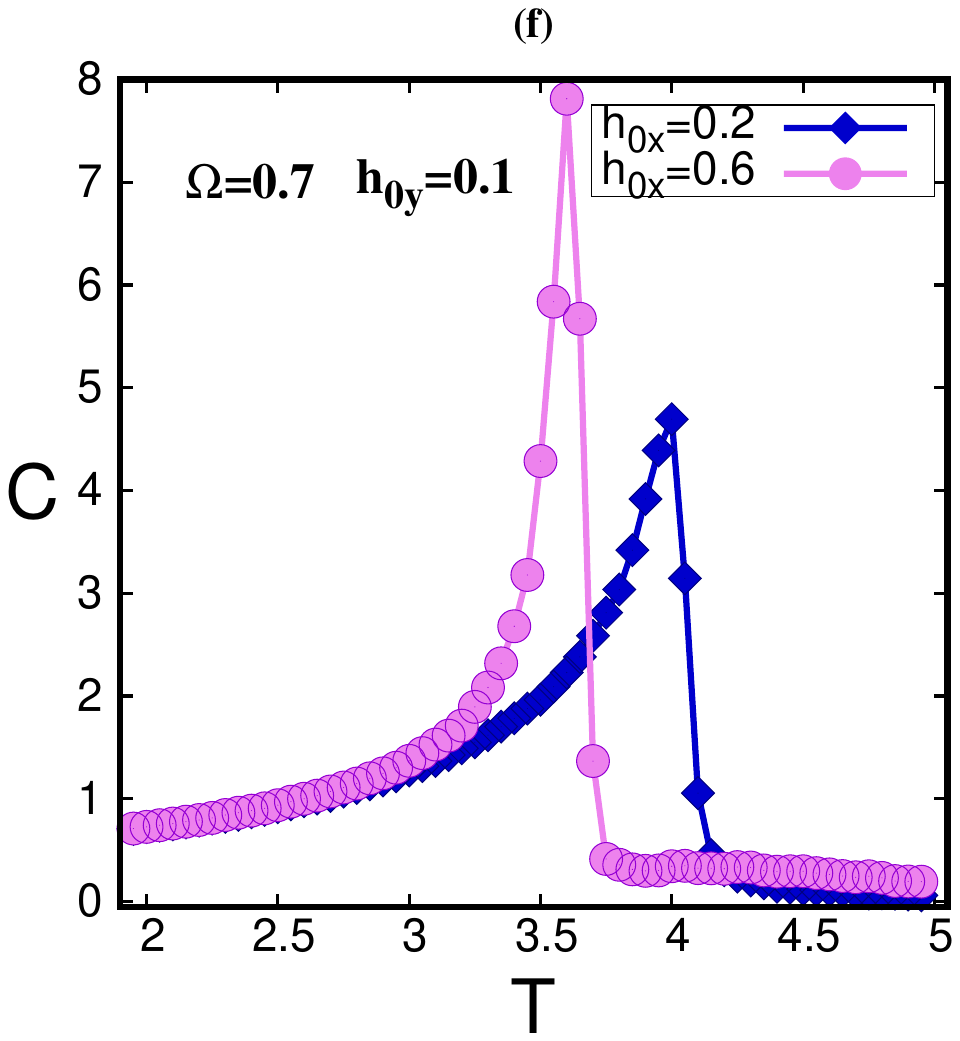}


\caption{ Temperature $(T)$ dependences of order parameter components \textcolor{blue}{$\langle Q_x \rangle$ (shown by solid symbols) and $\langle Q_y \rangle$ (shown by open symbols)}, variance of $x$-component of order parameter (Var $(Q_{x})$), and dynamic specific heat ($C$) for different anisotropy ($\Omega$) and different field amplitudes ($h_{0x}$ and $h_{0y}$). Left panel is for constant field amplitude and right panel is for constant anisotropy. Here, $L=20$, $f=0.01$ and $\lambda=10$.}
\label{fig:bilinear-observables}
\end{figure}

 In this section, we show the partial symmetry breaking of order parameters and the existence of dynamic phase transitions. Fig-\ref{fig:bilinear-observables} shows the typical thermal ($T$) variations of the order parameter components ($\langle Q_x \rangle$, 
 $\langle Q_y \rangle$), the variance of the order parameter (Var $(Q_x)$), and the dynamic specific heat ($C= {{d \langle E \rangle} \over {dT}}$, for different values of the field amplitude ($h_{0x}$) and various anisotropy strengths ($\Omega$). The amplitude of the y-component of the elliptically polarized field, the $h_{0y}=0.1$ is kept fixed throughout the simulation. As the system is cooled down from a high-temperature random state, $\langle Q_x \rangle$ gets a non-zero value 
 (from $\langle Q_x=0 \rangle$), but $\langle Q_y \rangle$ remains zero (apart from small, insignificant fluctuations). Fig-\ref{fig:bilinear-observables}(a) and Fig-\ref{fig:bilinear-observables}(b) show the variation of $\langle Q_x \rangle $ and $\langle Q_y \rangle$ as a function of temperature ($T$). There exists a critical temperature below which we get dynamical symmetry broken ordered phase, this temperature is called the dynamic transition temperature ($T_c$) or dynamic pseudocritical temperature. 
It may be noted here that this pseudo-critical temperature
 is the transition temperature for the finite sized system. The actual
 critical temperature is usually defined as the transition temperature in the thermodynamic limit ($L \to \infty$).
 The variance of $Q_x$ i.e Var $(Q_x)$ has been found to exhibit a sharp peak (in Fig-\ref{fig:bilinear-observables}(c) and  Fig-\ref{fig:bilinear-observables}(d)) at the transition temperature. This is a sign of the growth of critical correlations in dynamic ordering (associated with the large critical fluctuations). 
  \textcolor{blue}{The dynamic pseudo-critical temperature is derived from the temperature that maximizes the variance (Var $(Q_x)$).} The dynamic specific heat ($C$) is also studied as a function of temperature ($T$), which also gets sharply peaked near the transition point or dynamic pseudocritical temperature (in Fig-\ref{fig:bilinear-observables}(e) and  Fig-\ref{fig:bilinear-observables}(f)).

The dynamic transition temperature is found to vary with the field amplitude ($h_{0x}$, where the $y$-component is kept fixed at $h_{0y}=0.1$) of an elliptically polarized propagating magnetic field wave and the strength of bilinear exchange anisotropy ($\Omega$). From the left panel of the Fig-\ref{fig:bilinear-observables}, the transition temperature is found to increase as the strength of anisotropy increases for constant $h_{0x}=0.4$ (and for fixed $h_{0y}=0.1$). On the other hand,  the system is found to achieve a dynamic ordered phase at a lower transition temperature as the field amplitude ($h_{0x}$ only) increases for a fixed $\Omega=0.7$ (Fig-\ref{fig:bilinear-observables} right panel). In general, the dynamic pseudo-critical temperature $T_c$ has been observed to be dependent on the strength of anisotropy ($\Omega$) and the field amplitude of the $x$-component ($h_{0x}$) of an elliptically polarized propagating magnetic field wave. 

\begin{figure}[h!]

\includegraphics[angle=0,height=6cm,width=8cm]{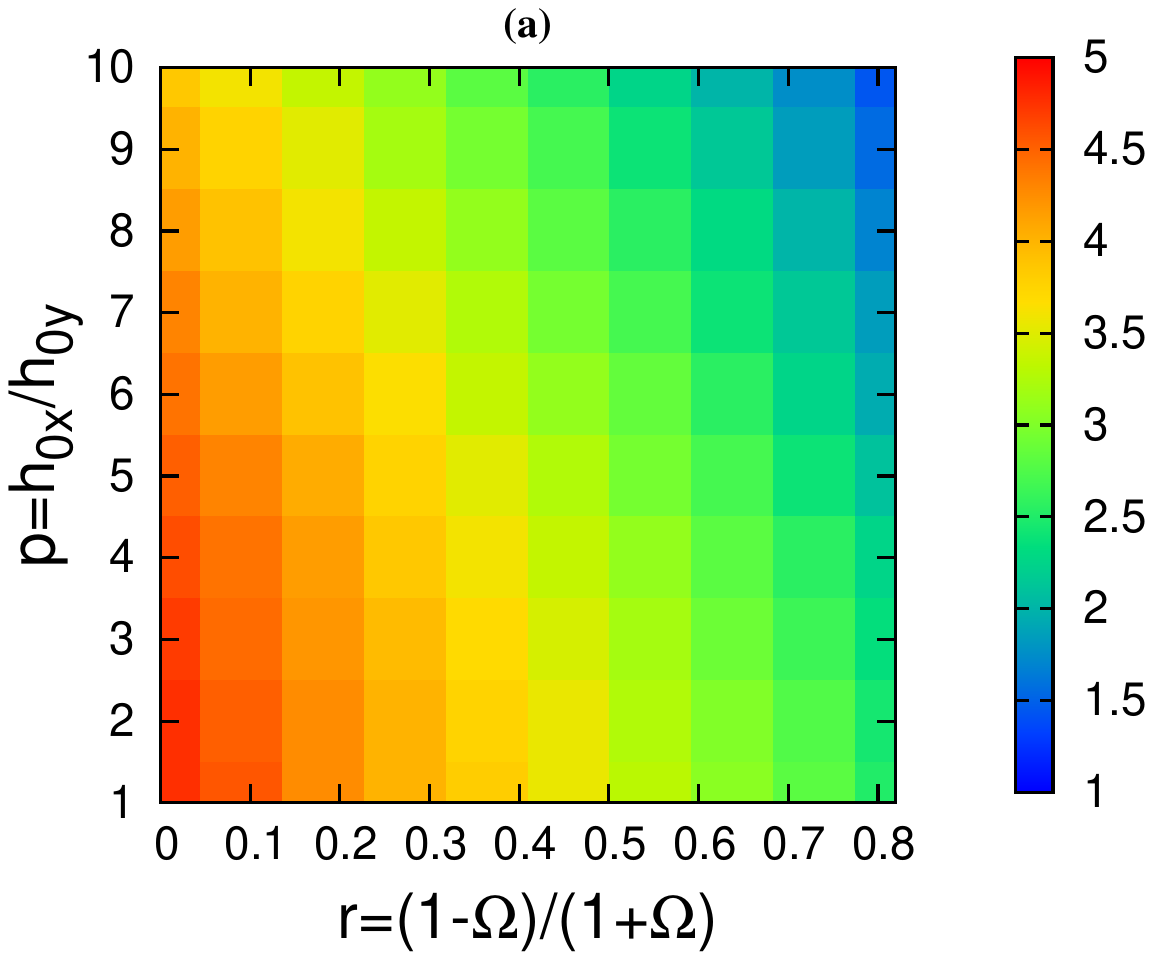}
\includegraphics[angle=0,height=6cm,width=8cm]{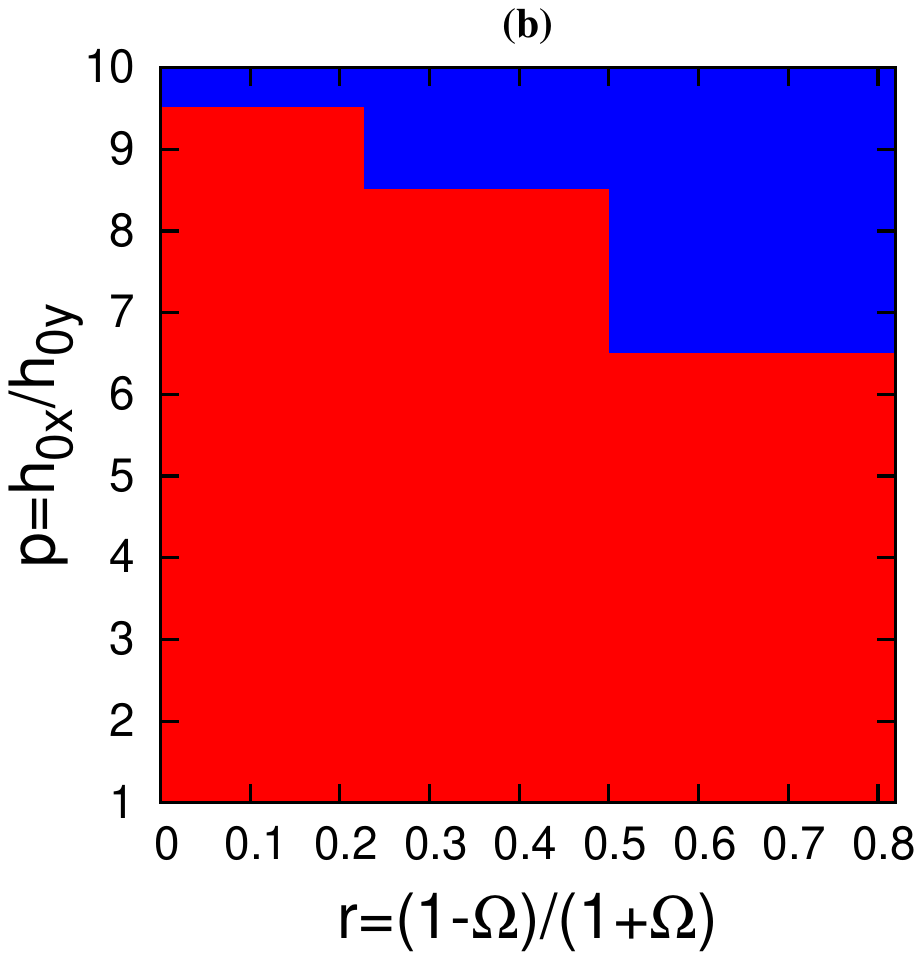}

\caption{(a) The comprehensive phase diagram (or the image plot of $T_c$) in $p-r$ plane where $p={{h_{0x}} \over {h_{0y}}}$ and $r={{1-\Omega} \over {1+\Omega}}$. The transition temperature is obtained from the position of the peak of Var $(Q_{x})$ plotted against the temperature ($T$). (b) The nature (continuous or discontinuous) of transition marked by different colors. The region marked by red color corresponds to the continuous or second order transition and the region of discontinuous or first order transition is marked by blue color. Here, $L=20$, $f=0.01$ and $\lambda=10$.}
\label{fig:bilinear-phase}
\end{figure}

Fig-\ref{fig:bilinear-phase}(a) shows the comprehensive phase diagram (or the image plot of the transition temperature), of such a nonequilibrium phase transition, in the plane described by $r=\frac{1-\Omega}{1+\Omega}$ and $p=\frac{h_{0x}}{h_{0y}}$. Actually, the pseudocritical temperature ($T_c$) is plotted here (in the image plot) as a function of $p$ and $r$. When constructing these phase diagrams, all the pseudo-critical temperatures ($T_c$ for a finite-sized system) are calculated from the peak positions of the thermal variation of Var $(Q_x)$. This is basically a surface, below which the system is in the ordered phase and above which the disordered phase is obtained.

It has also been observed that the {\it order} of such a non-equilibrium phase transition depends on the values of $h_{0x}$ (for a fixed value of $h_{0y}$) and $\Omega$. We have also shown the {\it order} of such a non-equilibrium transition (dependent on the $h_{0x}$ and $\Omega$) in a separate image plot in  Fig-\ref{fig:bilinear-phase}(b). Two different colors represent the different order (first or second) or nature (discontinuous or continuous) of the transition. The red color represents the second-order \textcolor{blue}{(continuous)} transition, whereas the blue color represents the first-order \textcolor{blue}{(discontinuous)} transition. Interestingly, the boundary of these two colors is the {\it tricritical line}, observed in such a non-equilibrium phase transition. Hence, our numerical results show the existence of tricritical behavior in XY ferromagnet driven by an elliptically polarized propagating magnetic field wave. The nature or order of such a non-equilibrium transition has been confirmed by the thermal variations of the fourth order Binder cumulant and the statistical distribution of the dynamic order parameter near the transition temperature ($T_c$).

\begin{figure}[h!]

\includegraphics[angle=0,height=5.5cm,width=6.5cm]{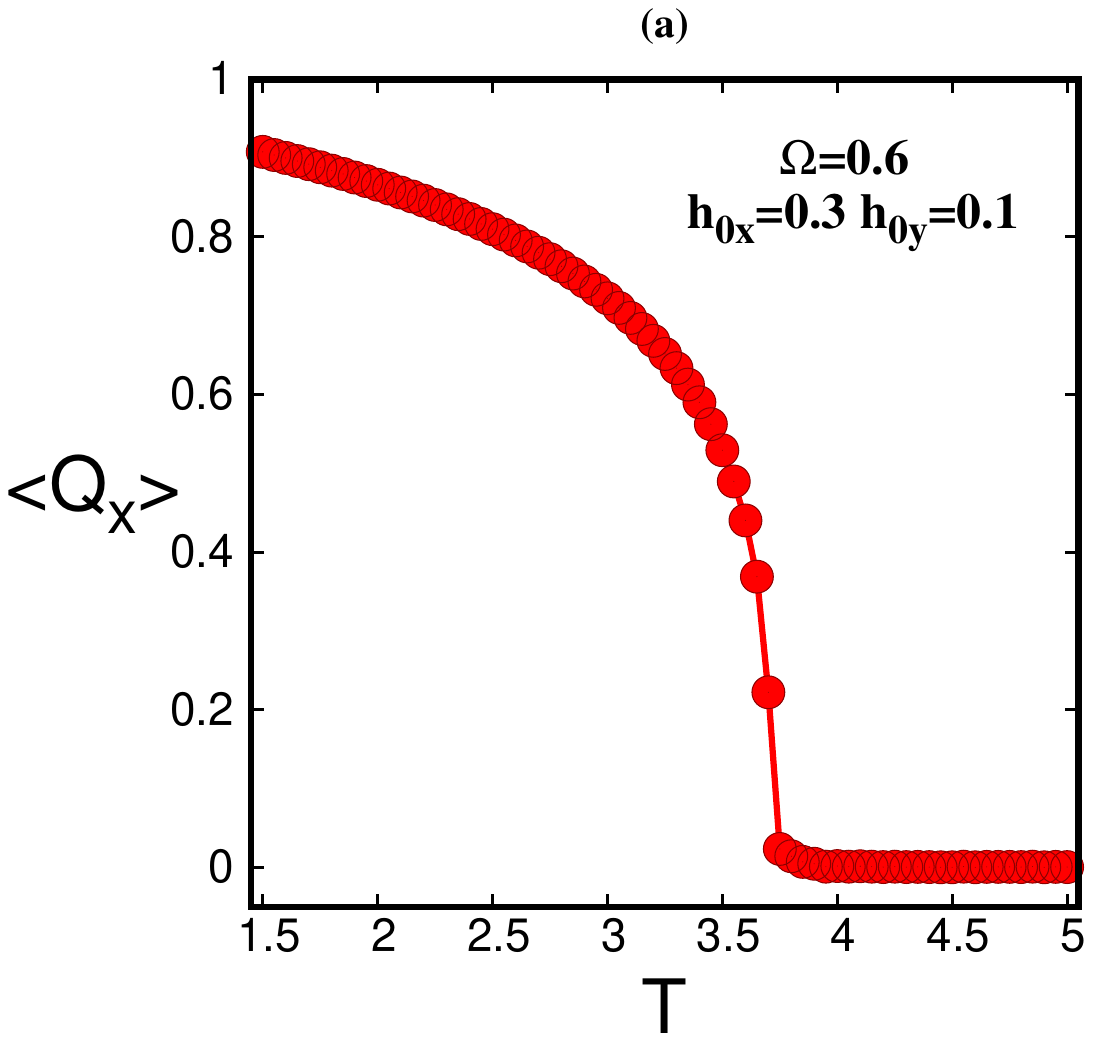}
\includegraphics[angle=0,height=5.5cm,width=6.5cm]{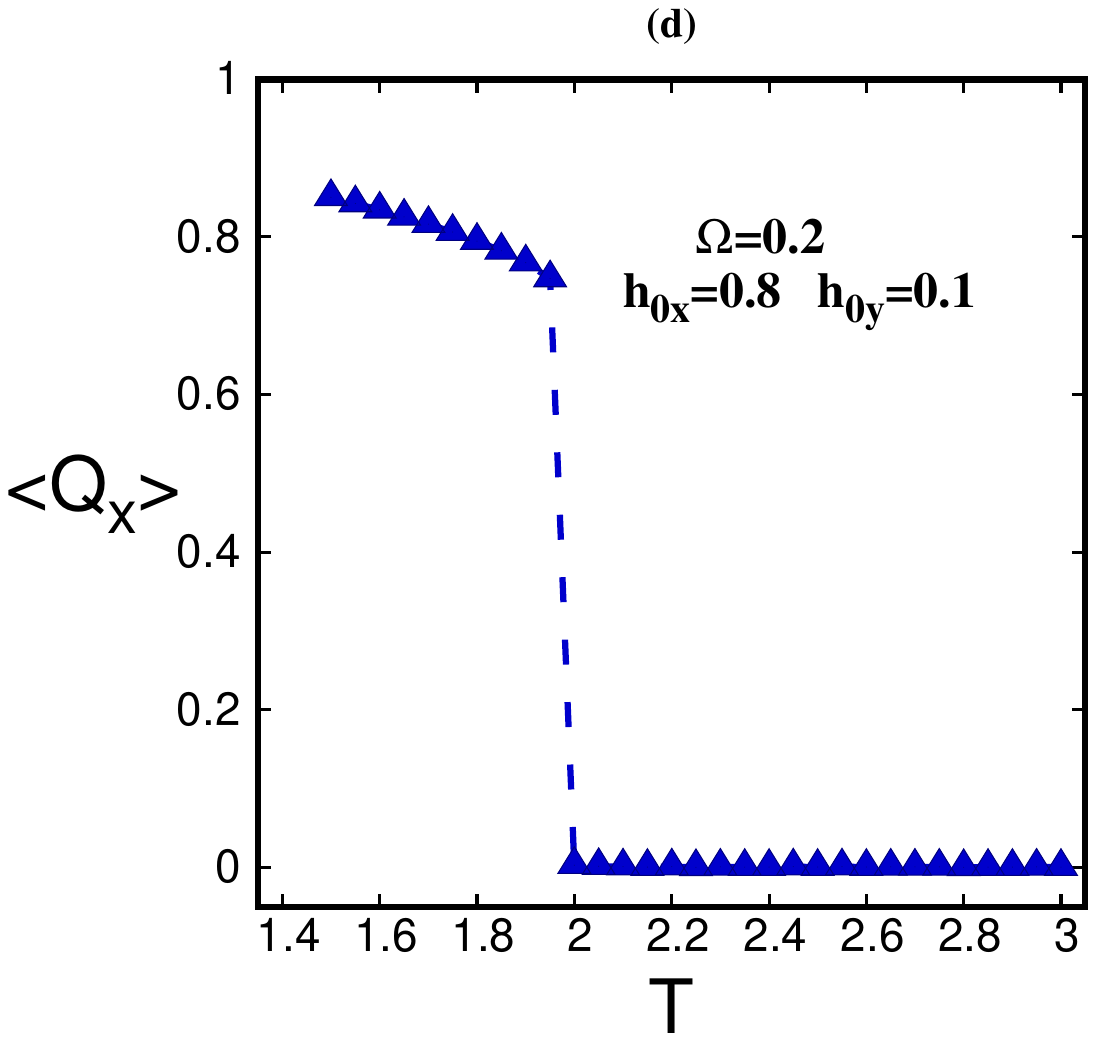}\\
\includegraphics[angle=0,height=5.5cm,width=6.5cm]{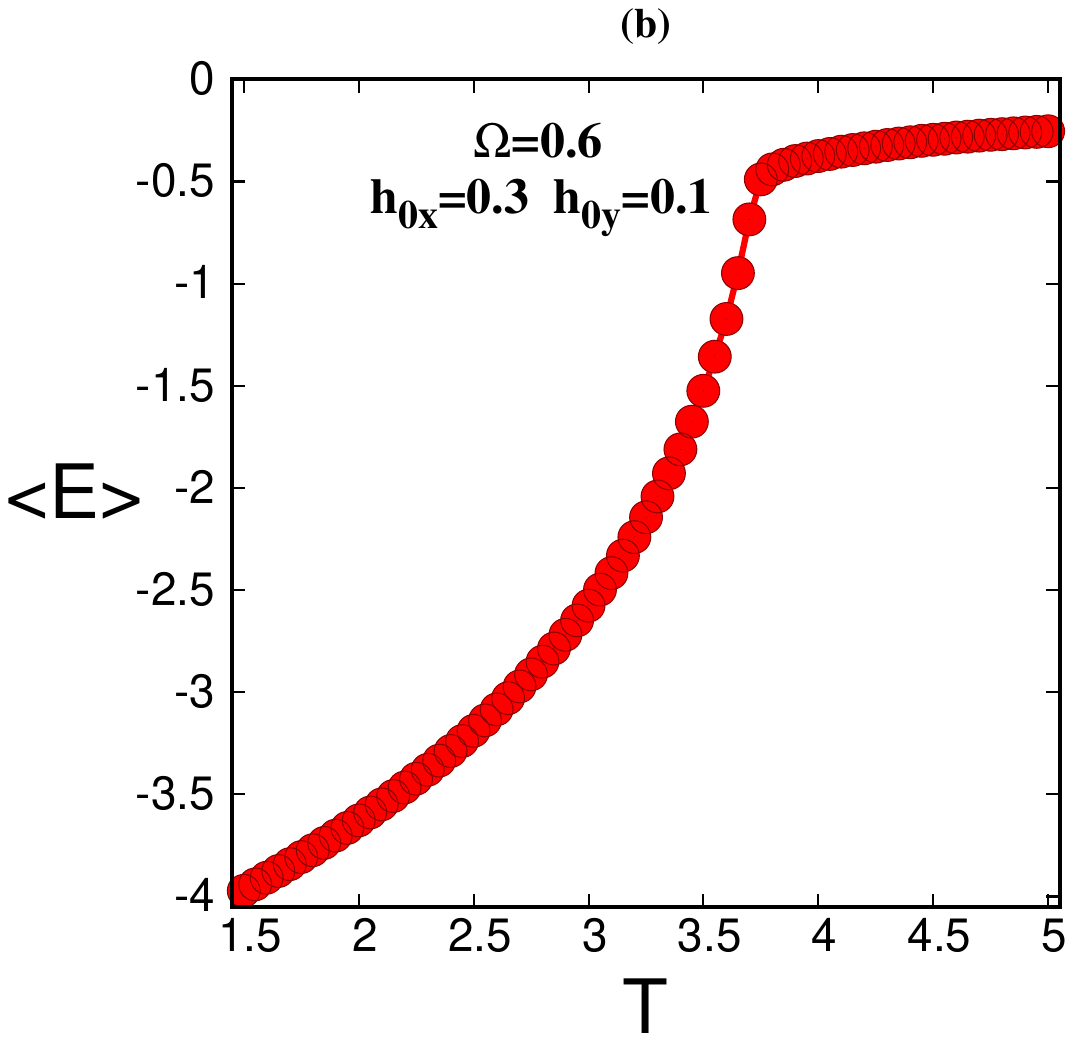}
\includegraphics[angle=0,height=5.5cm,width=6.5cm]{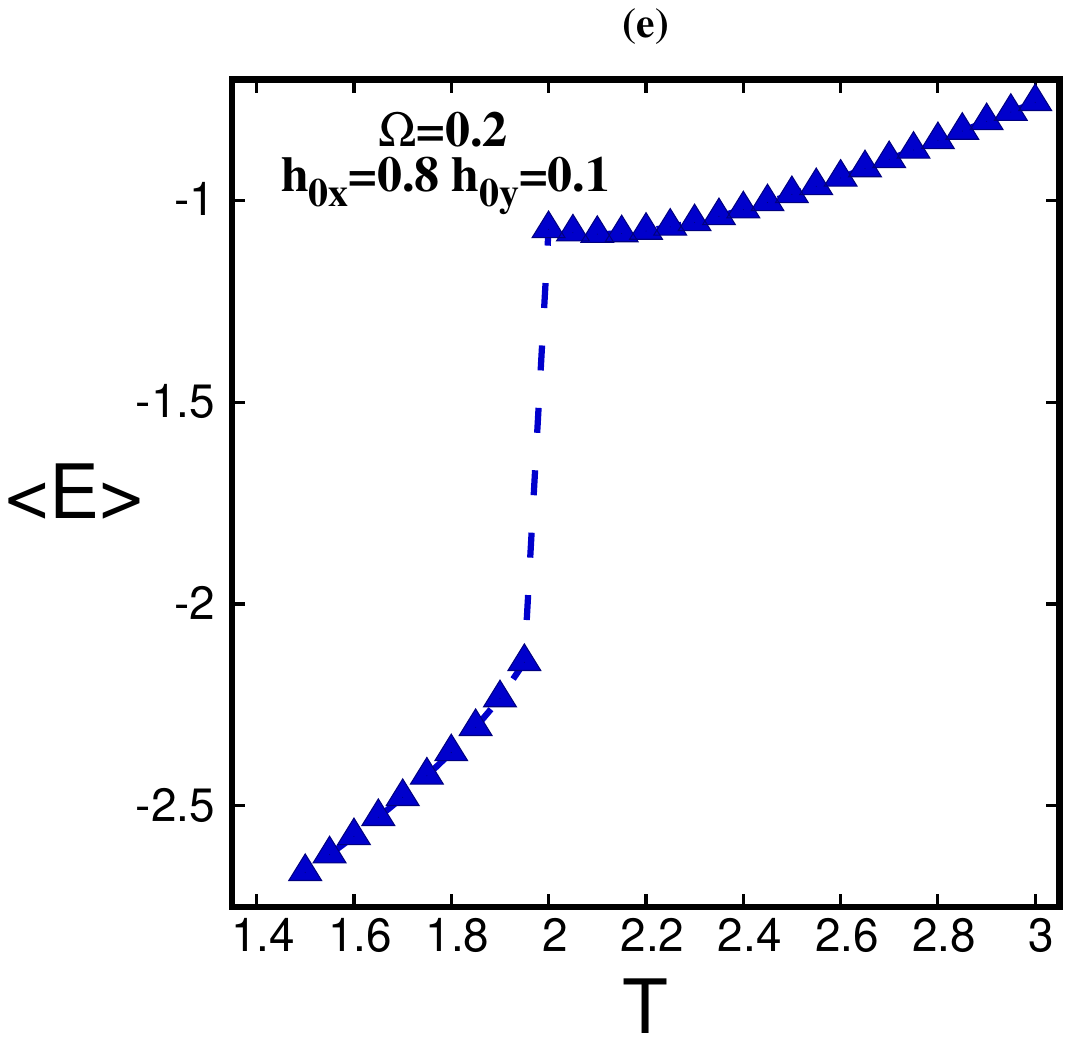}\\
\includegraphics[angle=0,height=5.5cm,width=6.5cm]{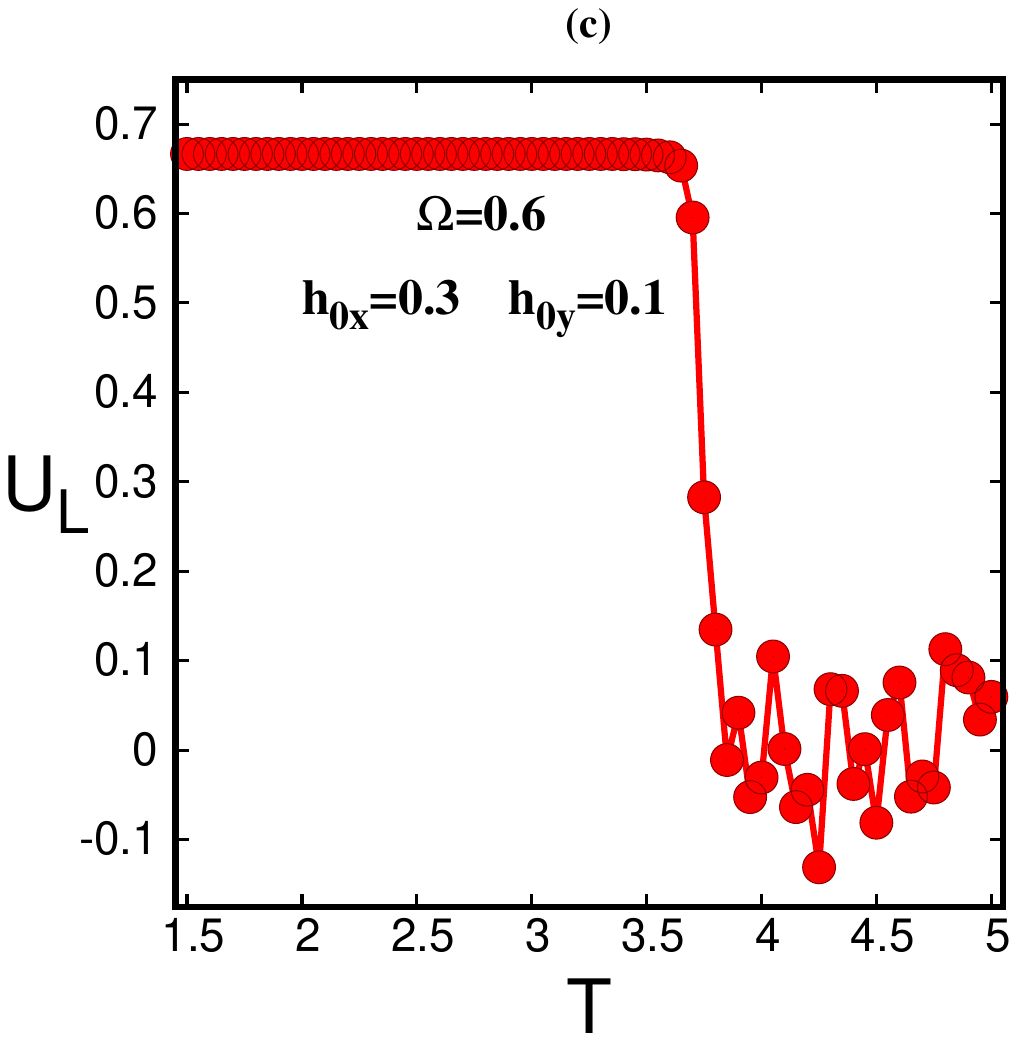}
\includegraphics[angle=0,height=5.5cm,width=6.5cm]{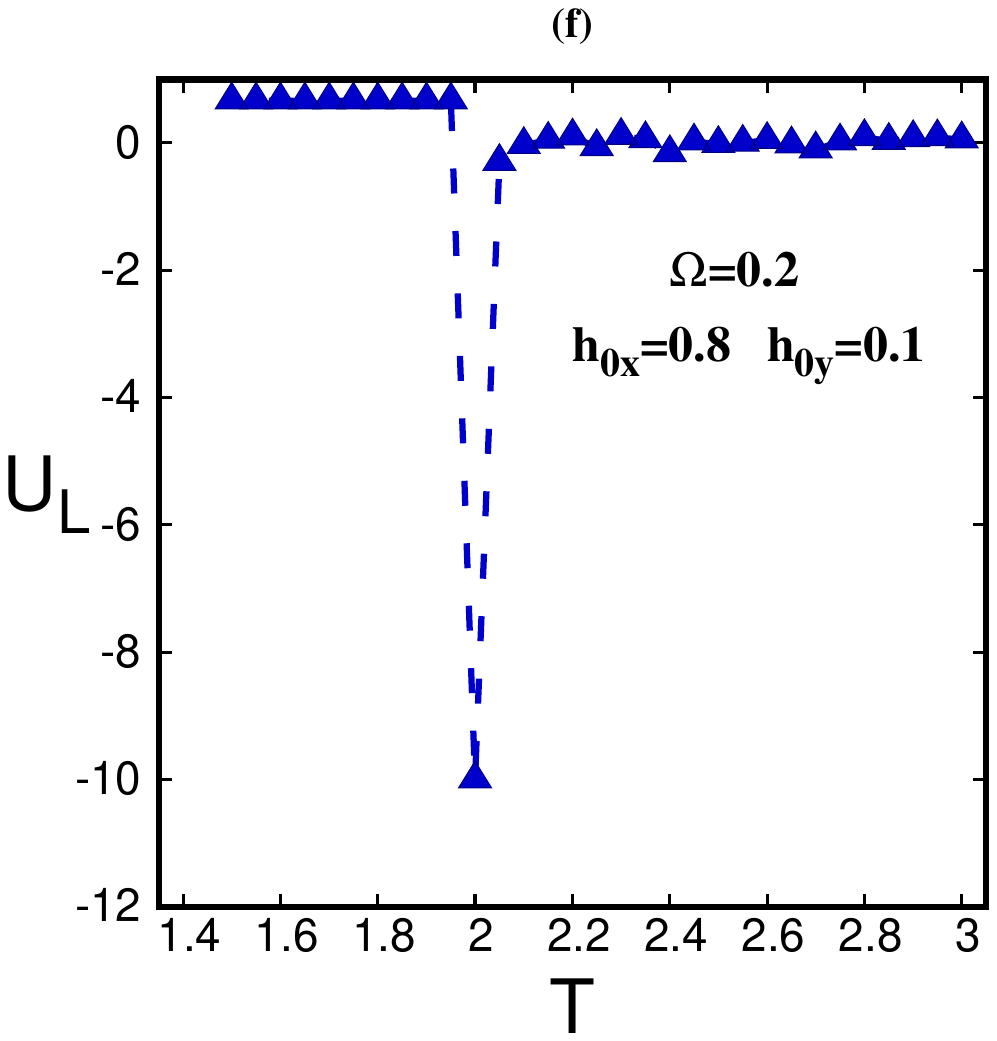}

\caption{Temperature ($T$) dependences of (a) the dynamic order parameter $(\langle Q_x \rangle)$, (b) dynamic energy density $(\langle E \rangle)$ and (c) the fourth order Binder cumulant $(U_{L})$ for a fixed set of values of the field amplitudes($h_{0x}=0.3$ and $h_{0y}=0.1$) and bilinear exchange anisotropy $(\Omega)$=0.6. Temperature ($T$) dependences of (d) the dynamic order parameter $(\langle Q_x \rangle)$, (e) dynamic energy density 
$(\langle E \rangle)$ and (f) the fourth order Binder cumulant $(U_{L})$ for another fixed set of values of the field amplitudes($h_{0x}=0.8$ and $h_{0y}=0.1$) and bilinear exchange anisotropy $(\Omega)$=0.2.
The thick (red) lines indicate continuous transition (displayed in the left panel) and the dashed (blue) lines indicate discontinuous transition (displayed in the right panel). Here, $L=20$, $f=0.01$ and $\lambda=10$.}
\label{fig:bilinear-order}
\end{figure}

To explain the order of transition, we first show (in Fig-\ref{fig:bilinear-order}) the temperature dependences of the order parameter ($\langle Q_x \rangle$)(Fig-\ref{fig:bilinear-order}(a)), energy density ($\langle E \rangle$)(Fig-\ref{fig:bilinear-order}(b)) and fourth order Binder cumulant ($U_{L}$)(Fig-\ref{fig:bilinear-order}(c)) at constant field amplitudes ($h_{0x}$ and $h_{0y}$) and anisotropy  ($\Omega$). For lower values of the field amplitude (lower $p$) ,  the order parameter ($\langle Q_x \rangle$) reduces smoothly as temperature ($T$) increases. The associated energy density of the system changes continuously throughout the transition (Fig-\ref{fig:bilinear-order}(b)). There is no abrupt jump in  energy in the vicinity of the transition point. This characteristic is also evident in the behavior of the fourth-order Binder cumulant. At lower values of the field amplitude $h_{0x}$ (lower $p$), the transition is found to be continuous or second order. In contrast, for higher field amplitudes, the corresponding changes happen discontinuously (the abrupt change in magnetization is dominated by the magnetic field instead of the thermal fluctuation). At the transition temperature, during a discontinuous transition (or first-order transition), there is a sudden change in both the order parameter(Fig-\ref{fig:bilinear-order}(d)) and the energy (Fig-\ref{fig:bilinear-order}(e)). In such transitions, the fourth-order moment ($\langle Q_{x}^4 \rangle$) is significantly higher than the square of the second-order moment ($3 \langle Q_{x}^2 \rangle^2$). Consequently, it results in a pronounced minimum (with a substantial negative value) of the fourth-order Binder cumulant $U_L$ at the transition point(Fig-\ref{fig:bilinear-order}(f)). This is the signature of a first-order or discontinuous transition. The transition changes its character (from continuous to discontinuous) for higher field amplitude and lower anisotropy.

\begin{figure}[h!]

\includegraphics[angle=0,height=15cm,width=7.5cm]{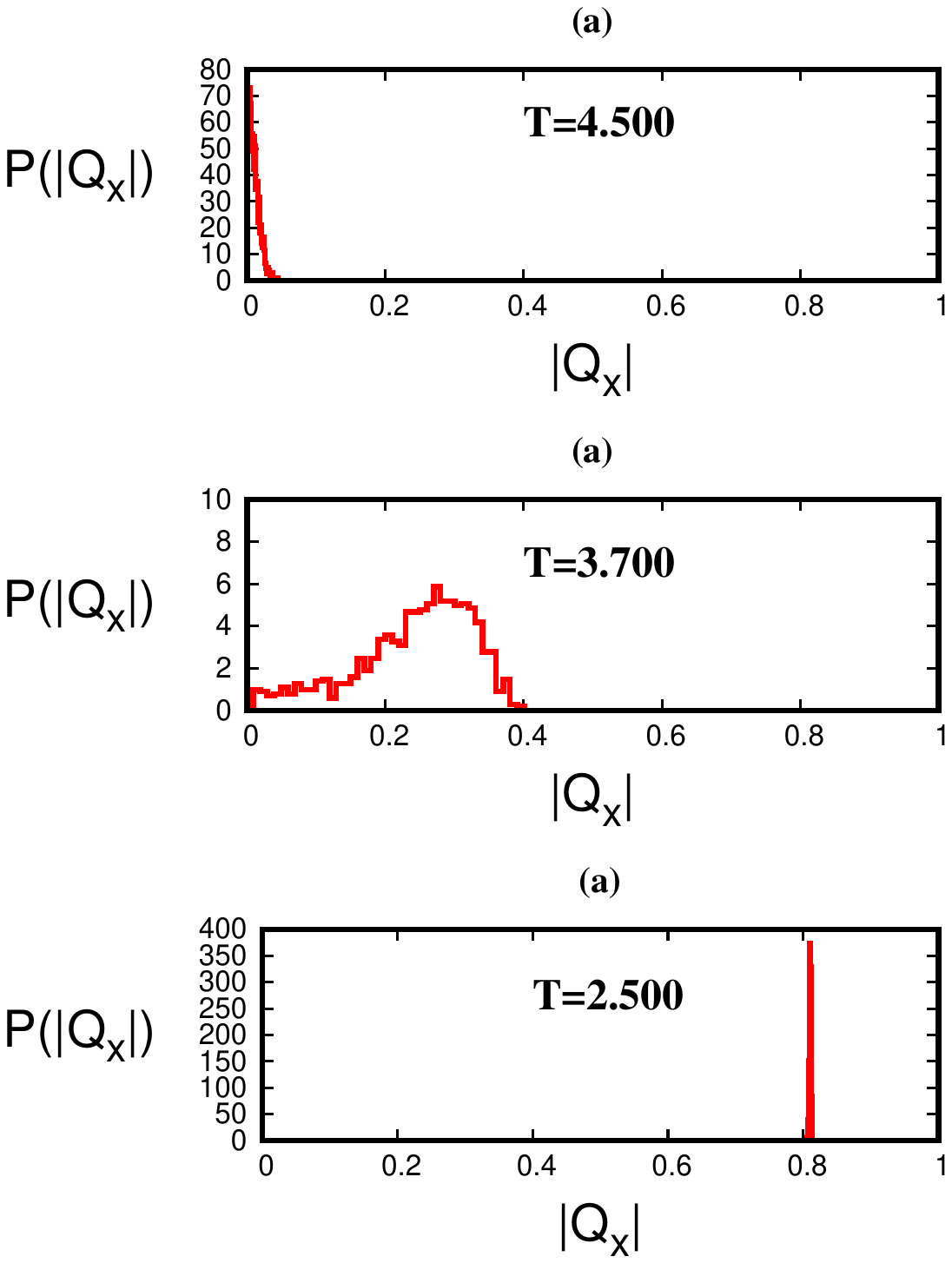}
\includegraphics[angle=0,height=15cm,width=7.5cm]{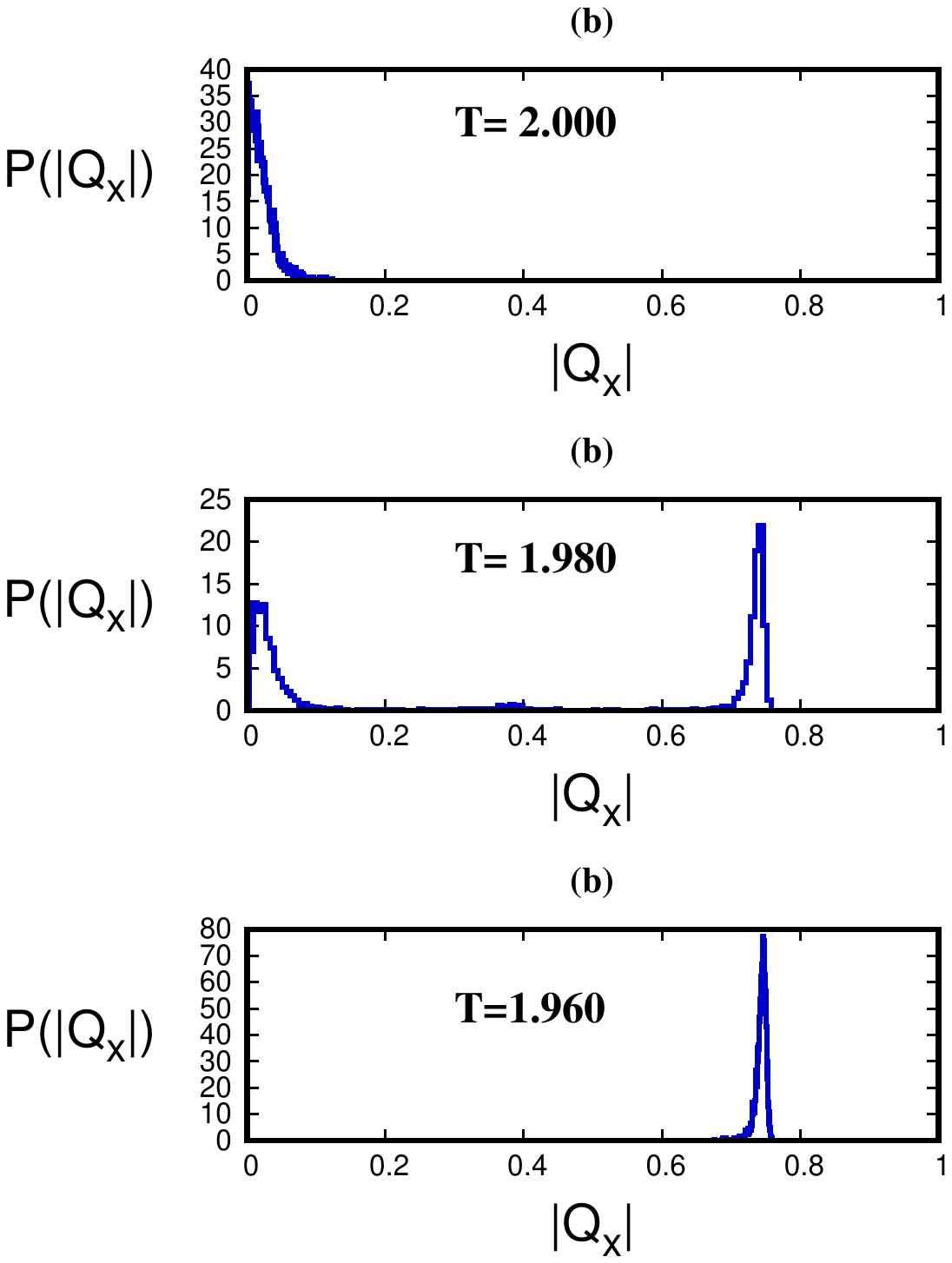}

\caption{The  probability density distribution ($P(|Q_x|)$ of the  order parameter $Q_x$ for different temperatures around the transition. The left panel (a) is for $h_{0x}=0.3$, $h_{0y}=0.1$,  $\Omega=0.6$ which exhibits the second order transition. The right panel (b) is for $h_{0x}=0.8$, $h_{0y}=0.1$, $\Omega=0.2$ which exhibits the first order transition.   Here, $L=20$, $f=0.01$ and $\lambda=10$.}
\label{fig:bilinear-dist-Qx}
\end{figure}

The statistical distribution (near the transition) of the order parameter $P(|Q_{x}|)$ and its temperature dependence can nicely expose the order (first or second order) or nature (discontinuous or continuous in nature)  of the transition. We consider only the positive branch ($|Q_x|$) of the order parameter. Fig-\ref{fig:bilinear-dist-Qx}(a) shows the statistical distribution of the order parameter for field amplitudes $h_{0x}=0.3$ and $h_{0y}=0.1$ with anisotropy $\Omega$=0.6 at three different temperature regimes. This distribution has been obtained from 3000 samples. At $T=4.50$( $>T_c$) a peak arises at 0, relieving the dynamically disordered phase. Near the transition temperature $T_{c}=3.70$ the peak shifts towards non-zero value of $Q_x$ (initiation of the dynamical ordering), and below the transition temperature $T_c=2.50$, the peak is centered around a higher value of the order parameter (well inside the ordered phase). This is usually observed in the case of any continuous or second-order phase transition. Fig-\ref{fig:bilinear-dist-Qx}(b) shows the distribution for higher field amplitude and lower anisotropy ($h_{0x}=0.8$, $h_{0y}=0.1$, $\Omega=0.2$) at three different temperatures. Above the transition temperature, a singular peak appears, around zero (indicating the dynamically disordered phase). Very near the transition temperature, an additional peak emerges, centered around a positive value of $Q_x$, concurrently with the zero-centered peak. This is the representation of the finite discontinuity of the dynamic order parameter at the transition temperature. Upon a slight decrease in temperature from the transition point, only the peak at positive $Q_x$ persists, while the zero-centered peak vanishes, resulting from the fully ordered region. This observation suggests a first-order or discontinuous nature of the phase transition. When the amplitude of an elliptically polarized magnetic field is high , the system becomes more susceptible to abrupt changes in its state. In the context of magnetic materials, the alignment of magnetic moments can suddenly switch directions, leading to a discontinuous change in the material's magnetic properties. In other words, the first-order transition is mostly driven by the magnetic field instead of thermal fluctuations.

\begin{figure}[h!]
    \centering
    \includegraphics[angle=0,height=5cm,width=4.5cm]{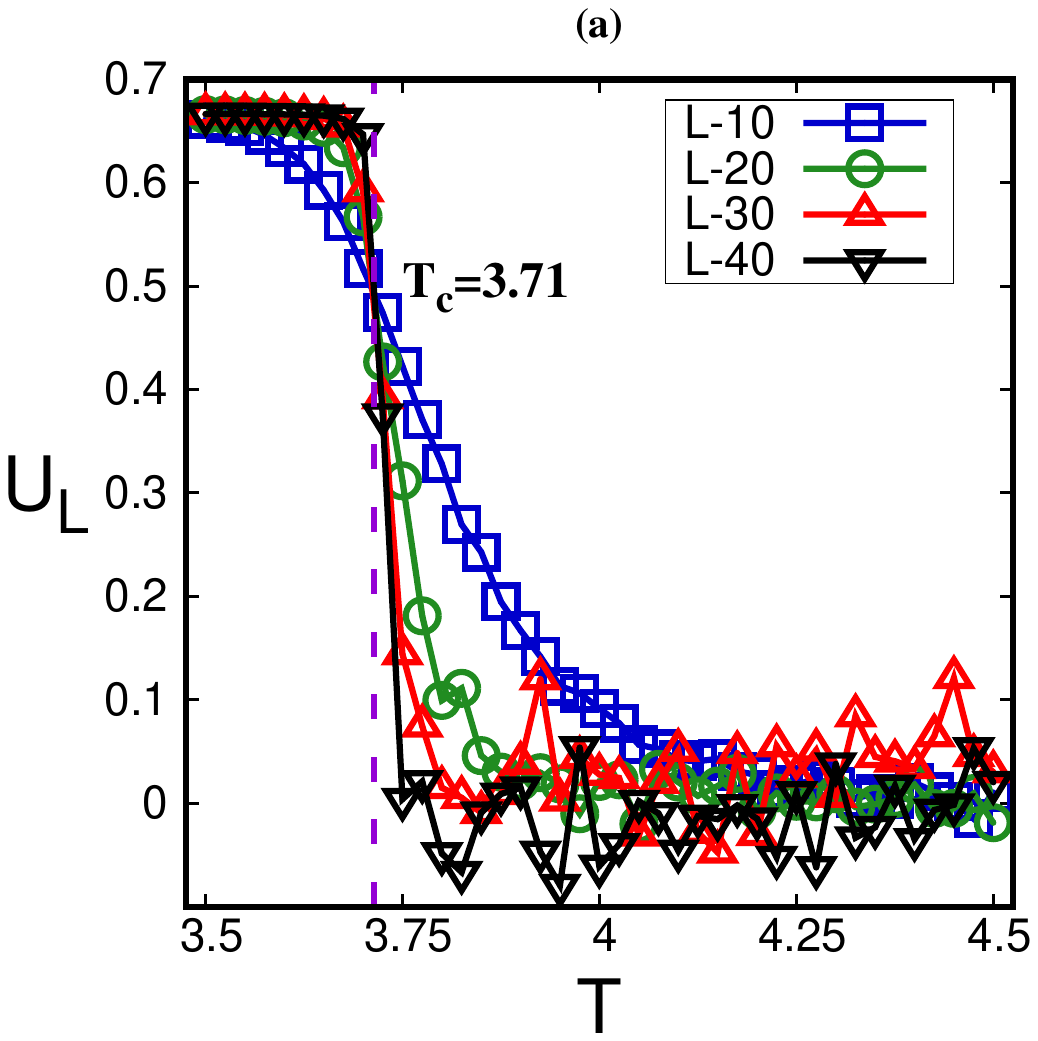}
    \includegraphics[angle=0,height=5cm,width=4.5cm]{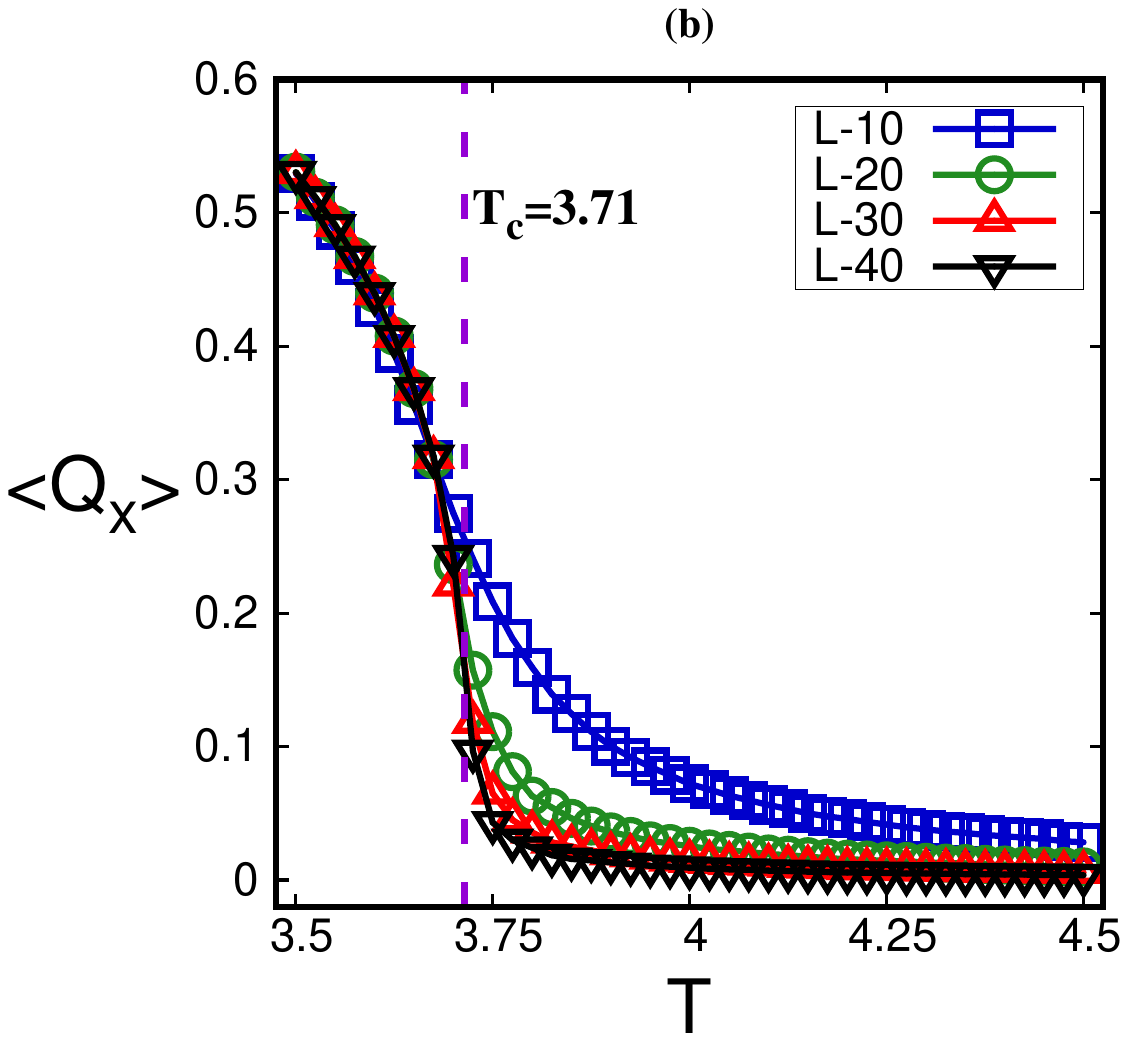}
    \includegraphics[angle=0,height=5cm,width=4.5cm]{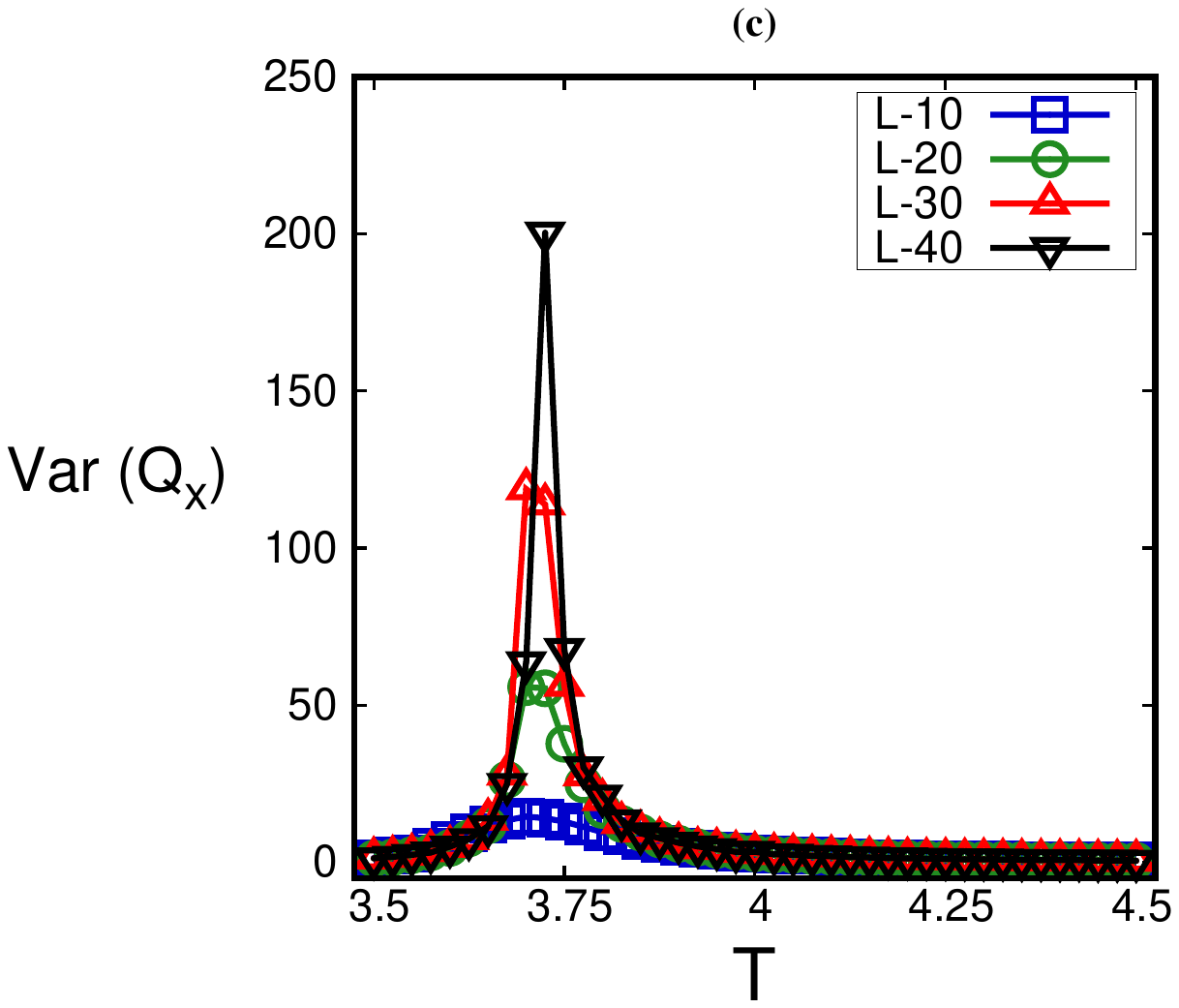}
    \caption{The (a) fourth-order Binder cumulant ($U_{L}$), (b) the dynamical order parameter ($\langle Q_x \rangle$) and (c) the Var $(Q_x)$ are plotted against the temperature ($T$) for different system sizes $L=10,20,30,40$  with anisotropy $\Omega$=0.6 and field amplitudes $h_{0x}=0.3$, $h_{0y}=0.1$. Here,  $f=0.01$ and $\lambda=10$.}
    \label{fig:bilinear-finite}
\end{figure}
\begin{figure}[h!]
    \centering
    \includegraphics[angle=0,height=6cm,width=6cm]{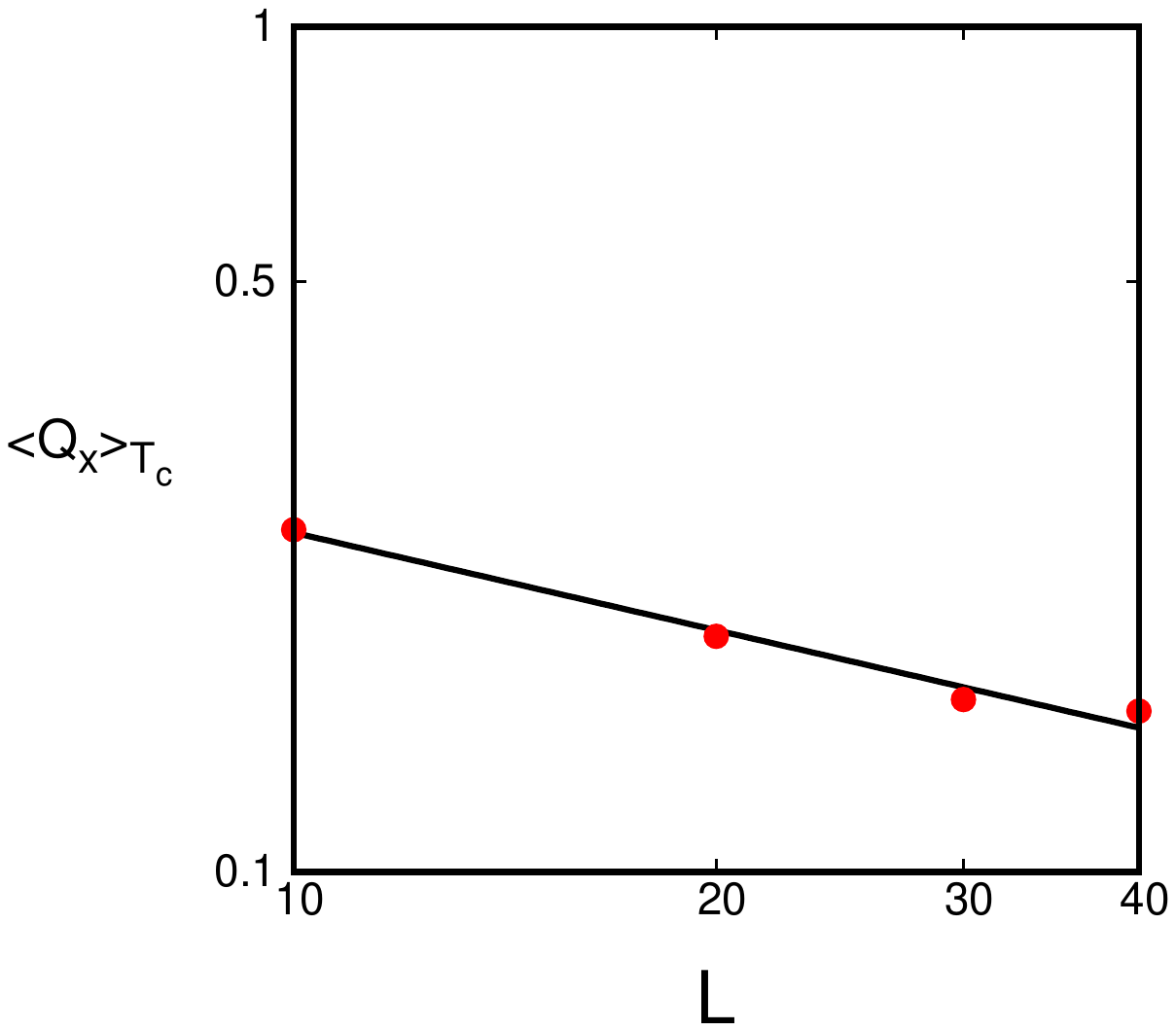}
    \includegraphics[angle=0,height=6cm,width=6cm]{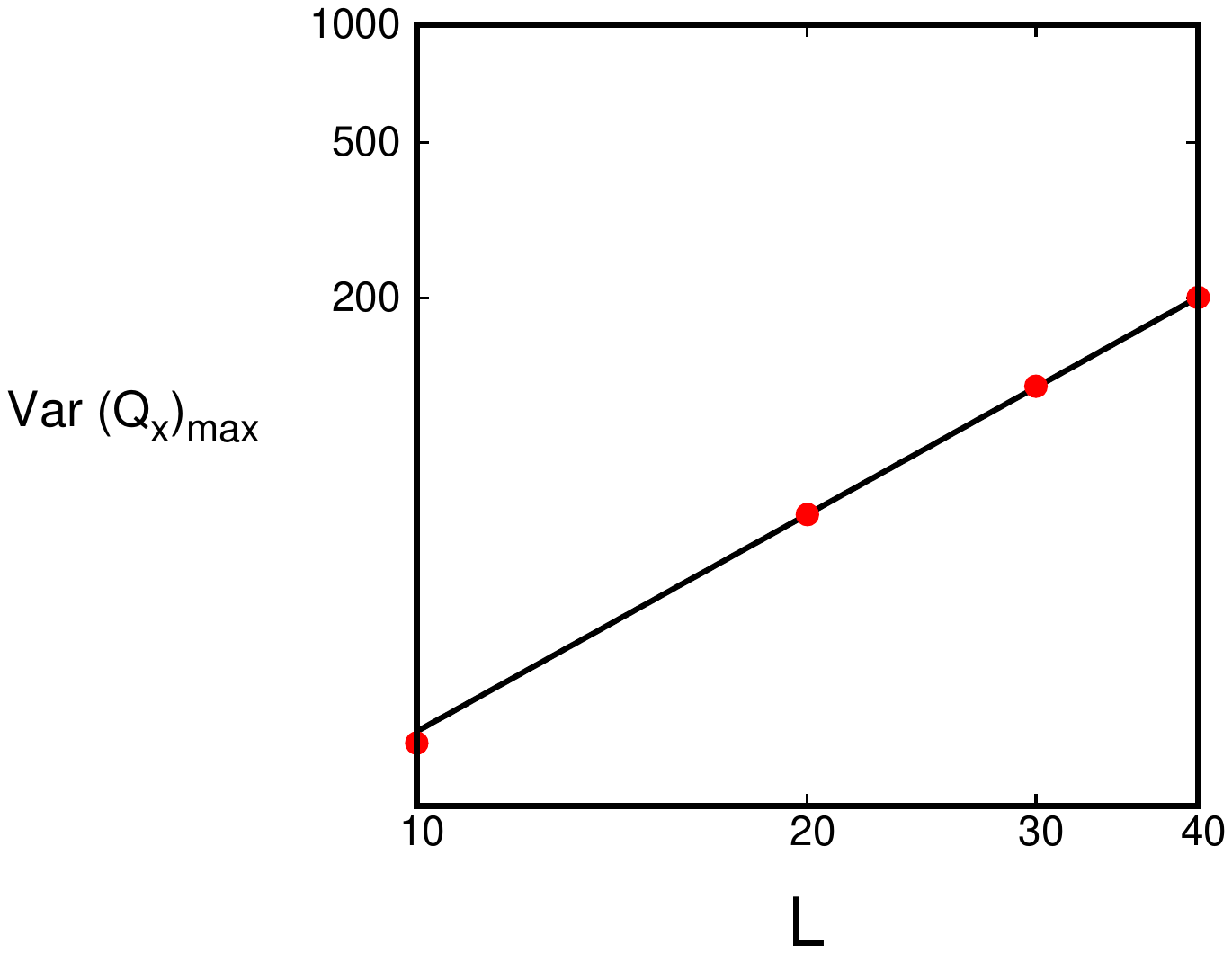}
    
    \caption{The scaling behaviour of (a) the dynamic order parameter $(\langle Q_{x} \rangle _{T_c} \sim L^{-{\frac{\beta}{\nu}}})$ at transition point and the scaling behaviour of (b) the maximum of the dynamic susceptibility Var $(Q_{x})_{max} \sim L^{\frac{\gamma}{\nu}}$. Results are shown in log-log scale. The critical exponents estimated $\frac{\beta}{\nu}$=$0.383\pm 0.032$; $\frac{\gamma}{\nu}$=$1.847\pm 0.019$. Here,  $f=0.01$ and $\lambda=10$. }
    \label{fig:bilinear-exponent}
\end{figure}

Any phase transition cannot be confirmed without a finite-size analysis. The existence of a  phase transition in the thermodynamic limit must be accompanied by the growth of critical correlation and the growth of critical fluctuations of the order parameter. Keeping this in mind, the finite size scaling analysis was performed for the magnetic field amplitudes $h_{0x}=0.3$, $h_{0y}=0.1$ (or $p=3.0$), with bilinear exchange anisotropy $\Omega=0.6$ or ($r=0.25$). We present our numerical data in Fig-\ref{fig:bilinear-finite} for the thermal ($T$) variation of the fourth-order Binder cumulant ($U_L$), the dynamic order parameter ($\langle Q_x \rangle$) and the variance (Var $(Q_x)$) of the dynamic order parameter with different lattice sizes ($L$). The vertical dashed line marks the non-equilibrium transition temperature $T_c=3.71$ in the thermodynamic limit. This is obtained from the point (temperature $T$) of the common intersection of the thermal ($T$) variation of the Binder cumulant ($U_L$) for different sizes ($L$) of the system. Certainly, the value of $\langle Q_x \rangle _{T_c}$ depends on the system size ($L$). In the thermodynamic limit ($L \to \infty$), $\langle Q_x \rangle$ must vanish with the scaling law $\langle Q_{x} \rangle _{T_c} \sim L^{-\frac{\beta}{\nu}}$. In Fig-\ref{fig:bilinear-exponent}(a) we present the values of $\langle Q_x \rangle$ at $T_c$ i.e., $\langle Q_x \rangle_{T_c}$, for different system sizes ($L$) in log-log plot. The solid line is the fitting of the data points with the proposed scaling form $\langle Q_{x}\rangle_{T_c} \sim L^{-\frac{\beta}{\nu}}$ with an estimate of the critical exponent $\frac{\beta}{\nu}=0.383 \pm 0.032$. We also investigate the finite size scaling behavior of the maximum value of the {\it dynamic susceptibility} i.e Var $(Q_x)_{max}$. This must diverge in the thermodynamic limit ($L \to \infty$) obeying the scaling law 
Var $(Q_x)_{max} \sim L^{\frac{\gamma}{\nu}}$. This is the 
signature of the growth of critical correlation (associated with the enormous growth of critical fluctuations) of the dynamic order parameter. We show the size dependence of Var $(Q_x)$ in log-log plot for the same set of parameter values ($h_{0x}$ and $\Omega$) in the Fig-\ref{fig:bilinear-exponent}(b) . The solid line is a fitting of the data supporting the scaling form Var $(Q_{x})_{max} \sim L^{\frac{\gamma}{\nu}}$ with the estimate of the exponent $\frac{\gamma}{\nu}=1.847 \pm 0.019$.

\begin{figure}[h!]
    \centering
    \includegraphics[angle=0,height=6cm,width=6cm]{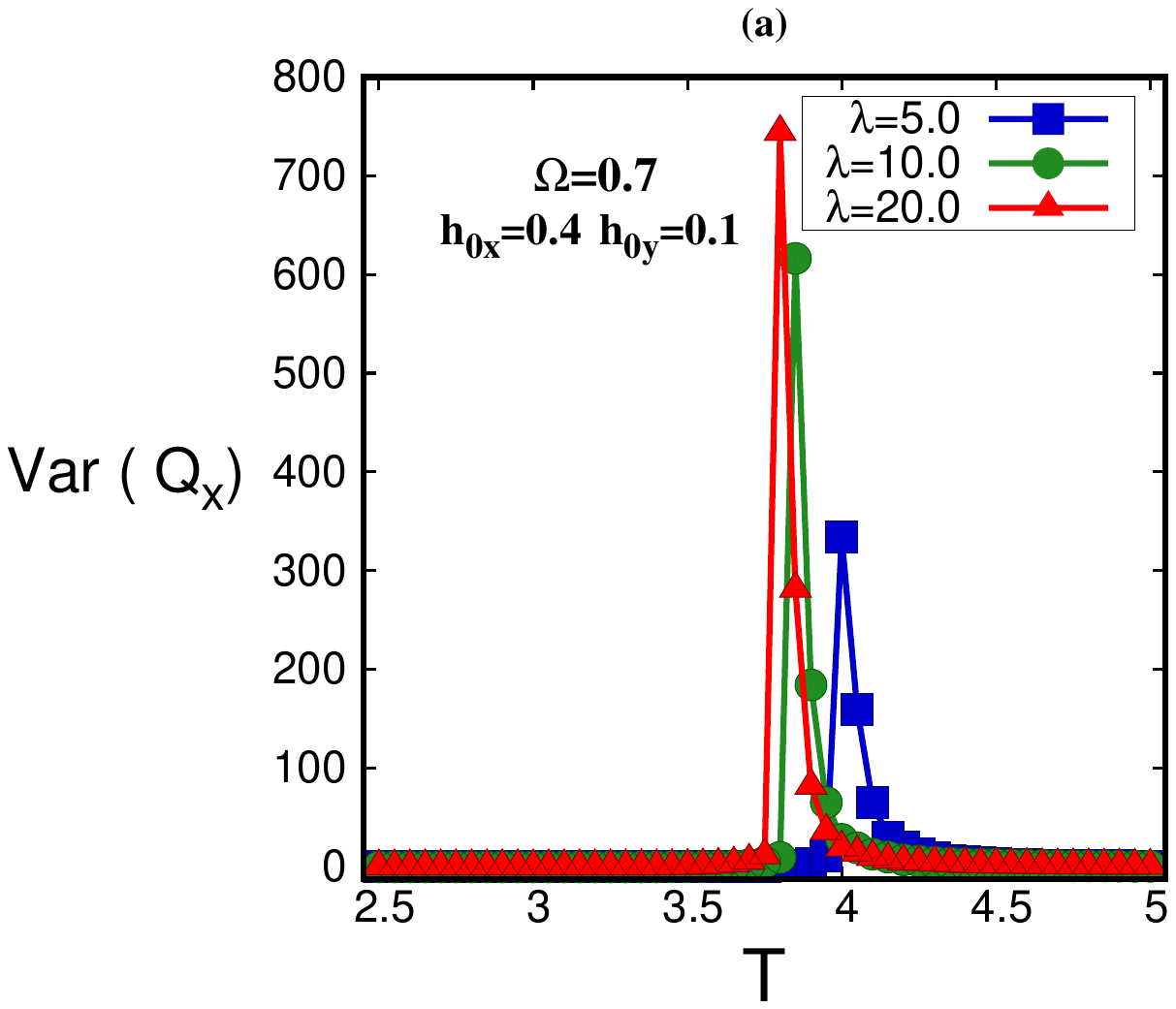}
    \includegraphics[angle=0,height=6cm,width=6cm]{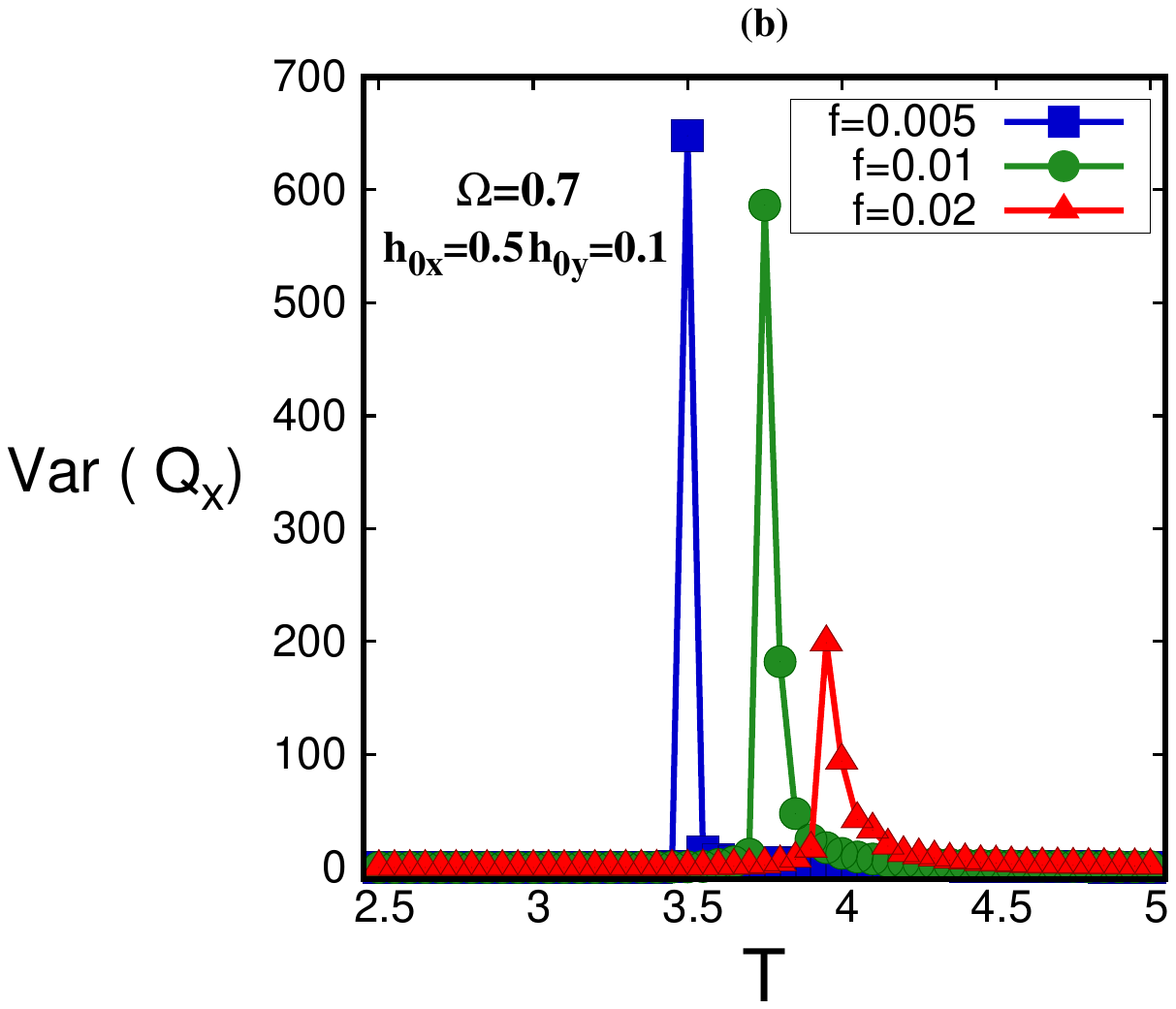}
    
    \caption{The (a) Var $(Q_x)$ is plotted against the temperature ($T$)
    for different values of wavelengths ($\lambda$) and (b) Var $(Q_x)$ is
    plotted against the temperature ($T$) for different values of frequencies ($f$). Here, $L=20$.}
    \label{fig:bilinear-freq-wave}
\end{figure}

In order to comprehensively conclude the study of bilinear anisotropy, it becomes crucial to comment on the variation of the transition temperature with respect to both the wavelength($\lambda$) and frequency ($f$) of the propagating field. Fig-\ref{fig:bilinear-freq-wave} provides the variation of $T_c$ with wavelength and frequency of the elliptically polarized magnetic field propagating along the z-direction. For constant anisotropy strength $\Omega=0.7$ and field amplitude $h_{0x}=0.5$, $h_{0y}=0.1$, the peak position of Var $(Q_{x})$ shifts towards the right if we increase frequency from $f=0.005$ to $f=0.02$. The transition temperature increases with the increase in frequency. On the other hand, as the wavelength increases, the transition takes place at a lower temperature. 

\vskip 0.5cm


\noindent {\bf B. Single site anisotropy ($D$):}\\

\begin{figure}[h!]

\includegraphics[angle=0, height=6cm,width=6.5cm]{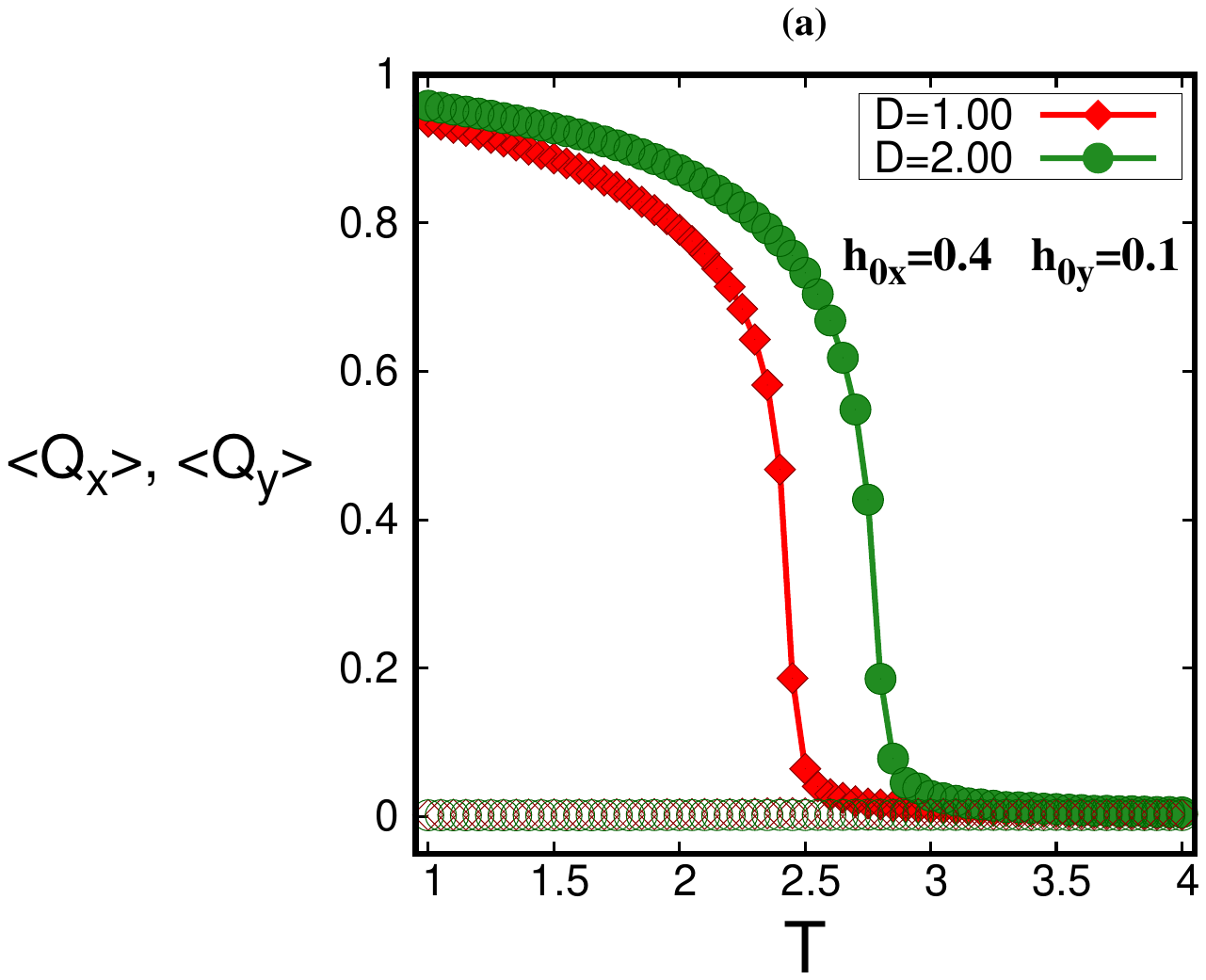}
\includegraphics[angle=0,height=6cm,width=6.5cm]{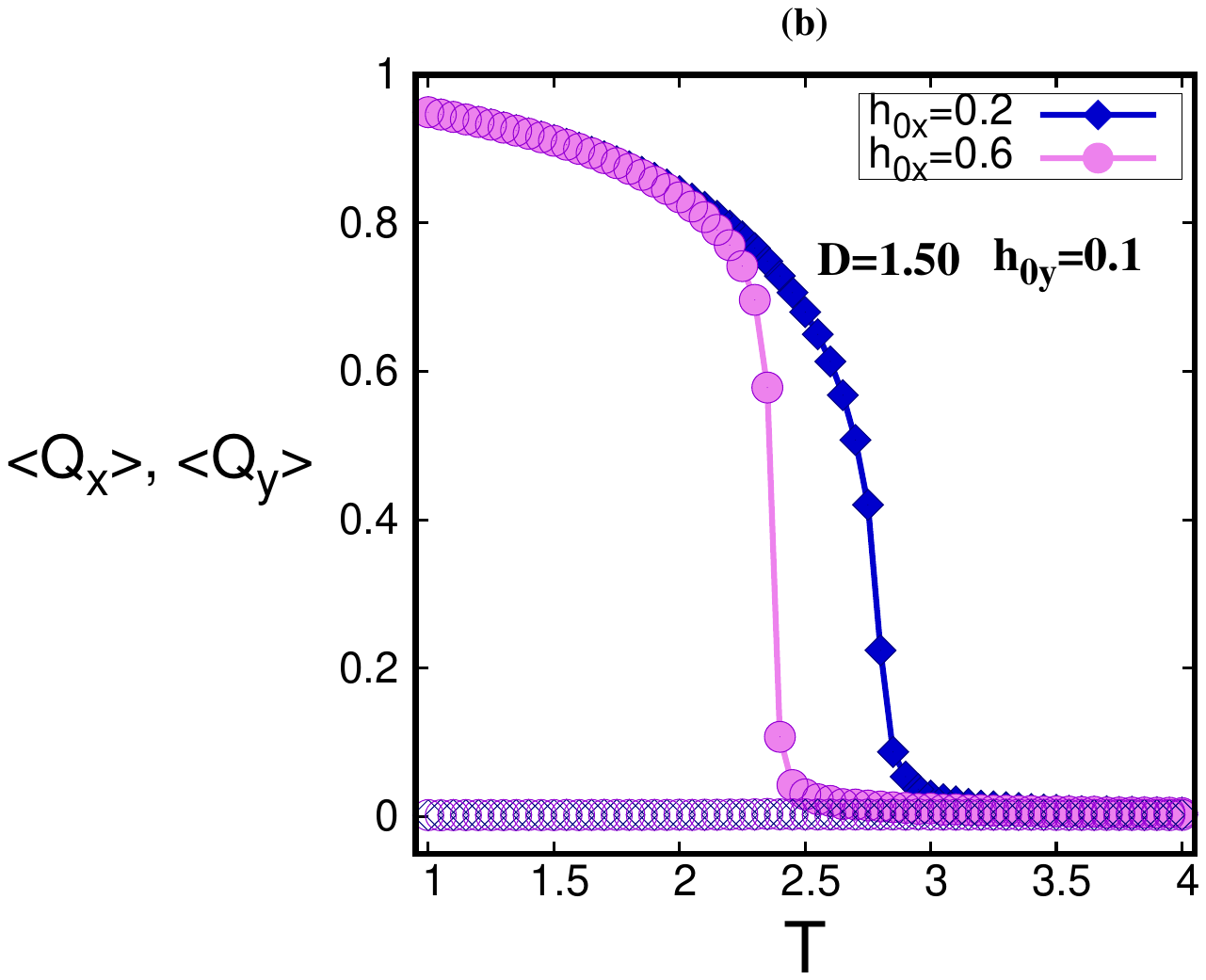}\\
\includegraphics[angle=0,height=6cm,width=6.5cm]{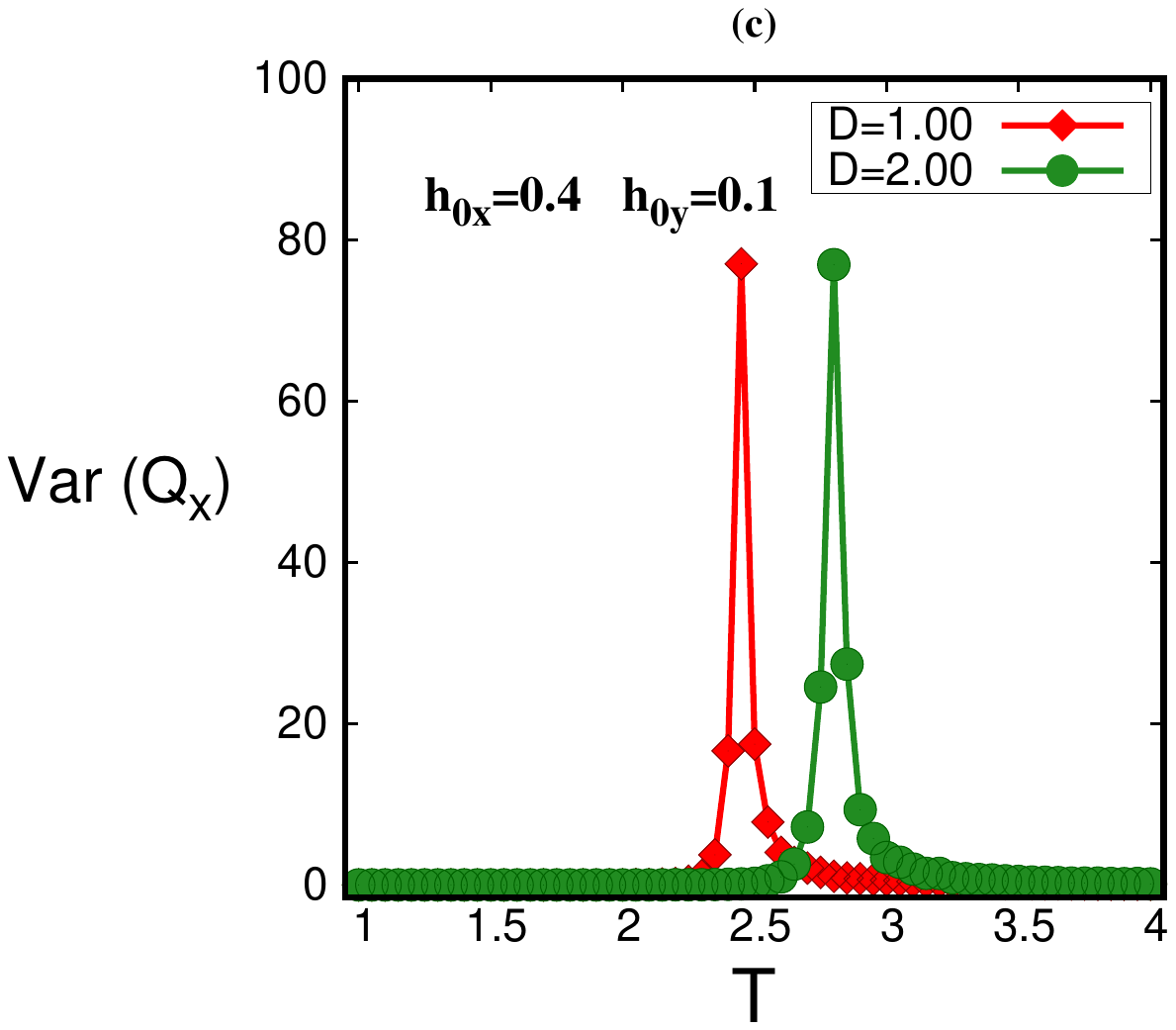}
\includegraphics[angle=0,height=6cm,width=6.5cm]{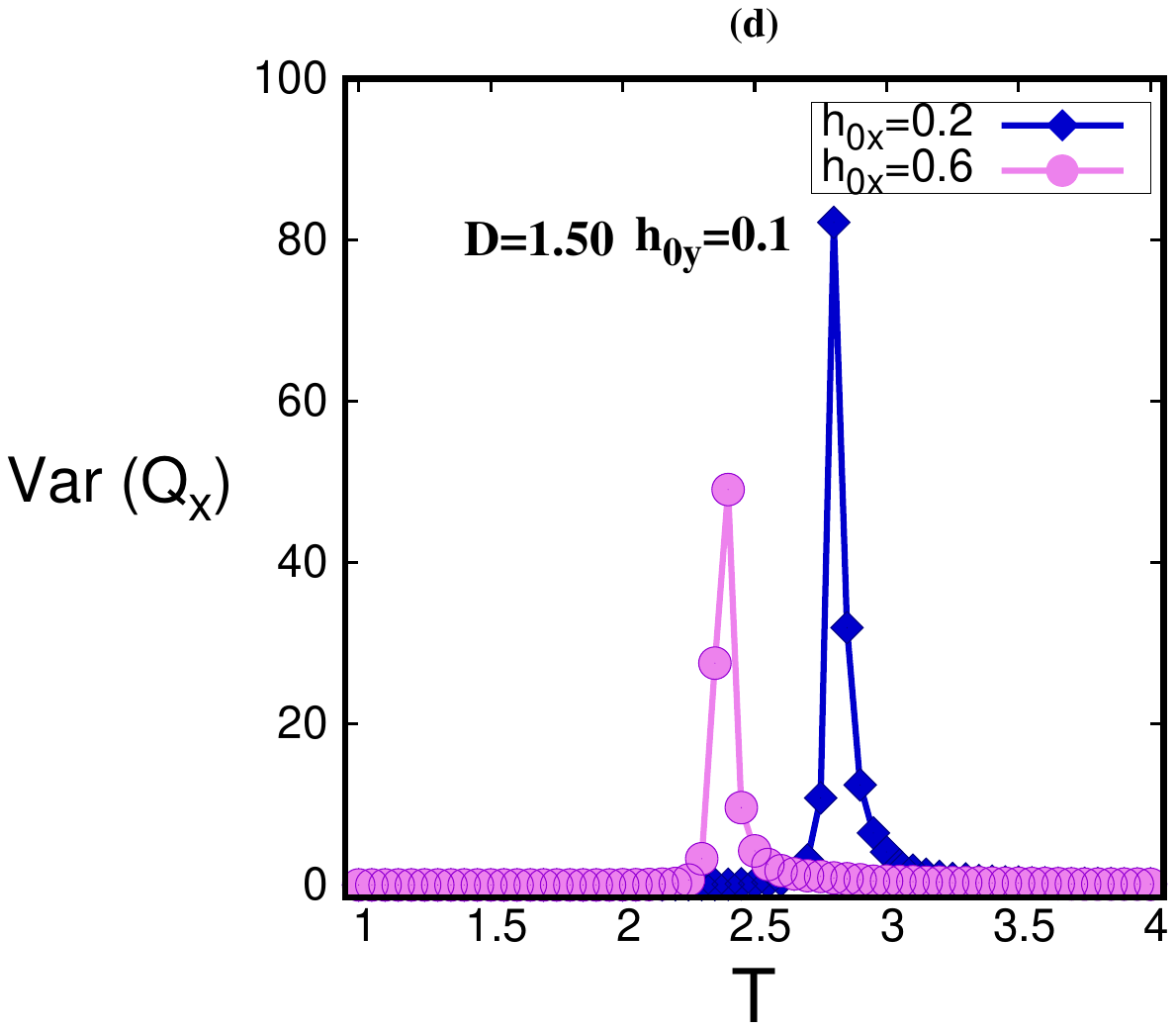}\\
\vskip 0.3cm
\includegraphics[angle=0,height=6cm,width=6.5cm]{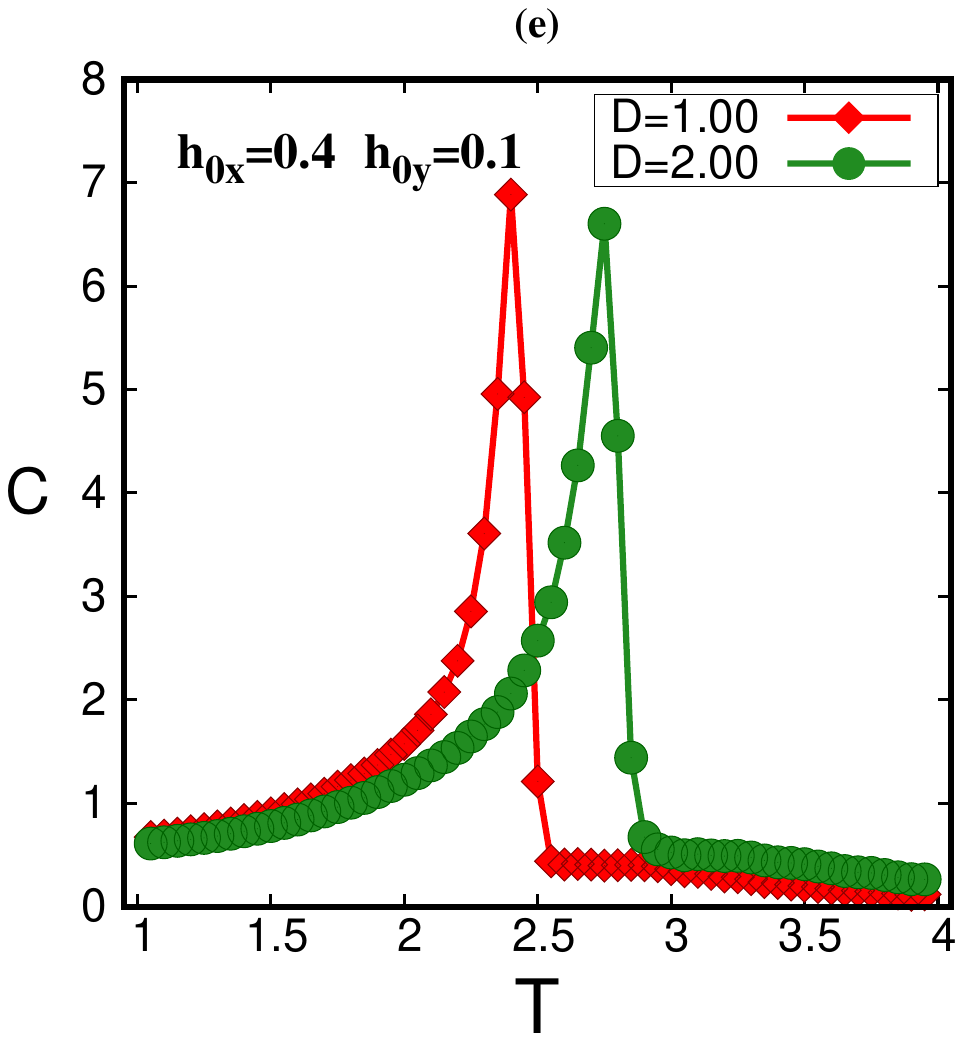}
\includegraphics[angle=0,height=6cm,width=6.5cm]{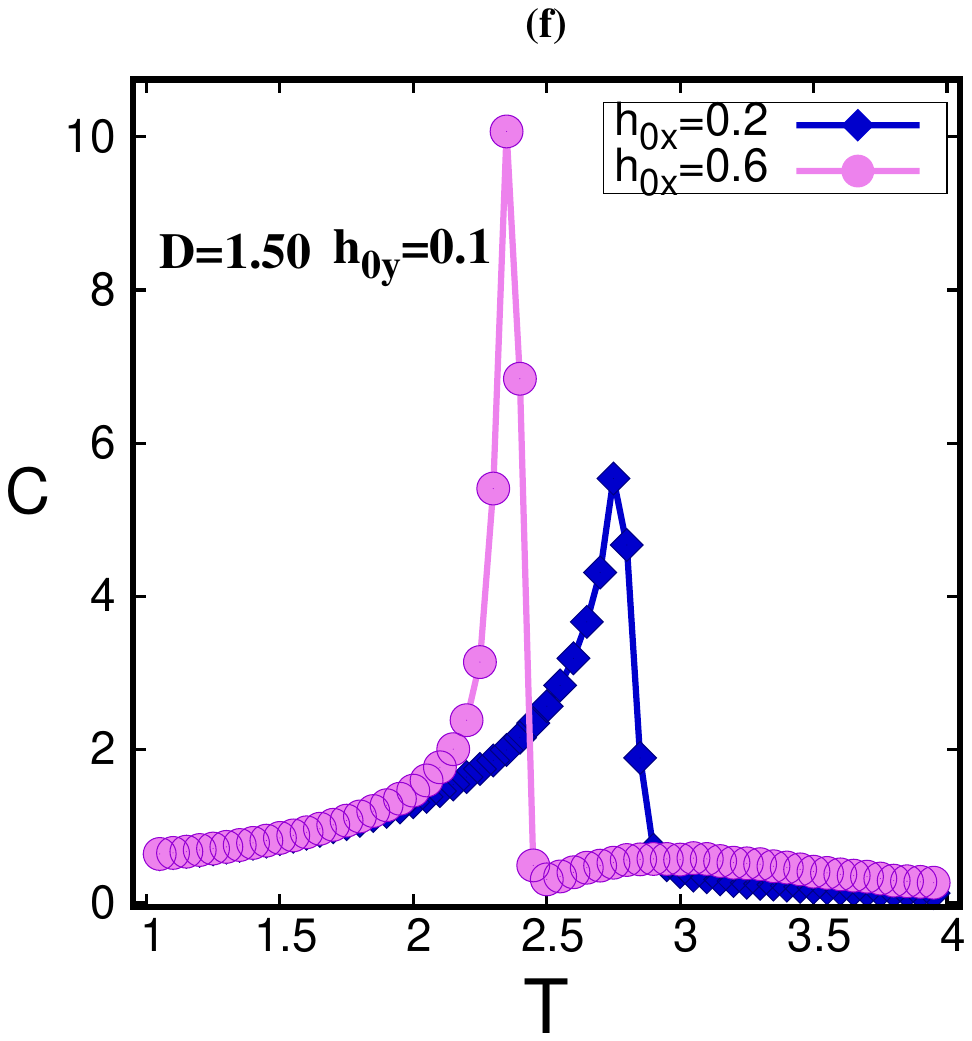}

\caption{Temperature $(T)$ dependences of components of the order parameter $\langle Q_x \rangle$ (shown by solid symbols) and $\langle Q_y \rangle$
(shown by open symbols)
(in (a) and (b)), variance of x component of order parameter 
(Var $(Q_{x})$) (in (c) and (d)), and dynamic specific heat ($C$)(in (e) and (f)) for different anisotropy ($D$) and for different field amplitudes $h_{0x}$ (fixed $h_{0y}=0.1$). Left panel is for constant field amplitudes and right panel is for constant anisotropy strength ($D$). Here, $L=20$, $f=0.01$ and $\lambda=10$.}
\label{fig:dtype-observables}
\end{figure}

 \noindent In this section, we report the influence of the single-site anisotropy ($D$) on the nonequilibrium behavior, found in the XY model in three dimensions. We show (in Fig-\ref{fig:dtype-observables}) the temperature dependences of the dynamic order parameter ($\langle Q_x \rangle$) and the variance of the order parameter Var $(Q_x)$ (dynamic susceptibility) and dynamic specific heat ($C$) for different values of the control parameter $D$, $h_{0x}$ (with fixed $h_{0y}=0.1)$. Like in the case of bilinear exchange anisotropy, here we also observed the partial symmetry-breaking of the order parameter. The $x$-component of the order parameter ($\langle Q_x \rangle$) gets dynamically ordered (at the transition temperature $T_c$) via the dynamic symmetry-breaking
 , while the $y$-component $\langle Q_y \rangle$ remains zero (dynamically symmetric for all temperatures)(in Fig-\ref{fig:dtype-observables}(a) and Fig-\ref{fig:dtype-observables}(b)). The variance (Var $(Q_x)$) of the order parameter ($Q_x$) gets sharply peaked at the transition point(in Fig-\ref{fig:dtype-observables}(c) and Fig-\ref{fig:dtype-observables}(d)), which diverges in the thermodynamic limit ($L \to \infty$). The pseudocritical temperature $T_c$ (or the transition temperature for the finite-sized system) has been estimated from the temperature, which maximizes the variance (Var $(Q_x)$). The dynamic
 specific heat $C$ has been shown as function of temperature in Fig-\ref{fig:dtype-observables}(e) and in Fig-\ref{fig:dtype-observables}(f).
 
\begin{figure}[h!]

\includegraphics[angle=0,height=6cm,width=8cm]{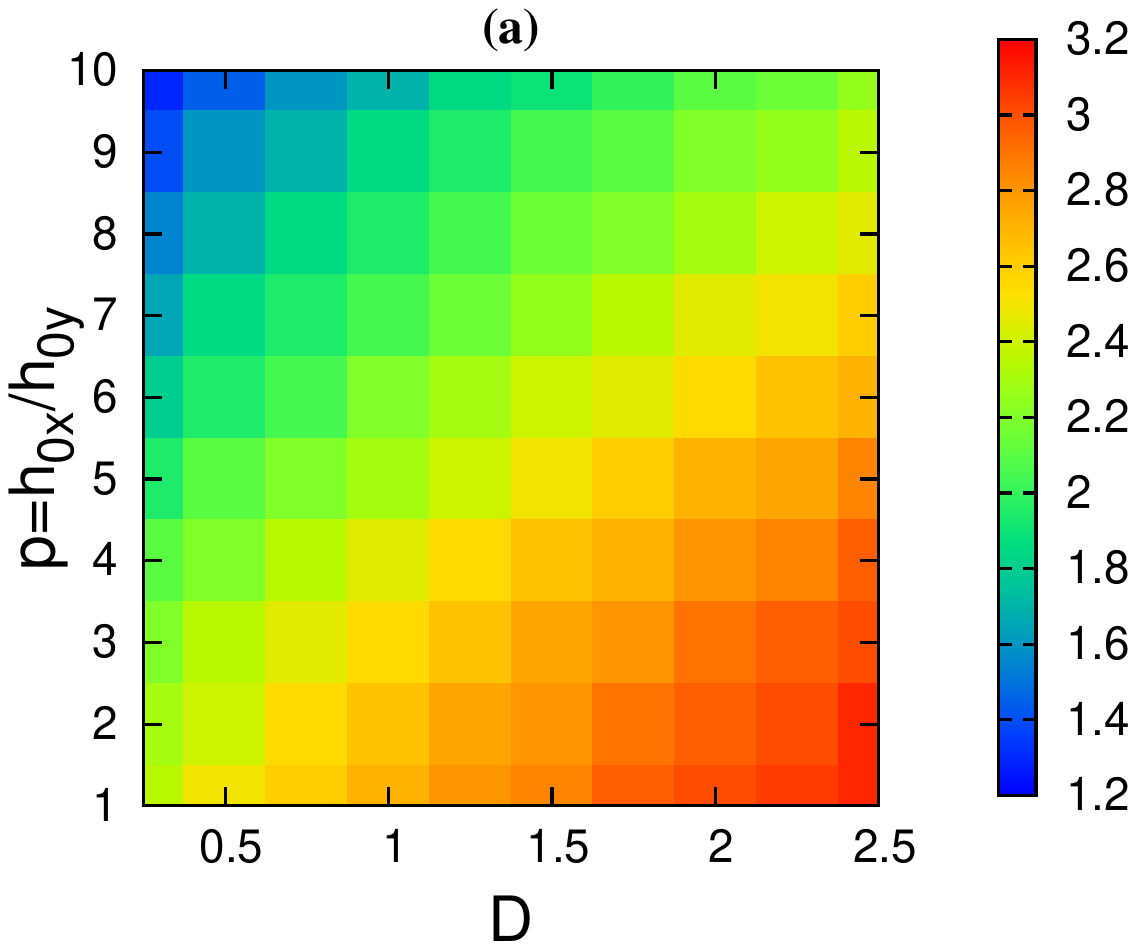}
\includegraphics[angle=0,height=6cm,width=8cm]{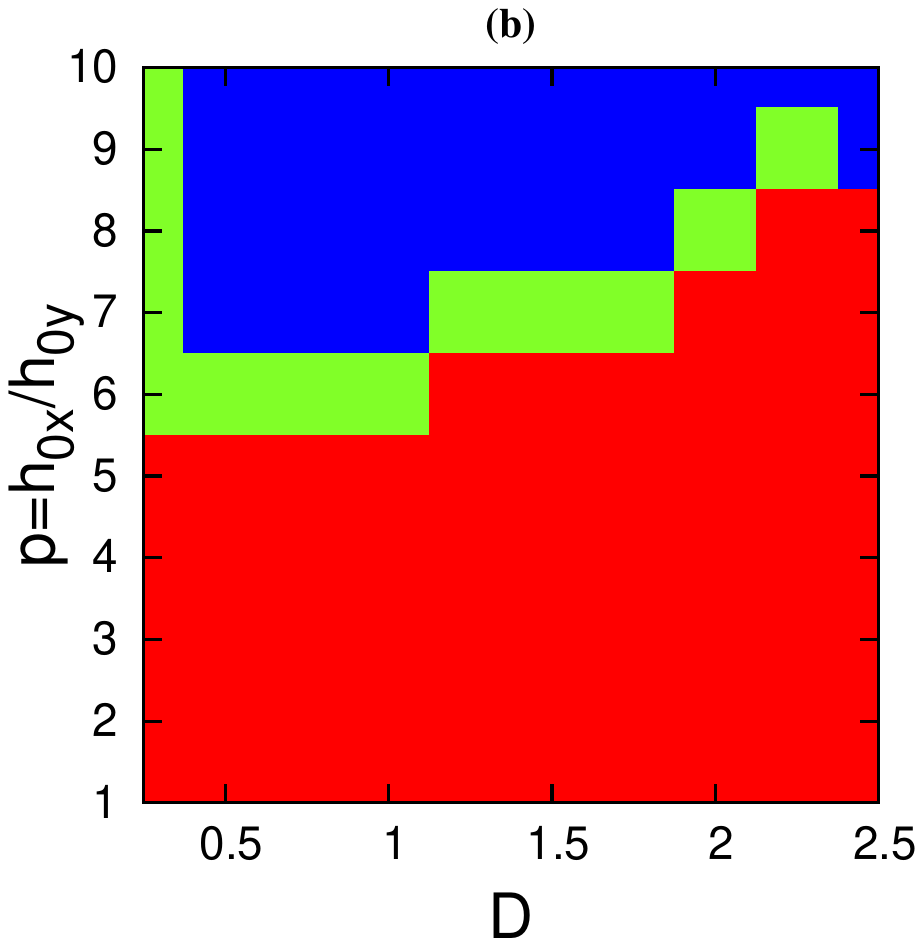}

\caption{(a) The comprehensive phase diagram (or the image plot of $T_c$) in $D-p({={h_{0x} \over {h_0y}}})$ plane. The transition temperature is obtained from the position of the peak of Var $(Q_{x})$ plotted against temperature ($T$). (b) The nature (continuous or discontinuous) or the order (first or second) of the transitions marked by different colors. The region of discontinuous or first order transition is shown in blue, the region of continuous or second order transition is shown in red and a narrow region of weakly first order transition is shown in green color. Here, $L=20$, $f=0.01$ and $\lambda=10$. }
\label{fig:dtype-phase}
\end{figure}

\begin{figure}[h!]

\includegraphics[angle=0,height=5cm,width=5cm]{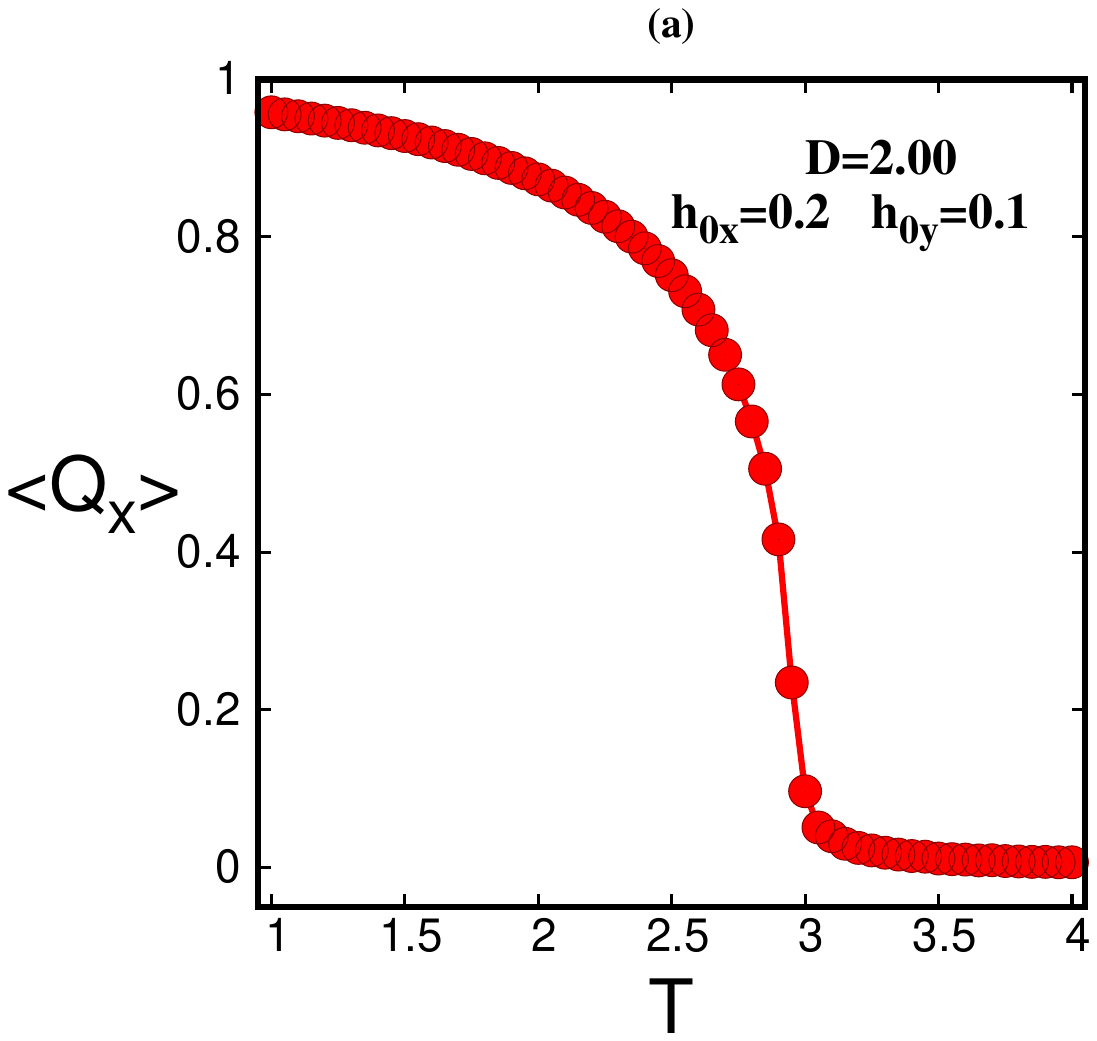}
\includegraphics[angle=0,height=5cm,width=5cm]{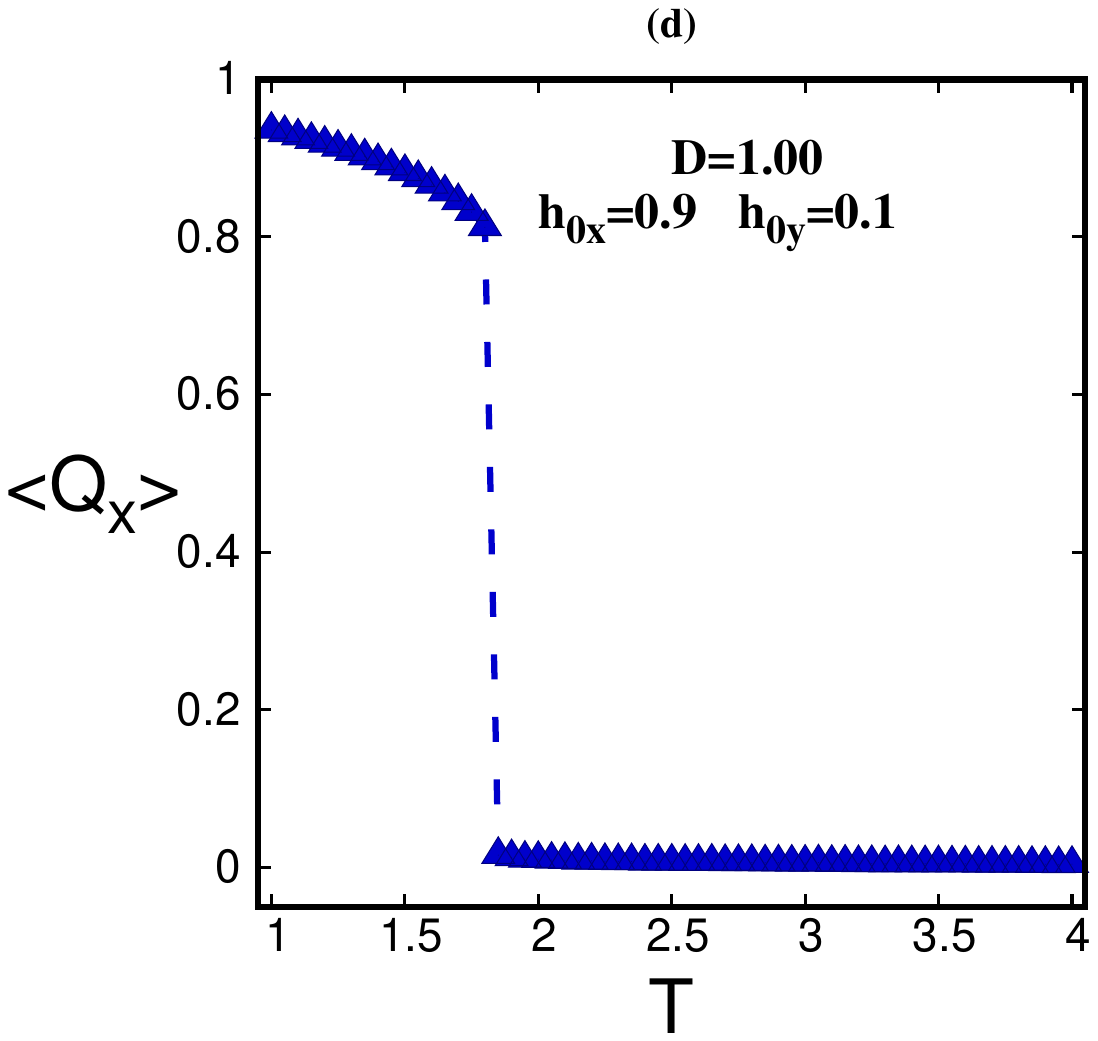}
\includegraphics[angle=0,height=5cm,width=5cm]{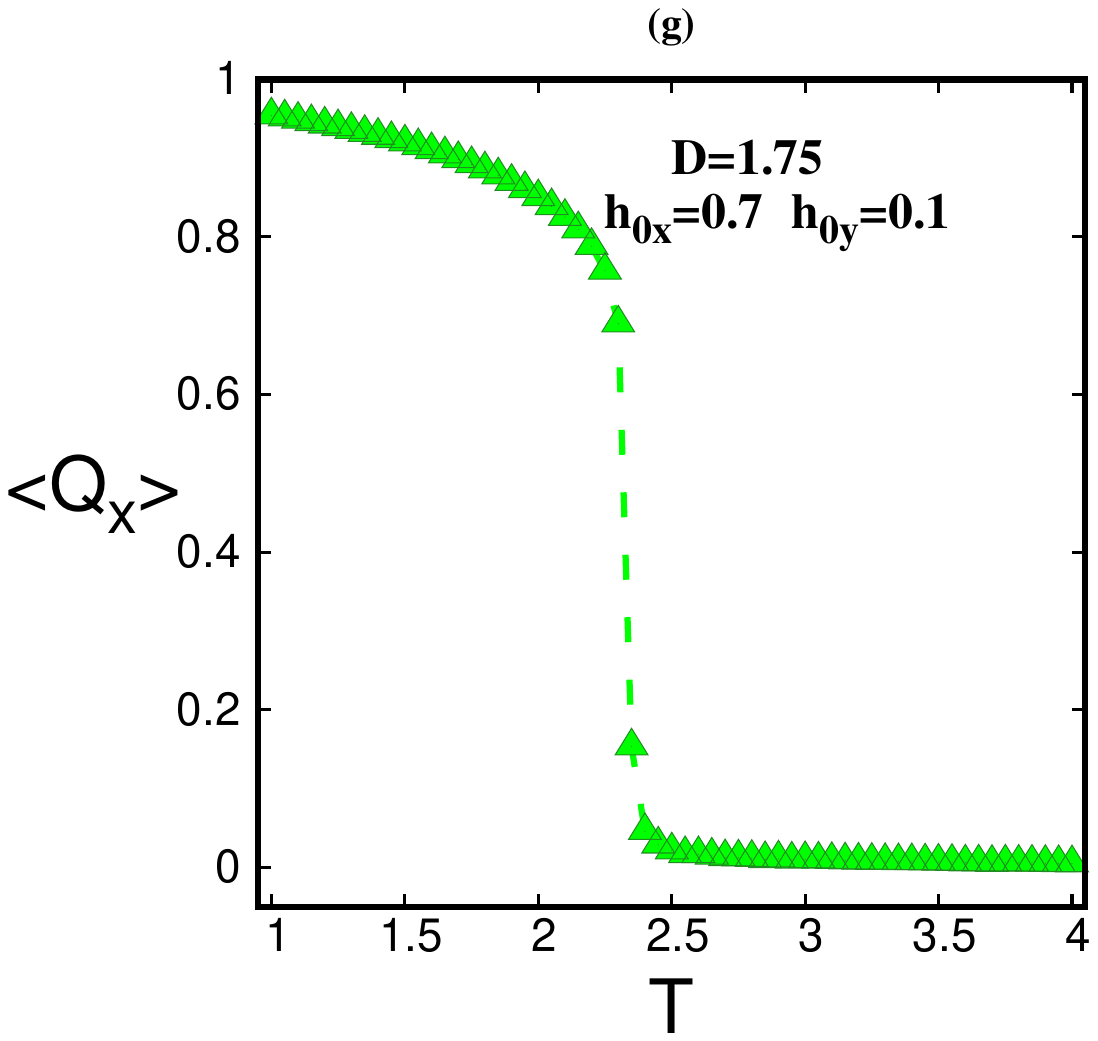}\\
\includegraphics[angle=0,height=5cm,width=5cm]{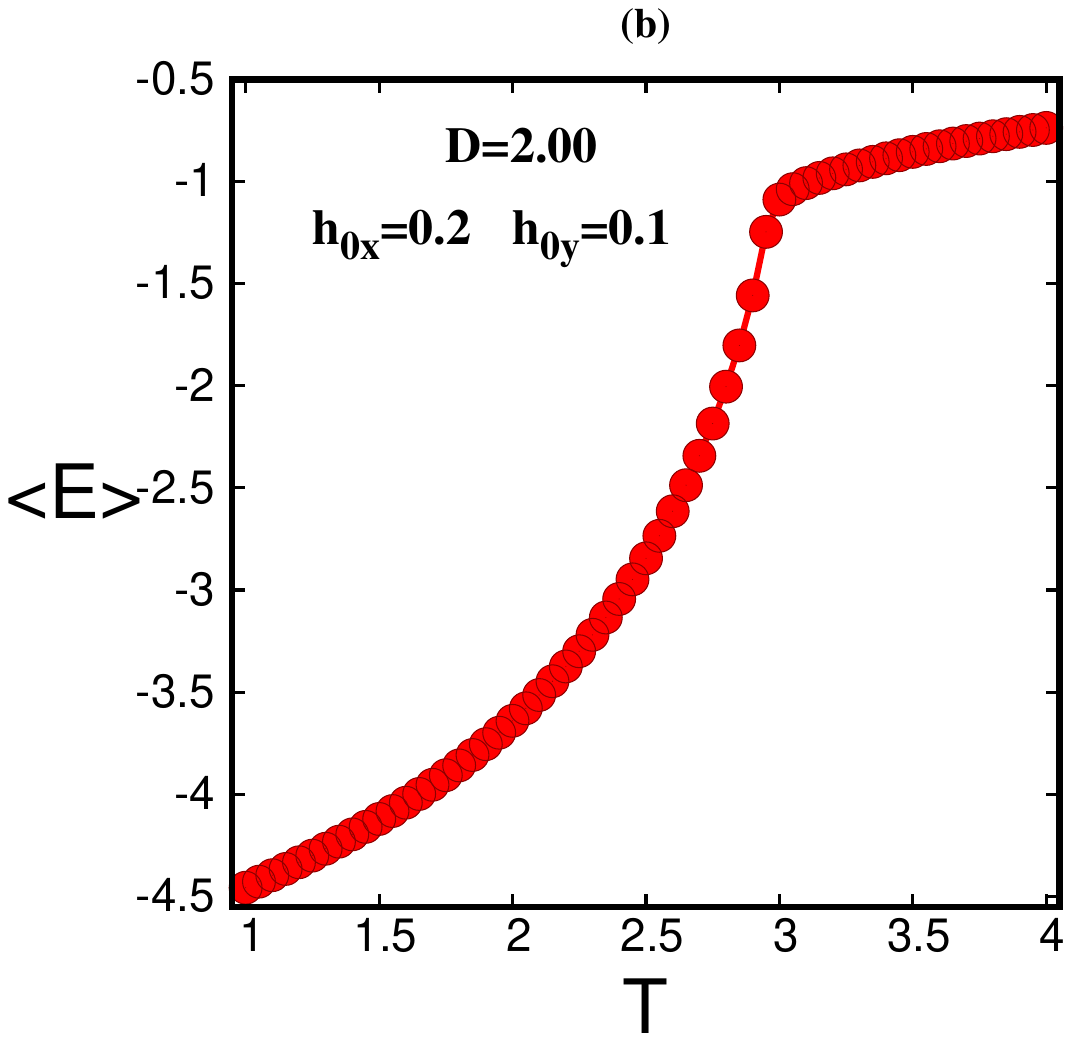}
\includegraphics[angle=0,height=5cm,width=5cm]{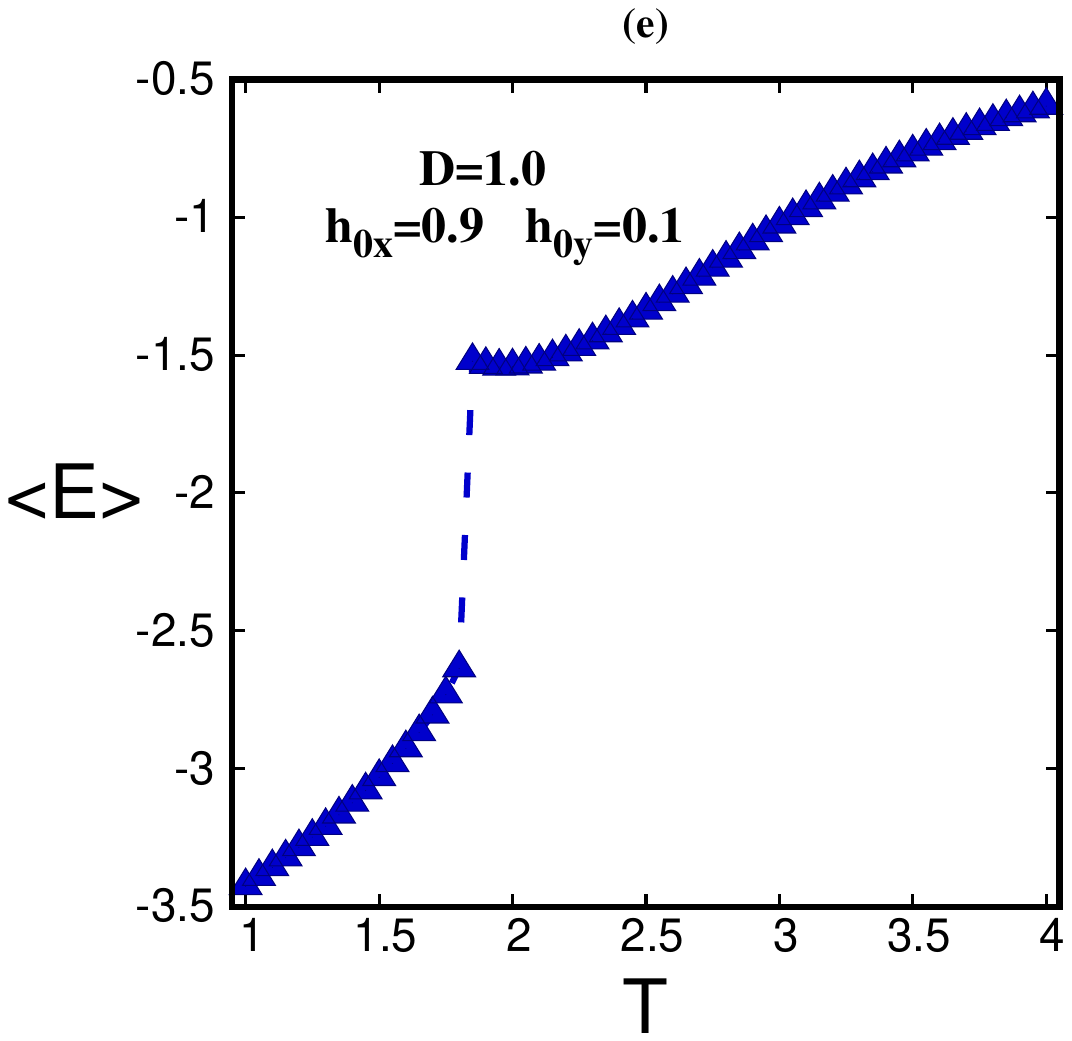}
 \includegraphics[angle=0,height=5cm,width=5cm]{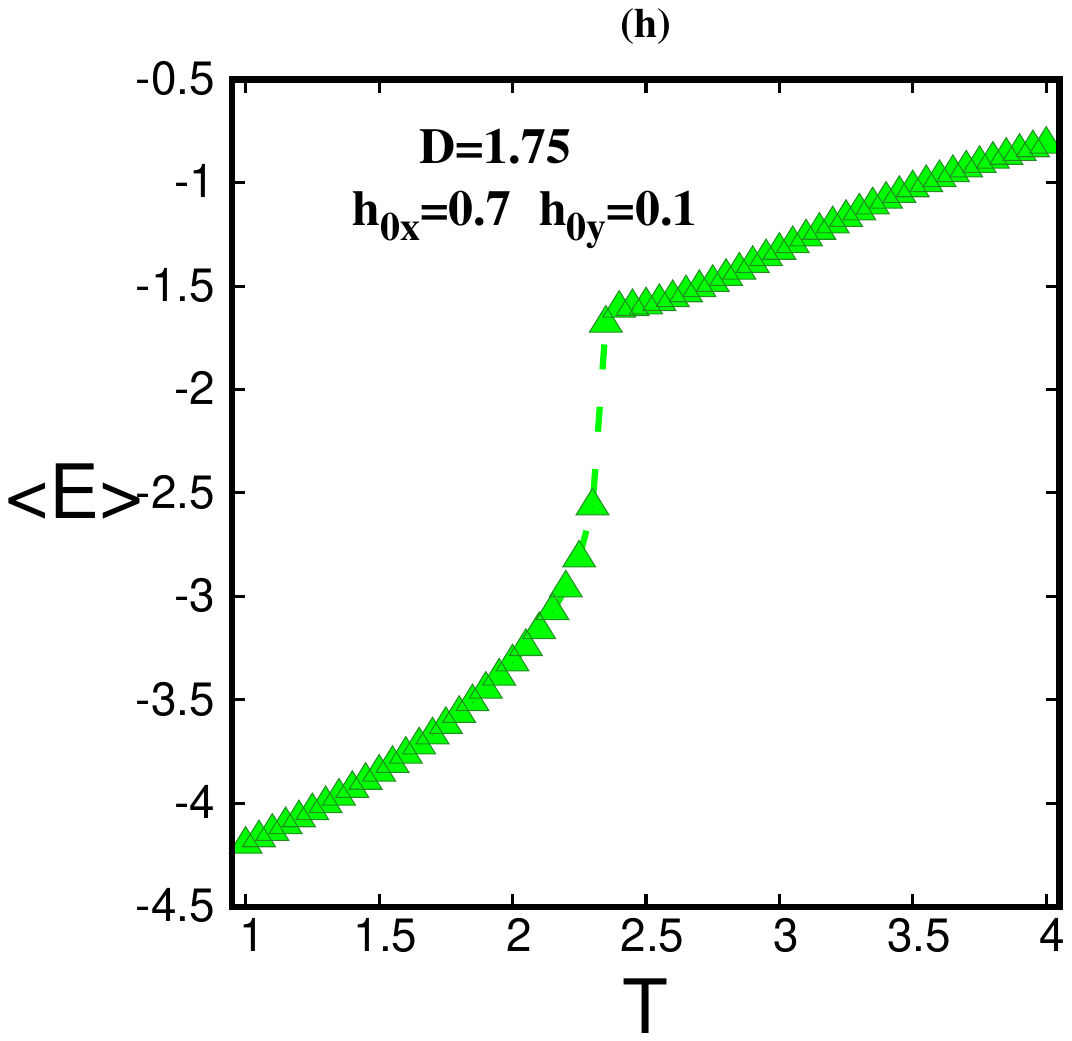}\\
\includegraphics[angle=0,height=5cm,width=5cm]{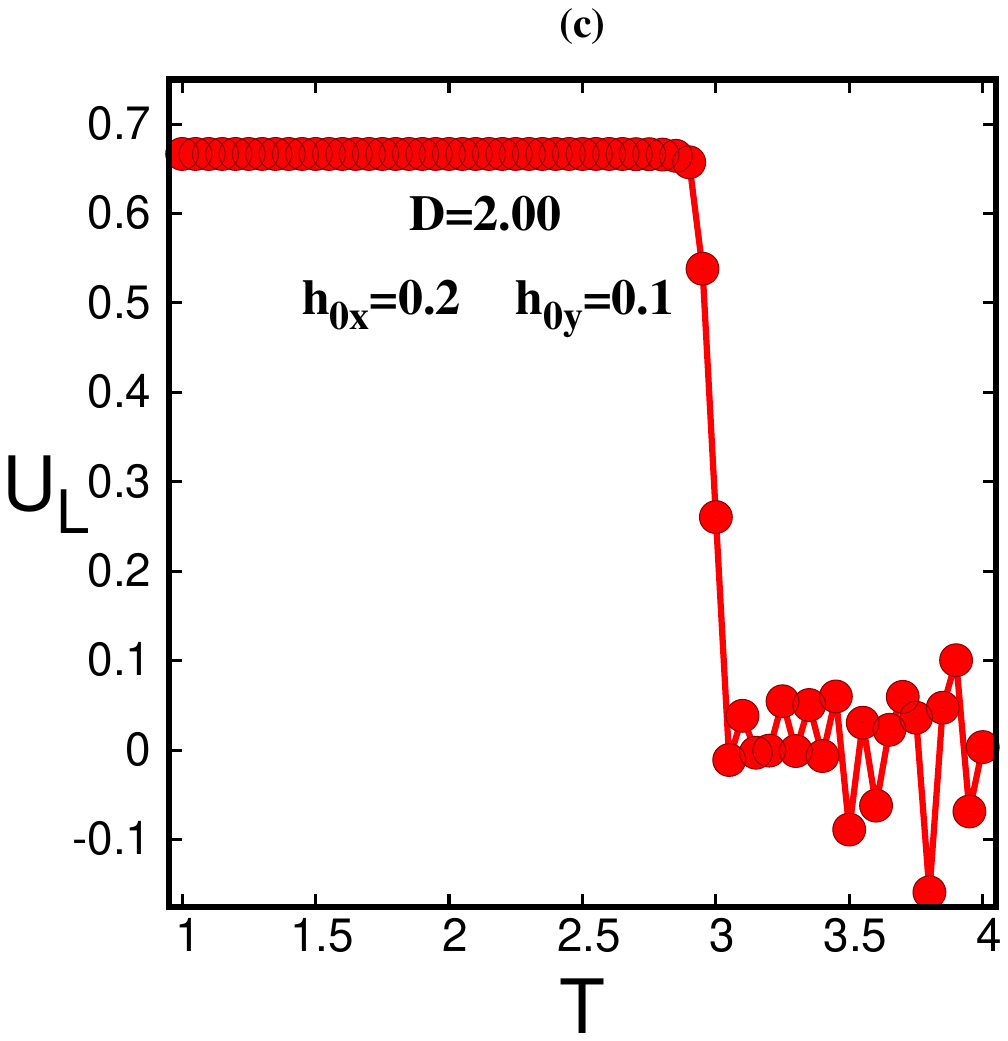}
\includegraphics[angle=0,height=5cm,width=5cm]{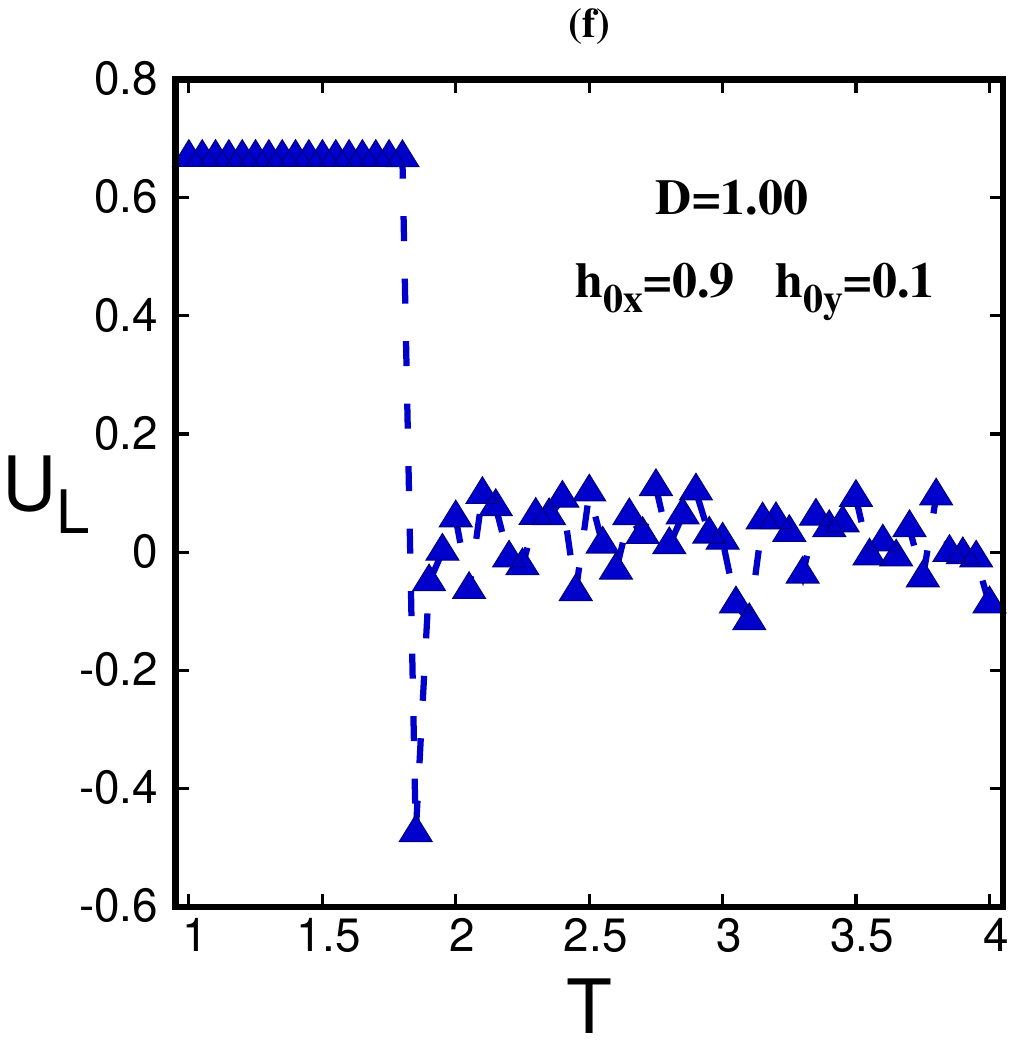}
 \includegraphics[angle=0,height=5cm,width=5cm]{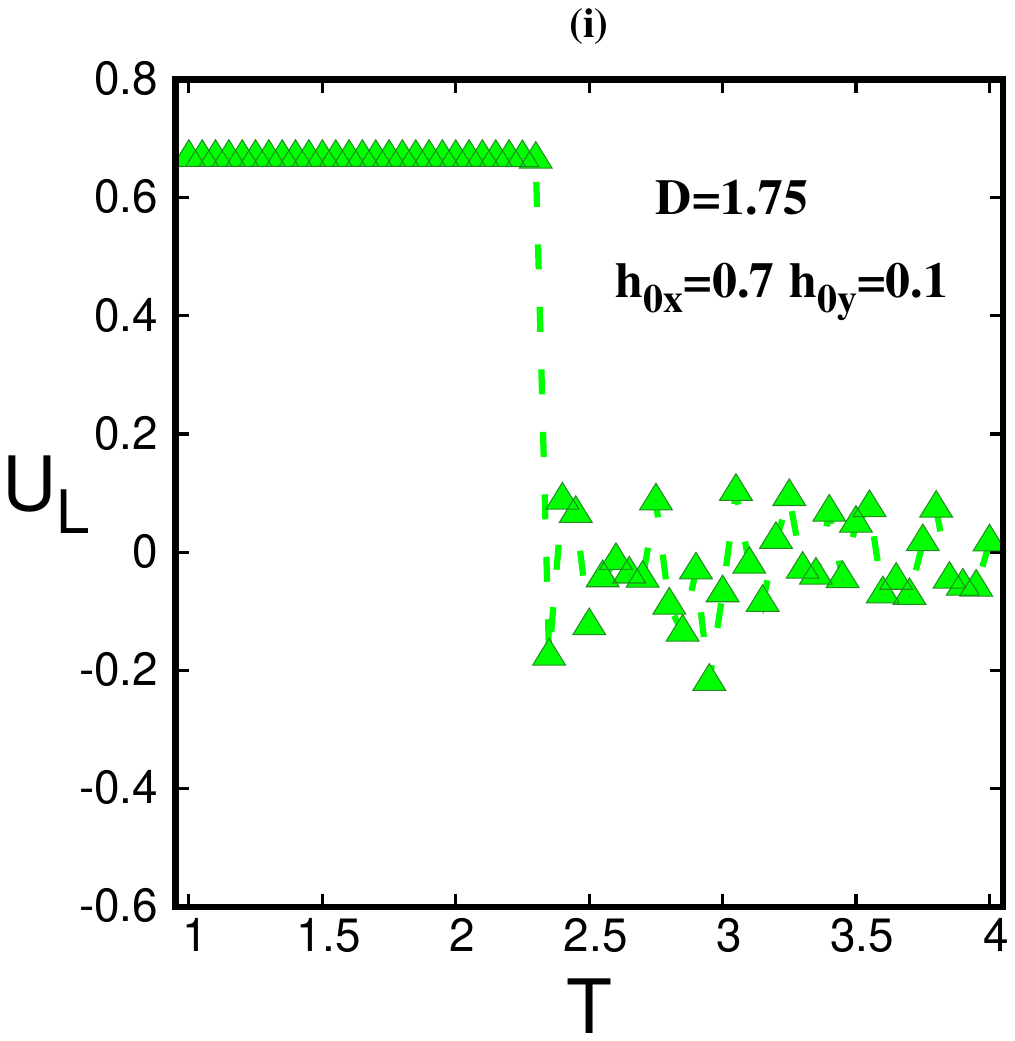}

\caption{Temperature ($T$) dependences of order parameter $(\langle Q_x \rangle)$
(in (a),(d) and (g)), dynamic energy density $(\langle E \rangle)$ (in (b),(e) and (h)) and fourth order Binder cumulant $({U_{L}})$ (in (c),(f) and (i)) for fixed set of values of field amplitudes ($h_{0x}$ and $h_{0y}$), single-site anisotropy  $(D)$.  In the left panel: The thick (red) lines indicate second order (continuous) transition. In the middle panel: the dashed (blue) lines indicate first order (discontinuous) transition. In the rightmost panel: The dashed (green) lines indicate weak first order transition. Here, $L=20$, $f=0.01$ and $\lambda=10$.}
\label{fig:dtype-order-transition}
\end{figure}

 Fig-\ref{fig:dtype-observables} represent the dependence of the transition temperature $T_c$ on the strength of single-site anisotropy ($D$) and also on the field amplitude $h_{0x}$. The y component of the polarized field $h_{0y}=0.1$ is kept constant here. We find that the critical temperature $T_c$ increases with an increase in anisotropy strength $D$ (Fig-\ref{fig:dtype-observables}(c)) for a fixed value of $h_{0x}$. On the other hand, the critical temperature decreases with the increase of field amplitude $h_{0x}$ (while keeping the y component $h_{0y}$ constant) for a fixed value of the strength of single site anisotropy ($D$)(Fig-\ref{fig:dtype-observables}(d)). The comprehensive phase diagram (actually the image plot of the pseudocritical temperature as function of $p$ and $D$) is depicted in the plane formed by the anisotropy strength ($D$) and $p$ (=${{h_{0x}} \over {h_{0y}}}$) (Fig-\ref{fig:dtype-phase}(a)).
 This diagram is called the phase diagram due to the following sense: here the image plot is nothing but a surface plot. Above the surface, the system is dynamically disordered and below the surface the nonequilibrium ordered phase is found. The transition temperature increases as the strength of single-site anisotropy increases. On the other hand, the increase in field amplitude($h_{0x}$) reduces the transition temperature. According to the values of field amplitude ($h_{0x}$) and anisotropy strength ($D$), here also we have observed the appearance of two types of transition (Fig-\ref{fig:dtype-phase}(b)). When the field amplitude is low, the order-disorder phases are accessed by the second-order (continuous) transition (red region). For higher field amplitudes, the order-disorder transition becomes discontinuous (blue region). The nature of transitions was scrutinized by the thermal variations of the order parameter ($Q_x$), corresponding Binder cumulant ($U_L$) as well as energy density ($\langle E \rangle$)(Fig-\ref{fig:dtype-order-transition}). For $D=2.00$ and $h_{0x}=0.2$ and $h_{0y}=0.1$, we observed that the order parameter ($Q_x$) changes continuously from zero to a finite value. Also, there is no abrupt jump in the energy curve, and $U_L$ grows monotonically from 0 to 2/3 indicating the transition is continuous (second-order). For higher field amplitudes $h_{0x}=0.9$ with $D=1.00$ and $h_{0y}=0.1$, there is a sudden change in both order-parameter and energy density. The pronounced minimum of the Binder cumulant supports that the order-disorder transition is first order. Interestingly, in the case of single site ($D$-type) anisotropy, between these two regions, we get an additional narrow region where the amounts of discontinuity of the order parameter and energy density are not so prominent (Fig-\ref{fig:dtype-order-transition}(g) and Fig-\ref{fig:dtype-order-transition}(h)). The energy gap during the order-disorder transition becomes smaller. It may be termed as, {\it weakly  first order} region (marked by green region in Fig-\ref{fig:dtype-phase}(b)).
 
\begin{figure}[h!]

\includegraphics[angle=0,height=15cm,width=5cm]{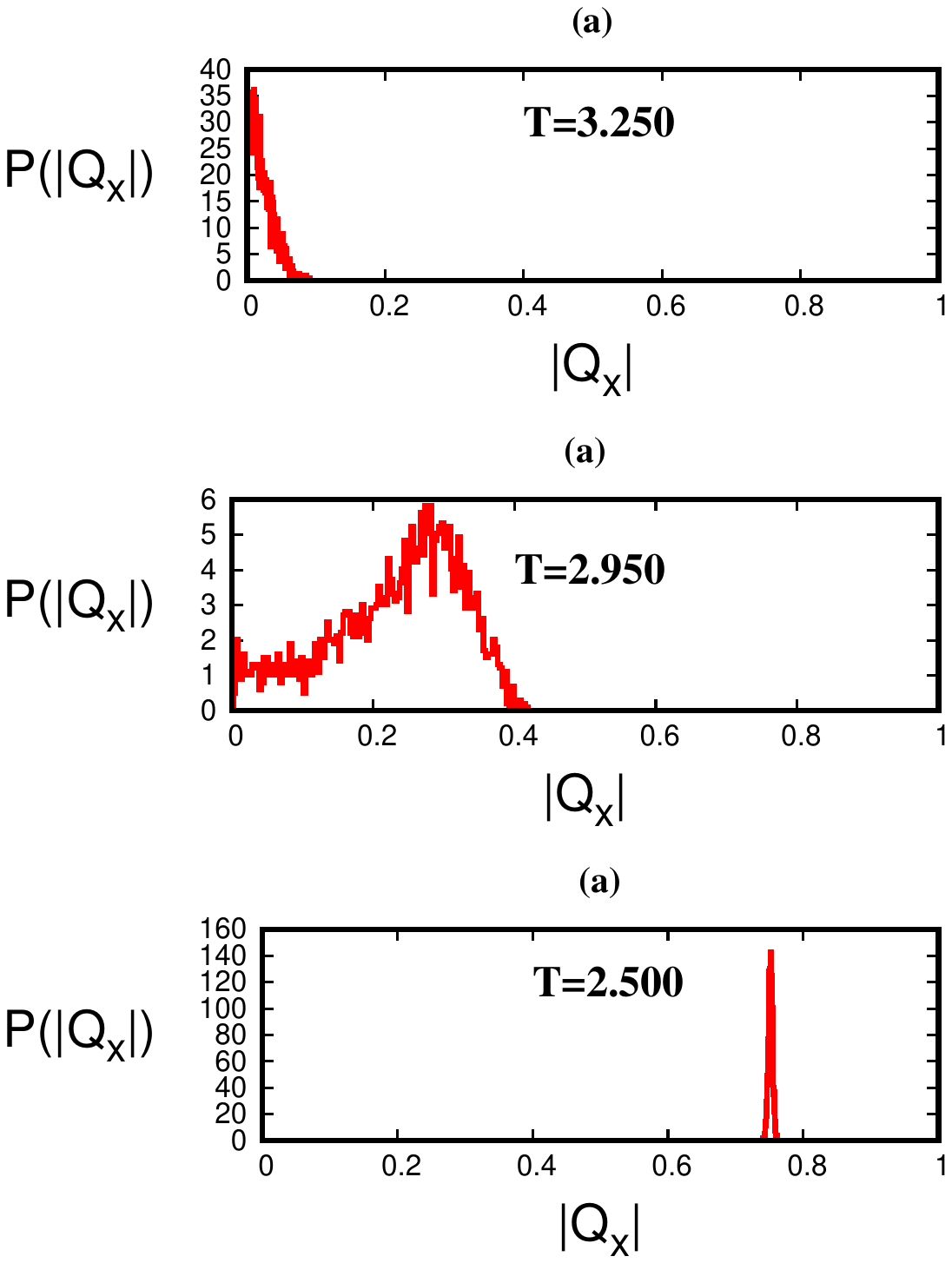}
\includegraphics[angle=0,height=15cm,width=5cm]{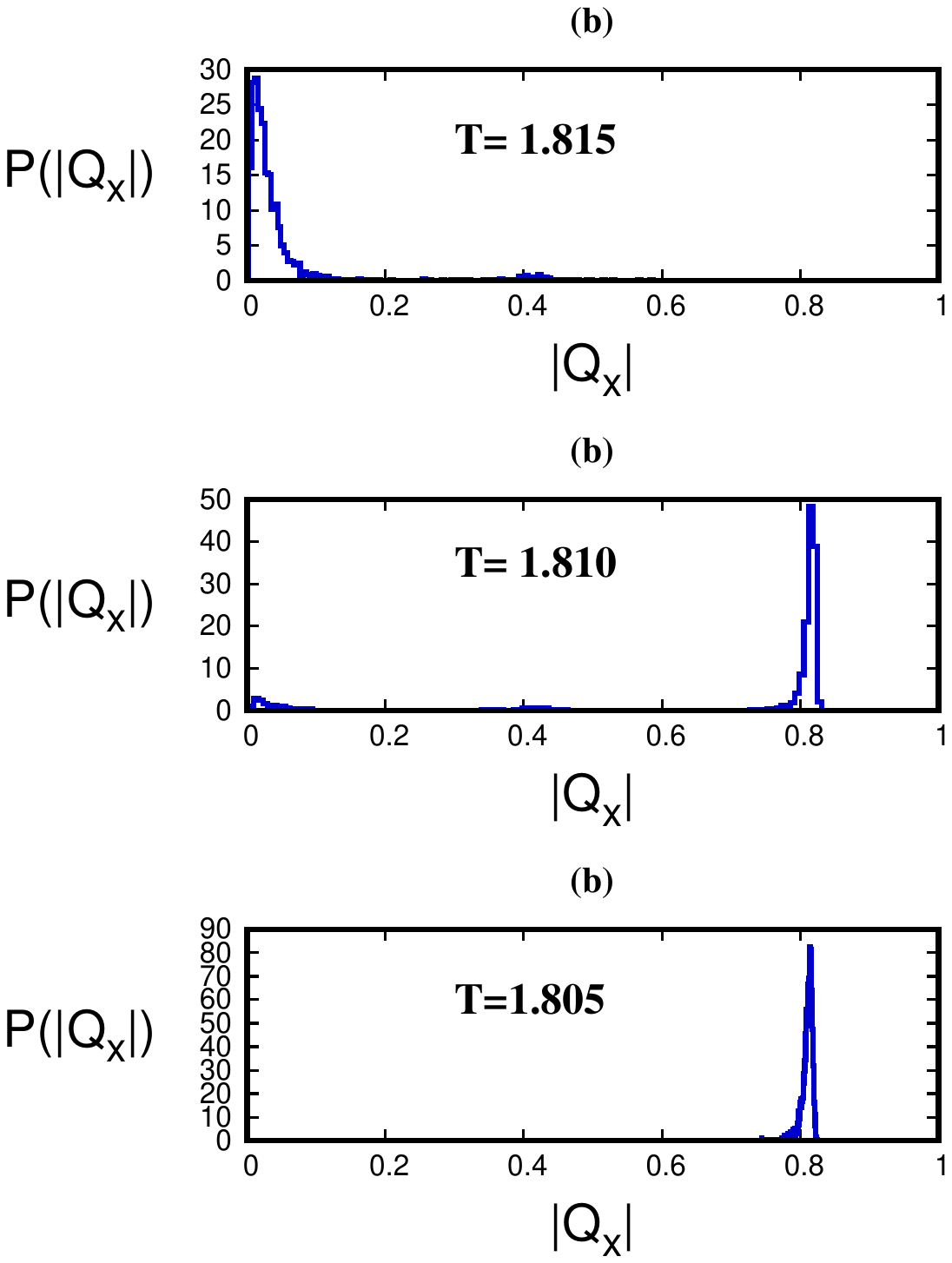}
\includegraphics[angle=0,height=15cm,width=5cm]{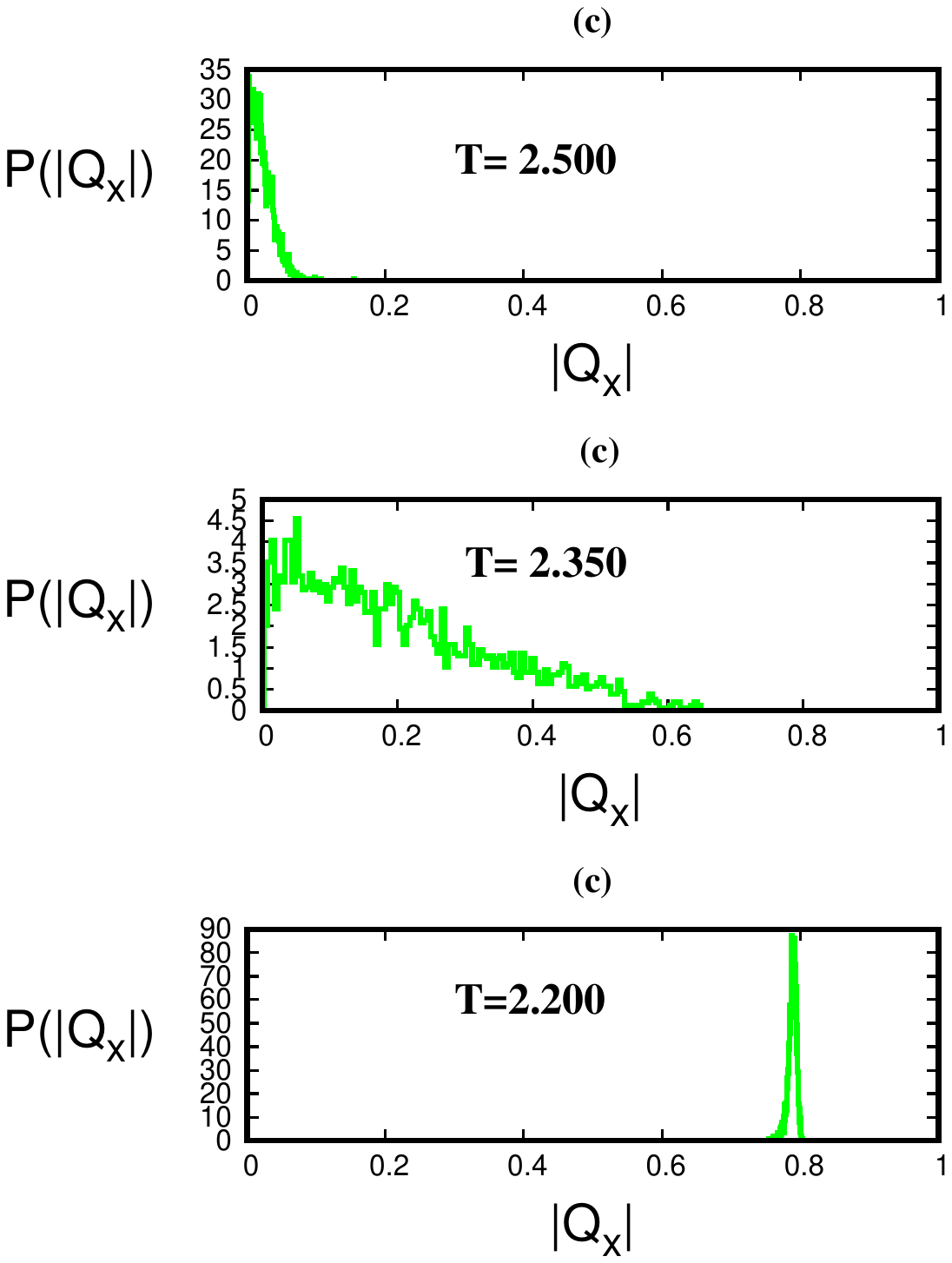}

\caption{The probability density distributions ($P(|Q_x|)$) of the  order parameter $Q_x$ at different temperatures around the transitions. The left panel (a) is for $D=2.00$, $h_{0x}=0.2$, $h_{0y}=0.1$ which corresponds to second order (continuous) phase transition. The middle panel (b) is for $D=1.00$,$h_{0x}=0.9$, $h_{0y}=0.1$ which corresponds to first order (discontinuous) phase transition. The rightmost panel (c) is for $D=1.75$,$h_{0x}=0.7$, $h_{0y}=0.1$ corresponds to the weak first order transition. Here, $L=20$, $f=0.01$ and $\lambda=10$.}
\label{fig:dtype-dist-Qx}
\end{figure}

 The order of the transition is confirmed by the distribution of the order parameter. The statistical distribution of the positive branch of the order parameter $P(|Q_x|)$ is obtained for the lower field amplitude as well as the high field amplitude regime. Fig-\ref{fig:dtype-dist-Qx}(a) shows the distribution of $Q_x$ for field amplitudes $h_{0x}=0.2$, $h_{0y}=0.1$ with anisotropy strength $D=2.00$. As the system is slowly cooled down from a high temperature, at $T=3.25$ ( away from the transition point), the distribution shows a peak around zero, which corresponds to the random orientation of spins. Near the transition temperature $T=2.95$, the system starts to get dynamically ordered. So the peak appears around the non-zero value of $Q_x$. The peak shifts towards a higher value of $Q_x$ well below the transition temperature $T=2.50$. Fig-\ref{fig:dtype-dist-Qx}(b) shows the distribution of the order parameter $P(|Q_x|)$ for higher field amplitude $h_{0x}=0.9$ at anisotropy strength $D=1.00$. At a temperature far from the transition point, all spins are randomly oriented, resulting in a dynamic order parameter of zero value. So there exists a single peak around zero at $T=1.815$. When the system approaches the dynamic phase transition point (for a lower temperature, $T=1.810$), an additional peak emerges, centered around the positive values of $Q_x$. This simultaneous existence of two peaks, in the distribution of the order parameter, is the sign of a first-order or discontinuous phase transition. This result is quite similar to that observed in  the case of bilinear exchange anisotropy. Only the peak at nonzero $Q_x$ survives when the temperature decreases slightly from the transition point. \textcolor{blue}{We have also shown the distribution of the order parameter (in the vicinity of the transition) for weak first order transition in 
 Fig-\ref{fig:dtype-dist-Qx}(c). In the case of single site anisotropy $(D)$ the weak first order transition is found. Please see the  Fig-\ref{fig:dtype-order-transition}(g) (green). It is clear from the figure, that the amount of discontinuity (in the order parameter $\langle Q_x \rangle$) is much less than that for the case of first order transition (Fig-\ref{fig:dtype-order-transition}(d)). The less amount of discontinuity,
 in the case of weak first order transition, is also found in the thermal
 variation of the dynamic energy density $\langle E \rangle$ (compare Fig-\ref{fig:dtype-order-transition}(e) and Fig-\ref{fig:dtype-order-transition}(h)). 
 This less amount of discontinuity, in the case of weak first order transition, is reflected in the distribution of order parameter (shown in Fig-\ref{fig:dtype-dist-Qx}).
For first order transition (Fig-\ref{fig:dtype-dist-Qx}(b)), the large amount of discontinuity is being manifested from the clear separation of two peaks (at $T=1.810$). The probability in bewteen these two peaks is zero. Though the height of zero centered peak is  quite small but it is distinctly identifiable. On the other hand, for weak first order transition (Fig-\ref{fig:dtype-dist-Qx}(c) at $T=2.350$), no such clear discontinuity is found. Widely distributed nonzero probability  is incapable of showing any significant amount of discontinuity. This means the amount of discontinuity is statistically small.}

\begin{figure}[h!]
    \centering
    \includegraphics[angle=0,height=5cm,width=4.5cm]{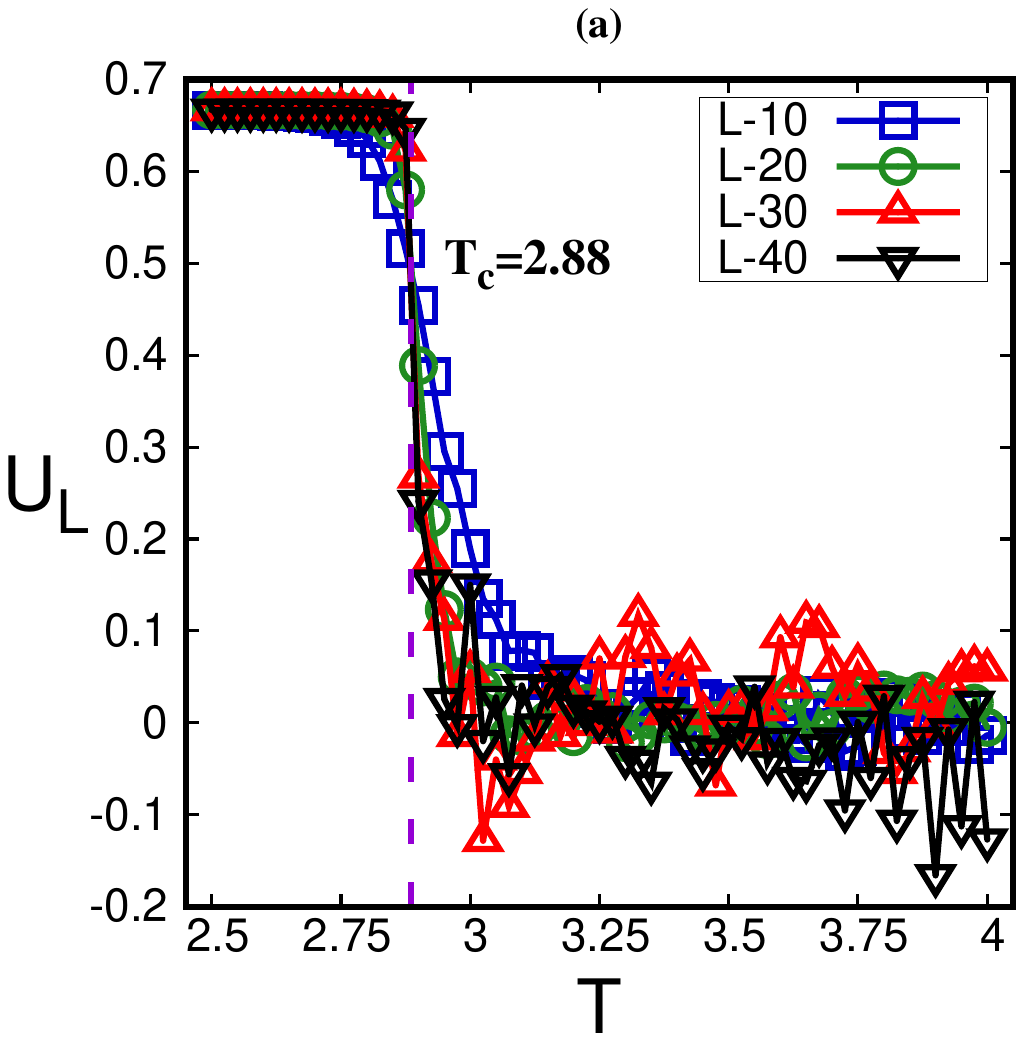}
    \includegraphics[angle=0,height=5cm,width=4.5cm]{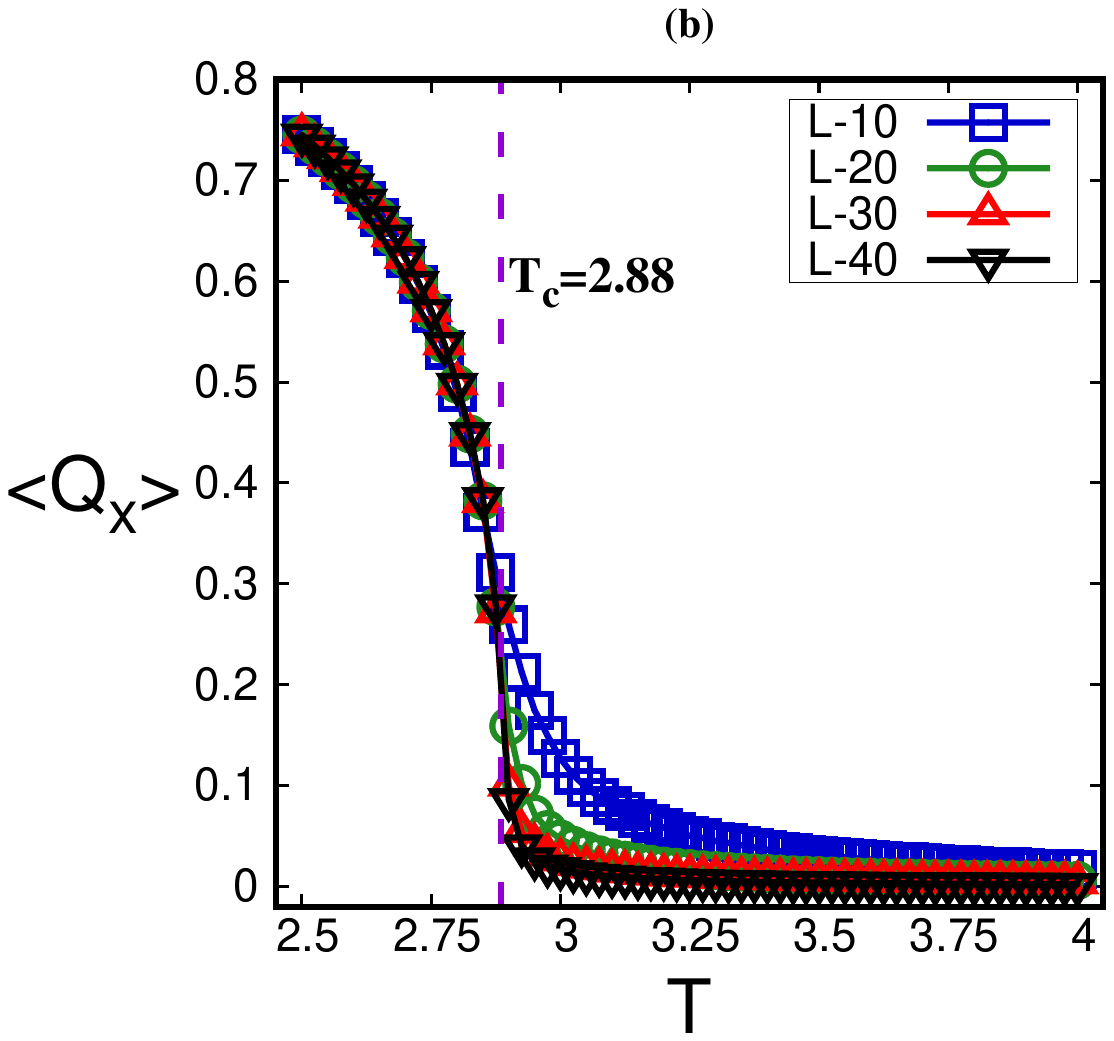}
    \includegraphics[angle=0,height=5cm,width=4.5cm]{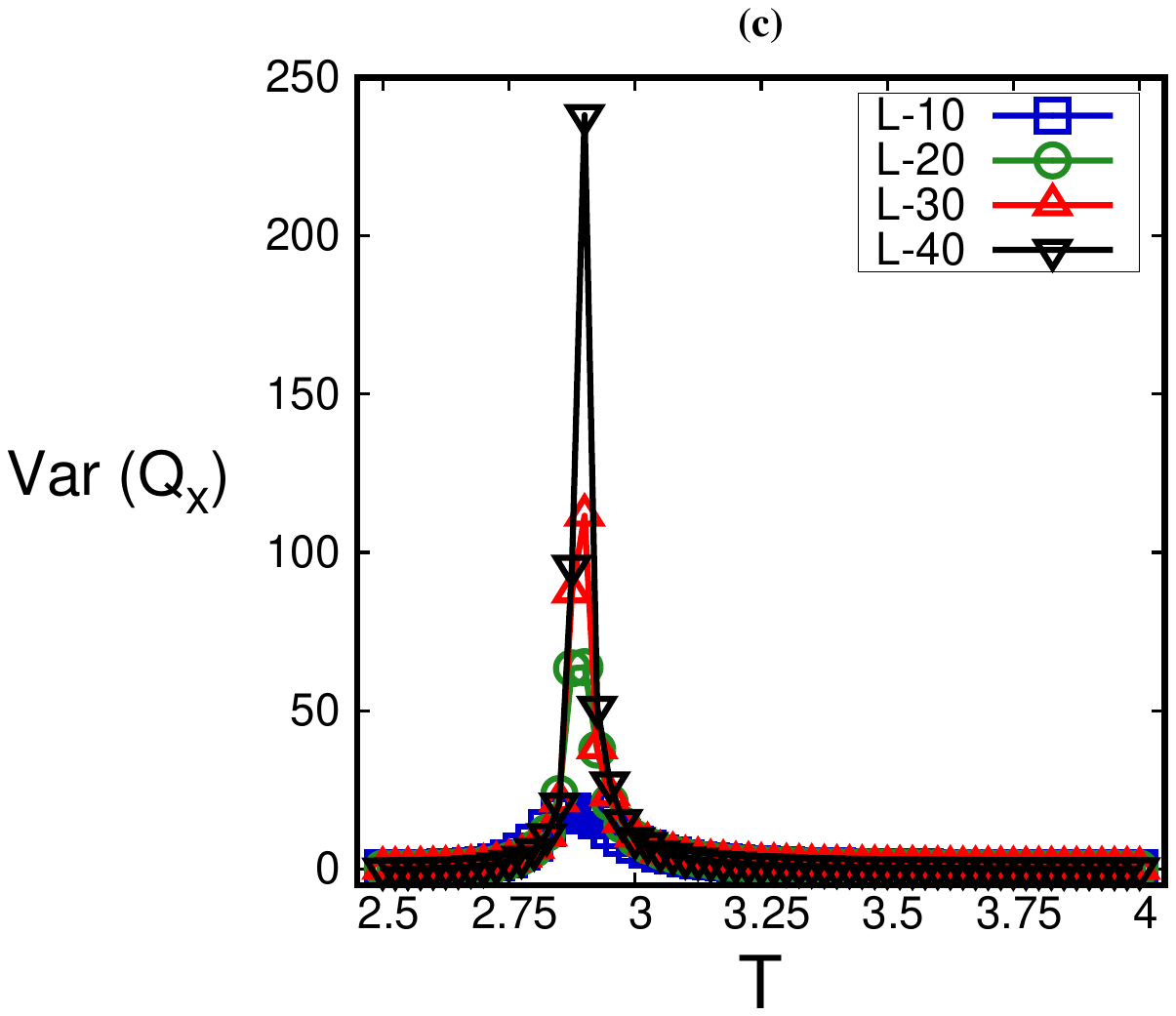}
    \caption{Temperature ($T$) dependences  of (a) the fourth-order Binder cumulant ($U_{L}$), (b) the dynamical order parameter ($\langle Q_x \rangle$) and (c) the Var $(Q_x)$  for different system sizes $L=10,20,30,40$. Here, the anisotropy $D$=2.00 and field amplitudes $h_{0x}=0.3$, $h_{0y}=0.1$, frequency $f=0.01$ and 
    the wavelength $\lambda=10$.}
    \label{fig:dtype-finite}
\end{figure}
\begin{figure}[h!]
    \centering
    \includegraphics[angle=0,height=6cm,width=6cm]{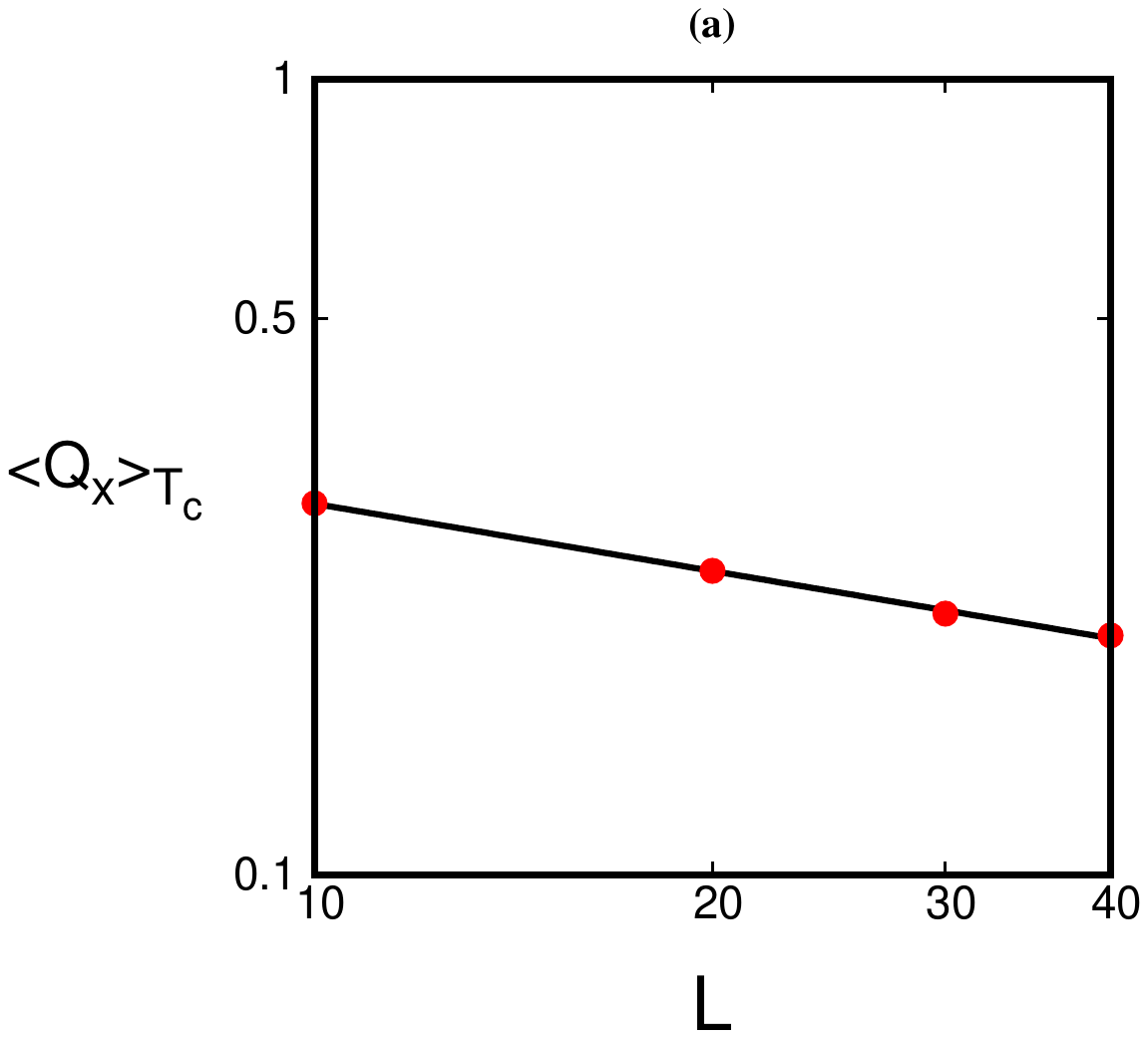}
    \includegraphics[angle=0,height=6cm,width=6cm]{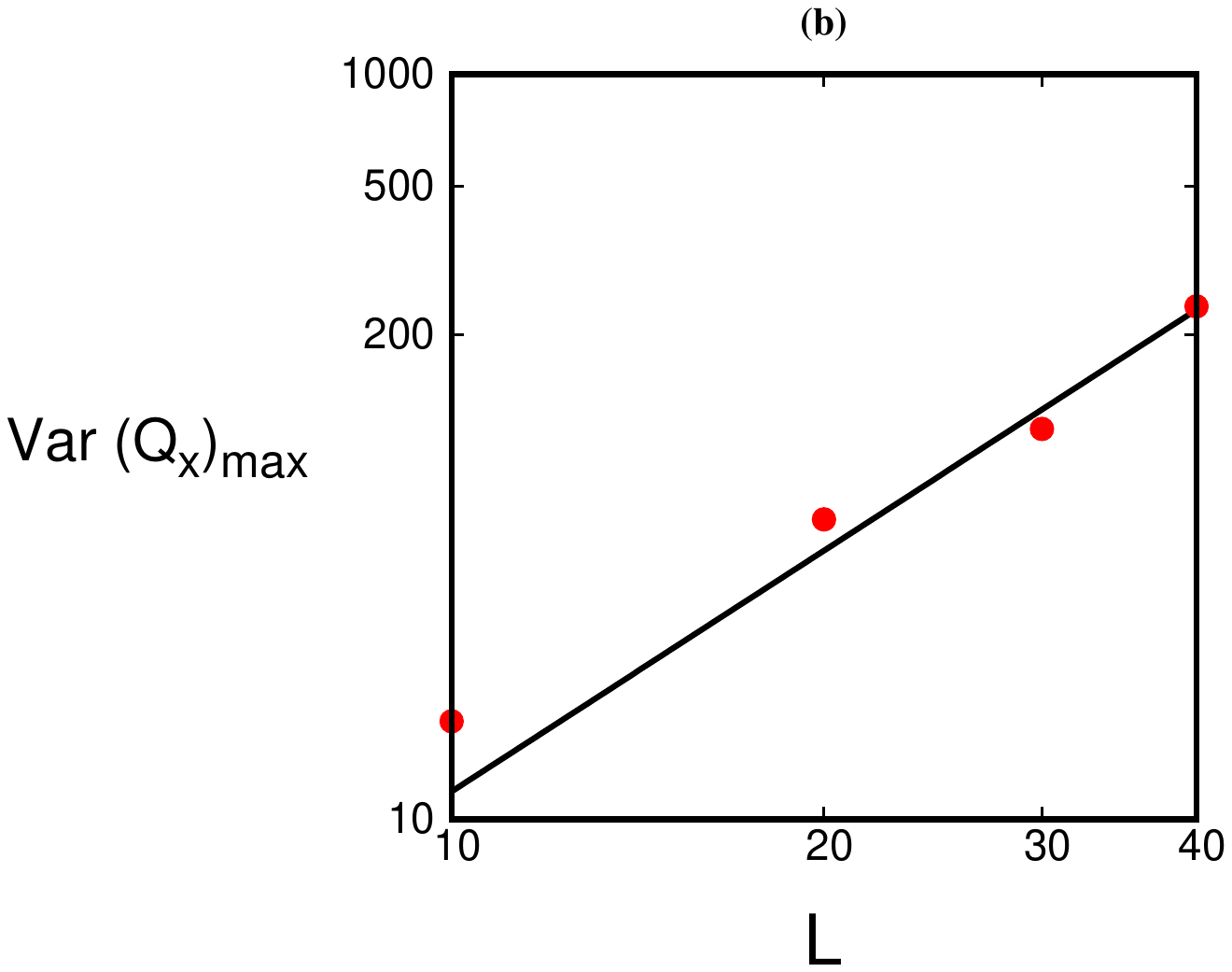}
    
    \caption{The scaling behaviour of dynamic order parameter (a) ($\langle Q_{x} \rangle _{T_c} \sim L^{-{\frac{\beta}{\nu}}}$) at transition point and the scaling behaviour of the maximum of (b) the dynamic susceptibility (Var $(Q_{x})_{max} \sim L^{{\frac{\gamma}{\nu}}})$. Results are shown in log-log scale. The critical exponents estimated $\frac{\gamma}{\nu}$=$0.281\pm 0.007$; $\frac{\gamma}{\nu}$=$2.148\pm 0.288$. Here, $f=0.01$ and $\lambda=10$.}
    \label{fig:dtype-exponent}
\end{figure}

Although the finite size scaling is analyzed for equilibrium phase transition, it can also be applied to systems far from equilibrium. We present the thermal variation of the $\langle Q_x \rangle$, Var($Q_x$) and the fourth order Binder cumulant $U_{L}$ for constant anisotropy strength $D=2.00$ and field amplitudes $h_{0x}=0.3$ and $h_{0y}=0.1$ in Fig-\ref{fig:dtype-finite} for different system sizes ($L=10,20,30,40$). The common intersection point estimates the critical temperature ($T_c=2.88$ here) at the thermodynamic limit ($L \rightarrow \infty$). This is the actual {\it critical} temperature for the non-equilibrium phase transition (in the case of single site anisotropy $D$). We also present in Fig-\ref{fig:dtype-finite}(b) and
Fig-\ref{fig:dtype-finite}(c) the finite-sized (for $L=10,20,30,40$ here) thermal variations of the dynamic order parameter ($\langle Q_{x} \rangle$) and corresponding dynamic susceptibility (Var $(Q_x)$) for the same set of parameter ($D$, $h_{0x}$ and $h_{0y}$) values. The height of the peak of the variance (Var $(Q_x)$) has been found to increase as the system size ($L$) is increased. We present (in Fig-\ref{fig:dtype-exponent}(a)) the values of $\langle Q_x \rangle $ at $T_c(=2.88$) for different system sizes ($L$) in the log-log scale. This has been denoted as $\langle Q_x \rangle _{T_c}$. The solid line fits with the scaling 
form $\langle Q_{x} \rangle _{T_c} \sim L^{-\frac{\beta}{\nu}}$ with the best estimate of the 
critical exponent $\frac{\beta}{\nu}=0.281\pm0.007$. We also show in Fig-\ref{fig:dtype-exponent}(b), the size ($L$) dependence of the height of the peak of Var $(Q_x)$, in the log-log scale. The solid line is fit of the scaling form Var $(Q_{x})_{max} \sim L^{\frac{\gamma}{\nu}}$ with the best estimate of the {\it susceptibility} exponent $\frac{\gamma}{\nu}=2.148\pm0.288$. 

\begin{figure}[h!]
    \centering
   \includegraphics[angle=0,height=6cm,width=6cm]{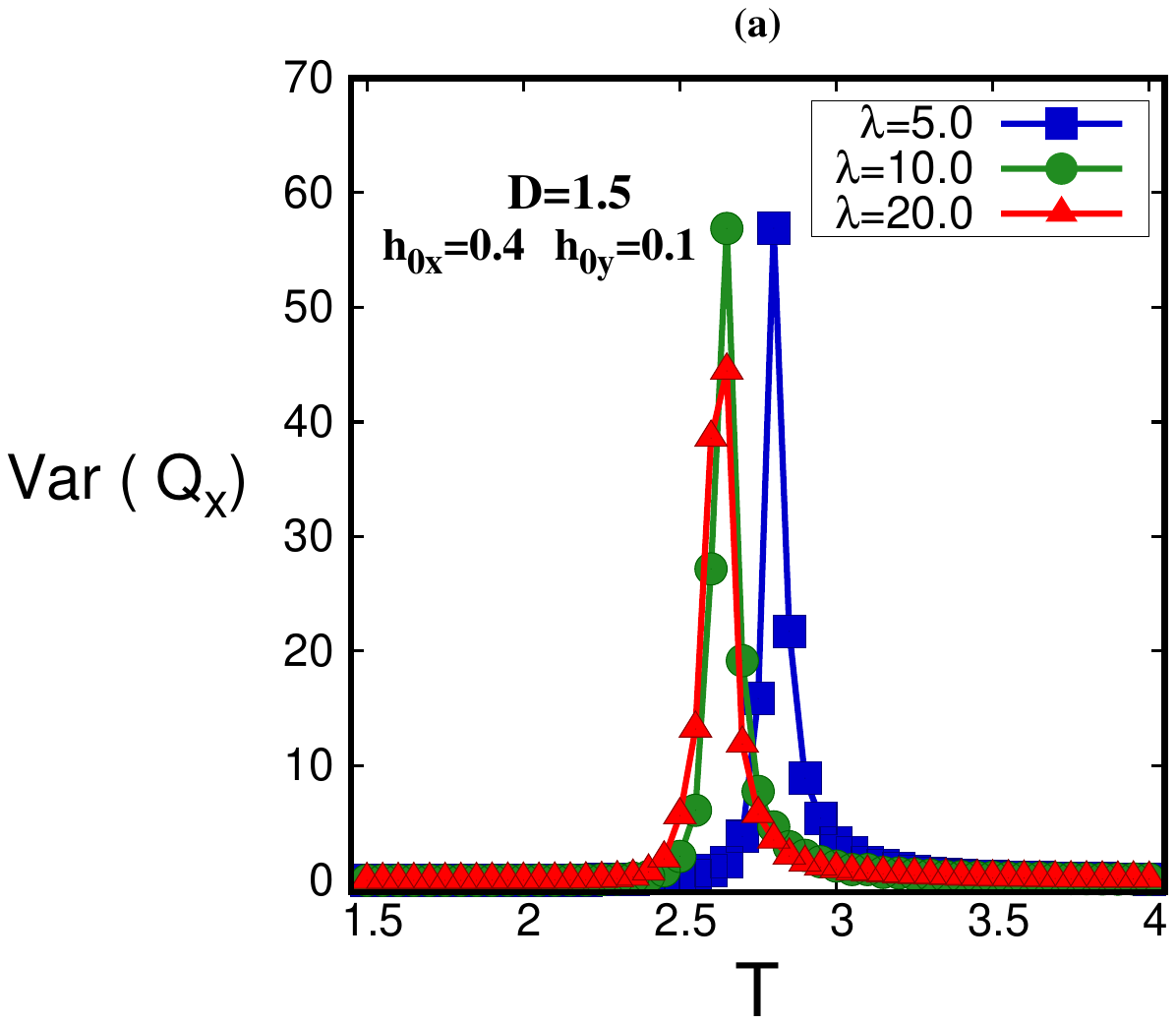}
  \includegraphics[angle=0,height=6cm,width=6cm]{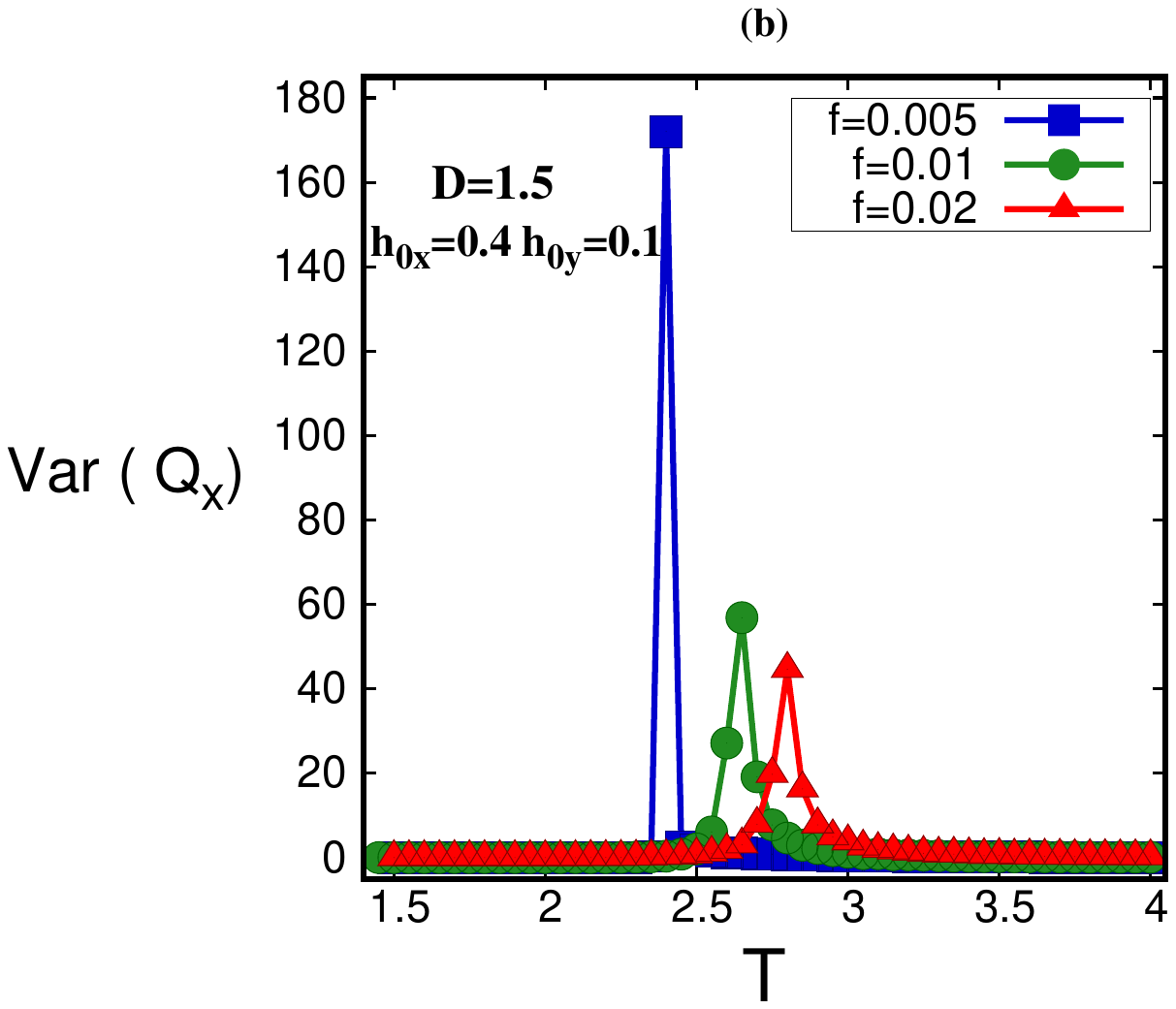}
    
    \caption {(a) The Var $(Q_x)$ is plotted against the temperature ($T$) for different values of wavelengths ($\lambda$). (b) The Var $(Q_x)$ is plotted against the temperature ($T$) for different values of frequencies ($f$). Here, $L=20$.}
    
   \label{fig:dtype-freq-wave}
\end{figure}

Let us conclude our discussions of the results for the single-site anisotropy ($D$)  by representing the variation of the transition temperature (peak position of Var $(Q_x)$)  with the frequency as well as with the wavelength of the perturbing field. For constant anisotropy strength $D=1.5$ and field amplitude $h_{0x}=0.4$, $h_{0y}=0.1$, the peak position of Var $(Q_{x})$ shifts towards right (higher temperature side) if we increase frequency from $f=0.005$ to $f=0.02$. The transition temperature increases with the increase in frequency. Conversely, the transition points shift to lower temperatures as the wavelength increases. These results are shown in Fig-\ref{fig:dtype-freq-wave}.

\vskip 0.8cm
\noindent {\bf IV. Summary:}

\vskip 0.5cm

\noindent In equilibrium statistical physics, the phase transition is a fascinating field of research. The phase transition may be discontinuous (first order) or continuous (second order) . But the scenario where both types of phase transitions occur is certainly interesting. This behavior is usually known as tricritical behavior. The tricritical behavior was seen in the metamagnetic phase transition that occurs in FeCl$_2$ sample\cite{fecl2} . Can one expect tricritical behavior in a nonequilibrium phase transition?

We have investigated the three-dimensional anisotropic XY ferromagnet driven by an elliptically polarized propagating magnetic field wave using the Monte Carlo simulation technique. This propagating magnetic field wave keeps the system far from equilibrium because of its spatio-temporal variation. Upon cooling the system from a high-temperature random phase, the time averaged magnetization component has been observed to take a non-zero value at the finite temperature, depending on the values of the strength of anisotropy and the amplitude of the polarized propagating magnetic field wave. We have shown the phase diagram as an image plot of pseudocritical temperature as a function of the strength of anisotropy and the amplitude of the elliptically polarized propagating magnetic field wave. In the case of the bilinear exchange kind of anisotropy, we have identified two distinct regions of continuous and discontinuous nonequilibrium transitions. An accompanying phase diagram clearly indicates two distinct regions of two different orders of the transitions. The boundary of such two distinctly different regions is the {\it tricritical line} on the surface separating the two nonequilibrium phases. The order of the transitions is confirmed by the variations of Binder cumulant and the statistical distributions of the order parameter near the transition temperature. The discontinuous transition is mainly observed for high values of the field amplitudes, where the stronger field drives the spin system mechanically, which wins over (discontinuous change) the thermal variations (continuous change). The transitions are confirmed (in the region of continuous transition) by finite size analysis. Our data obey the proposed scaling form $\langle Q_{x}\rangle _{T_c} \sim L^{-\frac{\beta}{\nu}}$ with the estimated critical exponent $\frac{\beta}{\nu}=0.383 \pm 0.032$. The proposed scaling form of Var $(Q_{x})_{max} \sim L^{\frac{\gamma}{\nu}}$ estimated the exponent $\frac{\gamma}{\nu}=1.847 \pm 0.019$.

We have also studied the non-equilibrium phase transition of such a driven XY ferromagnet with single-site anisotropy. We have seen an almost similar kind of behavior and shown the phase diagram by image plot of the pseudocritical temperature as a function of the strength of anisotropy and the amplitude of the elliptically polarized propagating magnetic field wave. Here also, we have identified two different regions of continuous and discontinuous transitions. But these two regions are separated by a very narrow region with a weakly first order transition.
The weak first order means the amount of discontinuity
of the order parameter at the transition is quite small. It may be mentioned here that the weak first order transition has recently been observed in the Ising model in frustrated square lattice\cite{gangat} challenging the previous results\cite{jin} of pseudo-first order transition.
The magnetically driven Ising model also gives rise to various interesting nonequilibrium responses \cite{wang1,jiang2}.  The finite size analysis shows the proposed scaling form $\langle Q_{x}\rangle_{T_c} \sim L^{-\frac{\beta}{\nu}}$ with the estimated critical exponent $\frac{\beta}{\nu}=0.281 \pm 0.007$. The proposed scaling form of Var $(Q_{x})_{max} \sim L^{\frac{\gamma}{\nu}}$ fits with the best estimate of the exponent $\frac{\gamma}{\nu}=2.148\pm0.288$. The critical exponents estimated for two different types of anisotropy prompted us to state that the universality classes of the non-equilibrium phase transitions are different for two different types of anisotropies.

It may be noted here, that the forms of the Hamiltonian (Eqn-1 and Eqn-2) are invariant under the transformations $x \to y$, $\Omega \to  -\Omega$ (in Eqn-1) or $D \to -D$ (in Eqn-2).
 For positive values of $\Omega$(Eqn-1) or $D$ (eqn-2), the 
x-direction is the favourable direction for ordering. As a result, here, the nonequilibrium ordering takes place with respect to the x-component of the order parameter. The same results are expected to observe in the y-component for negative values of $\Omega$ (eqn-1) or $D$ (eqn-2). We have considered here only the positive values of $\Omega$ and $D$.

It is important to note that, the susceptibility and the specific heat
did not get peaked at the same temperatures. 
We have to keep in mind that we are studying the nonequilibrium behaviours. The usual partition function formalism (in equilibrium Statistical Mechanics) is not applicable here. The susceptibility defined in the present study is not the conventional susceptibility in equilibrium case (the fluctuation is not divided by $k_BT$ here). Here, we are getting the transition where the fluctuation becomes maximum. On the other hand, the specific heat defined here is ${{d \langle E \rangle} \over {dT}}$ (which will not be equal to the fluctuation of energy divided by $k_BT^2$ as obtained in equilibrium statistical mechanics). That is why we should not expect that both susceptibility and specific heat will be peaked exactly at the same temperature. However, they are peaked here approximately at the same temperature (within the limit of maximum error in the estimation of pseudocritical temperature).

We have also studied the variations of the non-equilibrium pseudocritical temperatures with the wavelength ($\lambda$) and the frequency ($f$) of the elliptically polarized propagating magnetic field for both (bilinear exchange and single site) kinds of anisotropies. In both cases, the transition was found to take place (for fixed anisotropy, field amplitude, and frequency) at a lower temperature for longer wavelengths. The lower frequency (for fixed field amplitude, wavelength, and anisotropy) compels the system to transit at a lower temperature.
\vskip 0.5cm

\noindent {\bf Acknowledgements:} Olivia Mallick acknowledges MANF,UGC, Govt. of India for financial support. 

\noindent {\bf Data availability statement:} Data will be available on request to Olivia Mallick.

\noindent {\bf Conflict of interest statement:} We declare that this manuscript is free from any conflict of
interest. The authors have no financial or proprietary interests in any material discussed in this
article.

\noindent {\bf Funding statement:} No funding was received particularly to support this work.

\noindent {\bf Authors’ contributions:} Olivia Mallick-developed the code, collected the data, prepared the
figures, analysed the results, wrote the manuscript. Muktish Acharyya-conceptualized the problem,
developed the code, analysed the results, wrote the manuscript.


\newpage

\newpage

\newpage

\newpage
\newpage
\newpage

\newpage

\newpage

\newpage

\newpage
\newpage

\newpage
\newpage

\end{document}